\pdfoutput=1
\pdfsuppressptexinfo=\numexpr 1+2+4+8+16+32+64+128+256+512 \relax

\documentclass[a4paper,oneside,fontsize=10pt,parskip=half,fleqn,numbers=noenddot]{scrartcl}

\RequirePackage{scrhack}

\KOMAoptions{headings=standardclasses}

\RequirePackage[left={2.5cm},right={2cm},top={1.5cm},bottom={1.5cm},bindingoffset={0cm},includeheadfoot]{geometry}

\RequirePackage[utf8]{inputenc}
\RequirePackage[T1]{fontenc}

\RequirePackage{lmodern}
\RequirePackage[ngerman,english]{babel}

\RequirePackage{graphicx}

\RequirePackage[pdftoolbar=true,pdfmenubar=true,pdfstartview=FitV,pdfdisplaydoctitle=true,pdfencoding=auto,final]{hyperref}
\RequirePackage[open,openlevel=1,final]{bookmark}

\hypersetup{
	colorlinks			= true,
	allcolors			= black,
}

\RequirePackage{amsmath}
\RequirePackage{amsfonts}
\RequirePackage{amssymb}
\RequirePackage{amsthm}

\RequirePackage{abstract}

\RequirePackage[normal,nooneline,format=plain,singlelinecheck=off,labelfont={bf}]{caption}
\RequirePackage{floatrow}

\RequirePackage[singlespacing]{setspace}

\RequirePackage[final]{microtype}

\RequirePackage{csquotes}

\RequirePackage[backend=bibtex,style=alphabetic,hyperref=true]{biblatex}

\DeclareFieldFormat[article]{volume}{\textbf{#1}\setunit*{\space}}
\renewbibmacro*{volume+number+eid}{%
	\printfield{volume}%
	\setunit*{\addcolon}%
	\printfield{number}%
	\setunit{\addcomma\space}%
}

\AtEveryBibitem{\iffieldundef{journaltitle}{}{\clearfield{eprint}}}
\AtEveryCitekey{\iffieldundef{journaltitle}{}{\clearfield{eprint}}}
	
\AtEveryBibitem{\clearfield{issue}}
\AtEveryCitekey{\clearfield{issue}}
	
\renewbibmacro{in:}{
	\ifentrytype{article}{}{\printtext{\bibstring{in}\intitlepunct}}%
}
	
\DeclareFieldFormat[article]{pages}{#1}
	
\renewbibmacro*{journal}{%
	\iffieldundef{shortjournal}%
		{%
			\iffieldundef{journaltitle}%
			{}%
			{\printtext[journaltitle]%
				{%
					\printfield[titlecase]{journaltitle}%
					\setunit{\subtitlepunct}%
					\printfield[titlecase]{journalsubtitle}%
				}%
			}%
		}%
		{%
			\printtext[journaltitle]%
				{%
					\printfield[titlecase]{shortjournal}%
				}%
		}%
}

\DeclareCiteCommand{\fullcite}
{\usebibmacro{prenote}}
{\usedriver
	{\defcounter{maxnames}{12}}
	{\thefield{entrytype}}}
{\multicitedelim}
{\usebibmacro{postnote}}

\DeclareFieldFormat{eprint:biorxiv}{%
	bioRxiv\addcolon\space
	\ifhyperref
		{\href{https://doi.org/10.1101/#1}{%
			\nolinkurl{#1}%
			\iffieldundef{eprintclass}
				{}
				{\addspace\texttt{\mkbibbrackets{\thefield{eprintclass}}}}}}
		{\nolinkurl{#1}%
			\iffieldundef{eprintclass}
				{}
				{\addspace\texttt{\mkbibbrackets{\thefield{eprintclass}}}}}}
\DeclareFieldAlias{eprint:bioRxiv}{eprint:biorxiv}

\DeclareFieldFormat{eprint:medrxiv}{%
	medRxiv\addcolon\space
	\ifhyperref
		{\href{https://doi.org/10.1101/#1}{%
			\nolinkurl{#1}%
			\iffieldundef{eprintclass}
				{}
				{\addspace\texttt{\mkbibbrackets{\thefield{eprintclass}}}}}}
		{\nolinkurl{#1}%
			\iffieldundef{eprintclass}
				{}
				{\addspace\texttt{\mkbibbrackets{\thefield{eprintclass}}}}}}
\DeclareFieldAlias{eprint:medRxiv}{eprint:medrxiv}

\RequirePackage[noblocks]{authblk}

\newcommand{\orcidlink}[1]{\href{https://orcid.org/#1}{\includegraphics{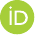}}}

\RequirePackage{xifthen}

\RequirePackage{xcolor}

\newcommand{\appendixfigures}{\setcounter{figure}{0}\renewcommand{\thefigure}{A\arabic{figure}}}

\newcommand{\appendixequations}{\setcounter{equation}{0}\renewcommand{\theequation}{A\arabic{equation}}}

\RequirePackage{array}
\newenvironment{grid}[2][t]{\begingroup\setlength{\tabcolsep}{0em}\begin{tabular}[#1]{#2}}{\end{tabular}\endgroup}

\RequirePackage{booktabs}

\RequirePackage{multirow}
\RequirePackage{tabularx}
\RequirePackage{xltabular}

\RequirePackage{enumitem}

\RequirePackage{siunitx}
\title{The influence of a transport process on the epidemic threshold}

\author[1,2]{Christian Kuehn \orcidlink{0000-0002-7063-6173}}
\author[1]{Jan Mölter \orcidlink{0000-0002-5964-6207}}

\affil[1]{Department of Mathematics, Technical University of Munich, Boltzmannstraße 3, 85748 Garching bei München,
Germany}
\affil[2]{Complexity Science Hub Vienna, Josefstädter Straße 39, 1080 Vienna, Austria}

\date{}

\RequirePackage{tikz}
\usetikzlibrary{calc}
\usetikzlibrary{shapes,arrows,decorations.pathmorphing}

\tikzstyle{vertex}=[coordinate,circle,inner sep=0,fill=.,minimum size=1.5mm]
\tikzstyle{open vertex}=[coordinate,circle,inner sep=0,draw=.,minimum size=1.5mm]

\DeclareSymbolFont{dsrom}{U}{dsrom}{m}{n}
\DeclareMathSymbol{\dsI}{3}{dsrom}{"31}

\DeclareSymbolFont{bbold}{U}{bbold}{m}{n}
\DeclareSymbolFontAlphabet{\mathbbold}{bbold}

\newcommand{\mathrelphantom}[1]{\ensuremath{\mathrel{\phantom{#1}}}}
\newcommand{\mathpunctuation}[1]{\ensuremath{\text{#1}}}

\newcommand{\reals}{\ensuremath{\mathbb{R}}}

\newcommand{\Rho}{\ensuremath{\mathrm{P}}}

\renewcommand{\Upsilon}{\ensuremath{\mathrm{Y}}}

\renewcommand{\emptyset}{\ensuremath{\varnothing}}
\newcommand{\interior}[1]{\ensuremath{\mathring{#1}}}
\newcommand{\closure}[1]{\ensuremath{\overline{#1}}}

\DeclareMathOperator{\trace}{tr}

\newcommand{\spectrum}[2][]{\ensuremath{\operatorname{spec_{#1}} #2}}

\DeclareMathOperator{\diag}{diag}

\newcommand{\indicator}[1]{\ensuremath{\dsI_{#1}}}

\newcommand{\inverse}[1]{\ensuremath{#1^{-1}}}
\newcommand{\transpose}[1]{\ensuremath{#1^{\top}}}

\newcommand{\suchthat}{\ensuremath{\, : \,}}

\newcommand{\parenth}[1]{\ensuremath{\left( #1 \right)}}
\newcommand{\tparenth}[1]{\ensuremath{( #1 )}}

\newcommand{\tbracket}[1]{\ensuremath{[ #1 ]}}

\newcommand{\anglebr}[1]{\ensuremath{\left\langle #1 \right\rangle}}
\newcommand{\tanglebr}[1]{\ensuremath{\langle #1 \rangle}}

\newcommand{\of}[1]{\ensuremath{\!\parenth{#1}}}
\newcommand{\tof}[1]{\ensuremath{\tparenth{#1}}}

\newcommand{\abs}[1]{\ensuremath{\left\vert #1 \right\vert}}

\newcommand{\norm}[2][]{\ensuremath{\left\Vert #2 \right\Vert_{#1}}}
\newcommand{\tnorm}[2][]{\ensuremath{\Vert #2 \Vert_{#1}}}

\newcommand{\restrict}[1]{\ensuremath{\!\left. \vphantom{x} \right|_{ #1 }}}

\newcommand{\set}[1]{\ensuremath{\left\lbrace #1 \right\rbrace}}
\newcommand{\tset}[1]{\ensuremath{\lbrace #1 \rbrace}}

\renewcommand{\complement}[1]{\ensuremath{#1^{c}}}

\newcommand{\binomial}[2]{\ensuremath{\binom{#1}{#2}}}

\newcommand{\varid}{\ensuremath{{\dsI}}}
\newcommand{\kroneckerdelta}[2]{\ensuremath{\delta_{#1, #2}}}

\newcommand{\derivative}[2][]{\ensuremath{\left.\partial_{#2}\ifthenelse{\isempty{#1}}{\right.}{\right\vert_{#2=#1}}}}
\newcommand{\e}{\mathrm{e}}

\renewcommand{\O}{\ensuremath{\mathcal{O}}}

\newcommand{\probabilitymeasure}{\ensuremath{\mathbb{P}}}

\newcommand{\probability}[2][]{\ensuremath{#1{\probabilitymeasure}\left[ #2 \right]}}
\newcommand{\expectation}[2][]{\ensuremath{#1{\mathbb{E}}\left[ #2 \right]}}

\newcommand{\texpectation}[2][]{\ensuremath{#1{\mathbb{E}}[ #2 ]}}

\newcommand{\condprobability}[3][]{\ensuremath{\probability[#1]{\left. #2 \vphantom{#3} \right\vert #3 }}}
\newcommand{\condexpectation}[3][]{\ensuremath{\expectation[#1]{\left. #2 \vphantom{#3} \right\vert #3 }}}

\newcommand{\link}[1]{\ensuremath{\overset{#1}{\sim}}}

\newcommand{\msclink}[1]{\href{https://zbmath.org/classification/?q=cc:#1}{#1}}

\begin{document}

\maketitle

\begin{abstract}
	By generating transient encounters between individuals beyond their immediate social environment, transport can have a profound impact on the spreading of an epidemic. In this work, we consider epidemic dynamics in the presence of the transport process that gives rise to a multiplex network model. In addition to a static layer, the (multiplex) epidemic network consists of a second dynamic layer in which any two individuals are connected for the time they occupy the same site during a random walk they perform on a separate transport network. We develop a mean-field description of the stochastic network model and study the influence the transport process has on the epidemic threshold. We show that any transport process generally lowers the epidemic threshold because of the additional connections it generates. In contrast, considering also random walks of fractional order that in some sense are a more realistic model of human mobility, we find that these non-local transport dynamics raise the epidemic threshold in comparison to a classical local random walk. We also test our model on a realistic transport network (the Munich \emph{U-Bahn} network), and carefully compare mean-field solutions with stochastic trajectories in a range of scenarios.
	
	\vspace{2\parsep}
	
	\textbf{Keywords} {epidemics on networks} $\cdot$ {multiplex network model} $\cdot$ {mean-field model} $\cdot$ {diffusive transport}
	
	\textbf{Mathematics Subject Classification 2020} \msclink{92D30} $\cdot$ \msclink{37N25} $\cdot$ \msclink{60J28}
\end{abstract}

\tableofcontents

\section{Introduction}

Transport processes such as the movement through a network of airlines on a global~\parencite{hufnagel2004forecast,colizza2007reaction} or a network of different public transport modes on a local scale~\parencite{balcan2011phase,ruan2015integrated} play a crucial role in the spread of an epidemic, and in recent years, there has been an increasing amount of work on that topic~\parencite{li2021modeling}. Events that transiently bring together people that would not otherwise meet and interact in the community impose a genuine risk and can imply a surge of infections driving the epidemic~\parencite{mccloskey2020mass,parnell2020covid,gilat2020covid}.

Traditional modelling of epidemics goes back to the seminal work of \textcite{kermack1927contribution}, who considered a population divided into different compartments, the susceptible (\enquote{$\mathrm{S}$}), the infected (\enquote{$\mathrm{I}$}), and the recovered (\enquote{$\mathrm{R}$}) (or removed -- in the sense that they do not contribute the epidemic spread anymore). The dynamics between the different compartments are governed by a set of ordinary differential equations that can be used to accurately describe the dynamics of many real epidemics in the last century~\parencite{kermack1927contribution,anderson1992infectious,chang2017estimation,dehning2020inferring,prodanov2021analytical}. The classical SIR-model has been subsequently extended to account for more features of epidemic spreading~\parencite{batista2021simulation}, ranging from the addition of single compartments, e.g. for individuals that have been exposed to the contagion and are carriers but not yet infectious~\parencite{li2001global,aleta2020evaluation} or for individuals that are in quarantine~\parencite{horstmeyer2022balancing}, to intricate models that include different disease progressions~\parencite{romano2020beyond,abrams2021modelling}, intervention and containment strategies~\parencite{dashtbali2021compartmental}, and age~\parencite{zhao2020five}. Using data-driven methods, these models can be used to make predictions for the course of an epidemic and to assess different mitigation strategies~\parencite{dehning2020inferring,parino2021modelling,ihme2021modeling}.
A fundamental assumption of these deterministic, compartmental models is population-wide random mixing, even though in a population every individual only has a finite number of contacts~\parencite{keeling2005networks}. Thus, an alternative and more realistic way to model an epidemic is as a stochastic process on a network of social relationships~\parencite{pastorsatorras2015epidemic,kiss2017mathematics}. Akin to the compartmental models, the nodes of the network, generally representing individuals in a population, can be in one of several discrete states, in the classical case, susceptible (\enquote{$\mathrm{S}$}), infected (\enquote{$\mathrm{I}$}), or recovered (\enquote{$\mathrm{R}$}). The disease is transmitted along the links of the network, e.g. representing social relationships, in a probabilistic way. While these network models are considerably more complex than the traditional, compartmental models, depending on the topology of the underlying network, their mean-field limits can often still be reduced to differential equations~\parencite{kiss2017mathematics}, which are more complex as they have to take the network topology into account. In particular, it has been shown that network structure can give rise to new phenomena, such as a vanishing epidemic threshold~\parencite{pastorsatorras2001epidemic}.

The effect of transport processes on epidemic spreading has so far mostly been considered in meta-population models~\parencite{hufnagel2004forecast,colizza2007reaction,balcan2011phase,colizza2007reaction,brockmann2013hidden,ruan2015integrated,linka2020outbreak,calvetti2020metapopulation,chang2021mobility}. In these models, one considers a set of communities in each of which the epidemic spreading is modelled by one of the classical compartmental models together with a network that connects these communities and governs the interactions between the epidemic in the different communities via a flux of the contagion. Importantly, it has been demonstrated that data-driven models of this kind can inform government policies and mitigation strategies~\parencite{chang2021mobility}. In another recent study, a framework for epidemic spreading on a network model of time-varying encounter networks was developed and used to deduce control policies for public transport~\parencite{mo2021modeling}.

In studying complex systems and dynamical systems on networks in general as well as epidemics and contact processes in particular, multilayer network structures have recently attracted an increasing amount of interest~\parencite{dedomenico2013mathematical,kivelae2014multilayer,boccaletti2014structure,dedomenico2016physics,bianconi2018multilayer}. As a special kind of these complex structures, multiplex networks combine several classical networks on top of the same set of nodes in a layered structure together with interlayer links between corresponding nodes in the individual layers~\parencite{boccaletti2014structure}. In the context of epidemic dynamics, multiplex structures have e.g. been used to study the spread across multiple layers~\parencite{saumellmendiola2012epidemic,ferrazdearruda2017disease}, the interplay between awareness of an epidemic and the epidemic itself~\parencite{granell2013dynamical}, or the competition of two contagions that spread in individual layers~\parencite{darabisahneh2014competitive,sanz2014dynamics}. Overall, multilayer network structures do generally give rise to a much richer phenomenology than their classical counterparts~\parencite{dedomenico2016physics}.

Transport dynamics are a central aspect of this work, especially those corresponding to human mobility~\parencite{barbosa2018human} and it has been recognised that the latter follows simple reproducible patterns~\parencite{gonzalez2008understanding}. One key observation in many studies has been that the distribution of jump-sizes tends to be heavy-tailed~\parencite{brockmann2006scaling,jiang2009characterizing}, which is specifically reminiscent of Lévy flights that have been a popular model for human mobility~\parencite{barbosa2018human,zaburdaev2015levy}. At the same time, it has also been shown that popular continuous-time random walk and Lévy-flight models do not satisfactorily describe certain features of human mobility such as the tendency to preferentially return to previously and recently visited sites as well as ultraslow diffusive exploration~\parencite{song2010modelling,barbosa2015effect}. These features correspond to non-Markovian dynamics and microscopic models that account for those have been shown to reproduce mobility patterns observed in data more accurately.

In this work, we introduce a model of epidemic dynamics on a network in the presence of a transport process that will give rise to a multiplex structure. In view of many epidemics for which recovery only leads to temporary immunity, we will consider SIRS-epidemic dynamics, develop a mean-field description and study it from a dynamical systems' point of view, considering its equilibria and how the transport process affects the epidemic threshold.

\section{Model and Results}

In the following, we will define the model and subsequently derive a mean-field description of first and second order, i.e. at the level of individuals and pairs, respectively. We will then go on and show that the mean-field solutions provide very good approximations for direct stochastic simulation of the network dynamics. We will characterise the long-term behaviour of the dynamics towards equilibrium in terms of an epidemic threshold and study the effect transport dynamics have on this threshold.

\subsection{The model}

We consider an epidemic in a population, whose individuals also take part in a transport process, moving through a transport network. In addition to the links every individual has to other individuals, e.g., through social relationships in the population, the transport process will generate additional transient links along which the contagion can spread from one individual to the other.

\begin{figure}[tb]
	\centering

	\includegraphics[scale=1]{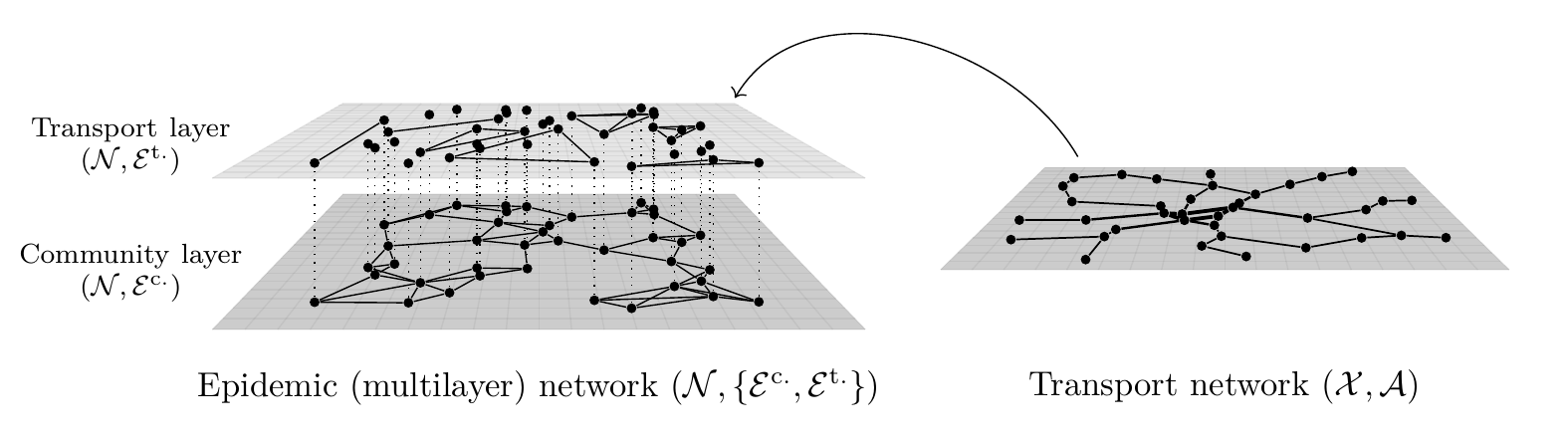}

	\caption{\textbf{A model combining epidemic dynamics on a network with a transport process.} The model consists of two networks, the epidemic network $\left(\mathcal{N}, \left\lbrace \mathcal{E}^{\text{c.}}, \mathcal{E}^{\text{t.}} \right\rbrace\right)$, a two-layer multiplex network structure, on the left and a transport network $\left(\mathcal{X}, \mathcal{A}\right)$ on the right. The two layers of the epidemic network are the community layer $\left(\mathcal{N}, \mathcal{E}^{\text{c.}}\right)$  and the transport layer $\left(\mathcal{N}, \mathcal{E}^{\text{t.}}\right)$. While the topology of the former is assumed to be static, the one of the latter is dynamically changing. The epidemic network as a whole supports the SIRS-epidemic dynamics, where the contagion can spread from one individual to another via either links in the community or the transport layer. At the same time, the individuals of the epidemic network are assumed to move through the transport network, performing a random walk so that whenever two of them occupy the same site they will generate by a link in the transport layer of the epidemic network that persists for as long as they both occupy the same site.}
	\label{fig:model-structure}
\end{figure}

The model the consists of two networks. On one side we have the epidemic network, a multiplex network structure of two layers, $\parenth{\mathcal{N}, \set{\mathcal{E}^{\text{c.}}, \mathcal{E}^{\text{t.}}}}$ with $\mathcal{N}$ the common set of nodes of the multiplex structure and $\mathcal{E}^{\text{c.}}$, $\mathcal{E}^{\text{t.}} \subseteq \mathcal{N} \times \mathcal{N}$ and on the other side the transport network $\parenth{\mathcal{X}, \mathcal{A}}$ with $\mathcal{A} \subseteq \mathcal{X} \times \mathcal{X}$. The two layers of the epidemic network, $\parenth{\mathcal{N}, \mathcal{E}^{\text{c.}}}$ and $\parenth{\mathcal{N}, \mathcal{E}^{\text{t.}}}$, are referred to as the community and transport layer, respectively~(Fig.~\ref{fig:model-structure}).
In terms of the topology, the community layer of the epidemic network is assumed to be a static, undirected, unweighted, simple network, while the topology of the transport layer is not fixed and subject to the transport process on the transport network (see below). The transport network is assumed to be a static, undirected, connected, weighted network, with (weighted) adjacency matrix $A$. Corresponding to the adjacency matrix, we define a transition matrix $\Rho = \inverse{K} A$ as a (right-) stochastic matrix where $K := \diag\tof{k\of{x}}_{x}$ with $k\of{x} = \sum_{x'} A\tof{x,x'}$ is the (weighted) degree-matrix. Regarding the notation, by $A\tof{x,x'}$ we mean the entry of $A$ corresponding at nodes $x$ and $x' \in \mathcal{X}$.

While the epidemic network, the multiplex structure, will support the epidemic dynamics, the transport network will support a transport process. More specifically, we will consider SIRS-epidemic dynamics across the multiplex epidemic network while at the same time letting the individuals perform a Poissonian node-centric continuous-time random walk on the transport network~\parencite{masuda2017random}. This process then defines the topology of the transport layer of the epidemic network. Whenever any two individuals occupy the same site in the transport network, they generate a link between them in the transport layer that persists until one of them leaves the site. As a result of that, the transport layer is a collection of complete networks on subsets of the population corresponding to the individuals at the different sites of the transport network.

To make all this more precise, with every individual $n \in \mathcal{N}$, we associate a state $\parenth{X^{n}_{t}, H^{n}_{t}} \in \mathcal{X} \times \set{\mathrm{S}, \mathrm{I}, \mathrm{R}}$ tracking its site in the transport network and its state of health in the epidemic network through time.
As for the dynamics, for every individual $n$,
\begin{itemize}
	\item $X^{n}_{t} : x \rightarrow x'$ with probability $\Rho\tof{x,x'}$ independently at a uniform, exponential rate $\mu$ and
	\item either $H^{n}_{t} : \mathrm{S} \rightarrow \mathrm{I}$ at an exponential rate $\beta^{\text{c.}}$ for every $n'$ such that $(n',n) \in \mathcal{E}^{\text{c.}}$ and, in addition, at an exponential rate $\beta^{\text{t.}}$ for any $n'$ such that $(n',n) \in \mathcal{E}^{\text{t.}}$, i.e. $X^{n}_{t} = X^{n'}_{t}$ with $n \neq n'$, provided that $H^{n'}_{t} = \mathrm{I}$, or $H^{n}_{t} : \mathrm{I} \rightarrow \mathrm{R}$ at an exponential rate $\gamma$, or $H^{n}_{t} : \mathrm{R} \rightarrow \mathrm{S}$ at an exponential rate $\sigma$.
\end{itemize}
The latter process describing SIRS-dynamics can be seen as a generalisation of the more frequently studied SIS- and SIR-dynamics. More specifically, we recover those in the limits $\sigma \to \infty$ and $\sigma \to 0$, respectively. We will make this precise once we have derived a mean-field description. Intuitively, however, $\frac{1}{\sigma}$ is the average time an individual spends in the state $\mathrm{R}$ once it has transitioned into this state. Thus, in the case of SIS-epidemic dynamics where upon recovery an individual becomes immediately susceptible again, this time can be thought of as arbitrarily small. In contrast, in the case of SIR-epidemic dynamics where an individual gains permanent immunity, it can be thought of as arbitrarily large. These two extremes correspond to the limits $\sigma \to \infty$ and $\sigma \to 0$, respectively. Moreover, it is worthwhile noting that we introduced different rates for infections via the community and the transport layer. This is to account for the fact that encounters in the community are generally different to the ones in public transport in terms of the risk of transmission of a contagion that depends on the external circumstances.

In addition to the epidemic parameters $\beta^{\text{c.}}$, $\beta^{\text{t.}}$, $\gamma$, and $\sigma$ for the infection rates in the community and transport layer, the recovery rate and the immunity loss rate, respectively, we have introduced parameters $\mu$ for the rate at which nodes move through the transport network to control their mobility. Also, we remark that since all of the transitions are subject to homogeneous Poisson-point-processes, the overall process is Markovian.

\subsection{Derivation of the model's mean-field description}

If we disregard the transport process and thus assume the transport layer of the epidemic network to be static, the mean-field description of the epidemic dynamics can be deduced via standard techniques after projecting the epidemic multiplex network down onto a single layer~\parencite{kiss2017mathematics}. Now, since all transitions are governed by homogeneous Poisson-point-processes, at any point in time at most one transition can occur. Therefore, we can take the mean-field description that we obtained by considering the transport process to be frozen in time and add the missing terms for a complete description by conversely considering the dynamics of the transport process in isolation with the epidemic frozen in time. Since the transport process dynamically changes only the links in the transport layer of the epidemic network and consequently does not affect the state of health of the individual nodes, it will manifest itself exclusively in the evolution of the expected number of pairs and, in general, higher-order motifs of the transport layer.

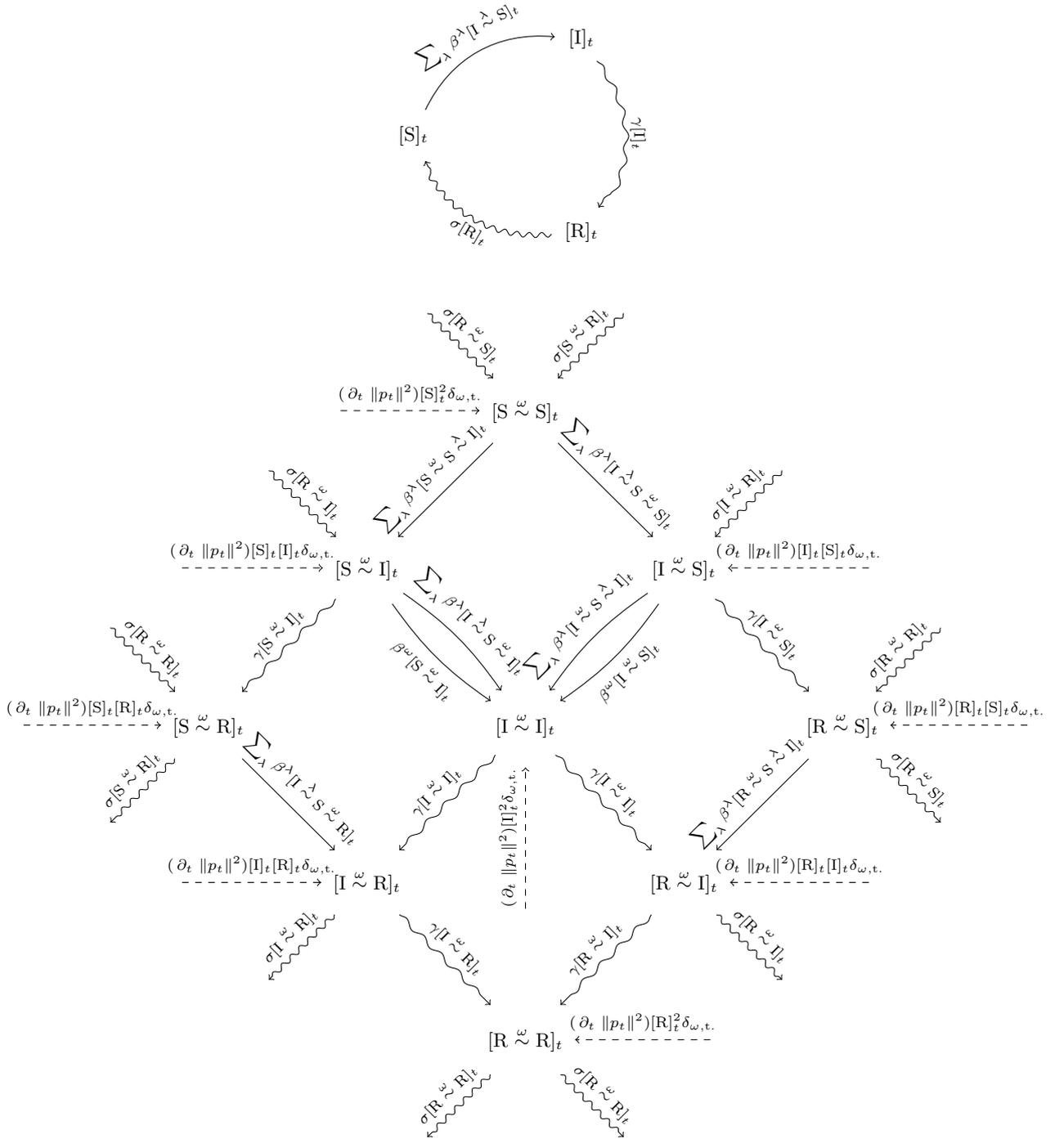
\begin{figure}[tbp]
	\centering

	\vspace{-1cm}

	\begin{grid}{>{\centering\arraybackslash}p{1\linewidth}}
		{
			\centering

\begin{tikzpicture}[scale=0.85,font=\scriptsize,wiggle/.style={decorate,decoration={snake,amplitude=.4mm,segment length=4mm,post length=0.5mm,pre length=0mm}},fast wiggle/.style={decorate,decoration={snake,amplitude=.4mm,segment length=2mm,post length=0.5mm,pre length=0mm,aspect=1}}]

	\useasboundingbox (-2.75,-2.75) rectangle (2.75,2.75);
					   
	\node[shape=circle,font=\small] (S) at (-2.121,0) {$\tbracket{\mathrm{S}}_{t}$};
	\node[shape=circle,font=\small] (I) at (1.061,1.837) {$\tbracket{\mathrm{I}}_{t}$};
	\node[shape=circle,font=\small] (R) at (1.061,-1.837) {$\tbracket{\mathrm{R}}_{t}$};
	
	\draw[->] (S) edge[bend left=35] node[sloped,above] {$\sum_{\lambda} \beta^{\lambda} \tbracket{\mathrm{I}\link{\lambda}\mathrm{S}}_{t}$} (I);
	\draw[->] (I) edge[bend left=35,wiggle] node[sloped,above] {$\gamma \tbracket{\mathrm{I}}_{t}$} (R);
	\draw[->] (R) edge[bend left=35,fast wiggle] node[sloped,below] {$\sigma \tbracket{\mathrm{R}}_{t}$} (S);
	
\end{tikzpicture}
		}
		\\
		{
			\centering

\begin{tikzpicture}[scale=0.85,font=\scriptsize,wiggle/.style={decorate,decoration={snake,amplitude=.4mm,segment length=4mm,post length=0.5mm,pre length=0mm}},fast wiggle/.style={decorate,decoration={snake,amplitude=.4mm,segment length=2mm,post length=0.5mm,pre length=0mm,aspect=1}}]

	\useasboundingbox (-8.5,-8.5) rectangle (8.5,8.5);
                       
	\node[shape=circle,font=\small] (SS) at (0,6) {$\tbracket{\mathrm{S}\link{\omega}\mathrm{S}}_{t}$};
	\node[shape=circle,font=\small] (SI) at (-3,3) {$\tbracket{\mathrm{S}\link{\omega}\mathrm{I}}_{t}$};
	\node[shape=circle,font=\small] (SR) at (-6,0) {$\tbracket{\mathrm{S}\link{\omega}\mathrm{R}}_{t}$};
	\node[shape=circle,font=\small] (IS) at (3,3) {$\tbracket{\mathrm{I}\link{\omega}\mathrm{S}}_{t}$};
	\node[shape=circle,font=\small] (II) at (0,0) {$\tbracket{\mathrm{I}\link{\omega}\mathrm{I}}_{t}$};
	\node[shape=circle,font=\small] (IR) at (-3,-3) {$\tbracket{\mathrm{I}\link{\omega}\mathrm{R}}_{t}$};
	\node[shape=circle,font=\small] (RS) at (6,0) {$\tbracket{\mathrm{R}\link{\omega}\mathrm{S}}_{t}$};
	\node[shape=circle,font=\small] (RI) at (3,-3) {$\tbracket{\mathrm{R}\link{\omega}\mathrm{I}}_{t}$};
	\node[shape=circle,font=\small] (RR) at (0,-6) {$\tbracket{\mathrm{R}\link{\omega}\mathrm{R}}_{t}$};

	\coordinate[shape=circle] (SI*) at (5,-5) ;
	\coordinate[shape=circle] (IS*) at (-5,-5) ;
	\coordinate[shape=circle] (IR*) at (5,5) ;
	\coordinate[shape=circle] (RI*) at (-5,5) ;
	\coordinate[shape=circle] (RS*+) at (-2,8) ;
	\coordinate[shape=circle] (SR*+) at (2,8) ;
	\coordinate[shape=circle] (RR*-) at (-8,2) ;
	\coordinate[shape=circle] (RR*+) at (8,2) ;
	\coordinate[shape=circle] (RS*-) at (-2,-8) ;
	\coordinate[shape=circle] (SR*-) at (2,-8) ;
	\coordinate[shape=circle] (SS*-) at (-8,-2) ;
	\coordinate[shape=circle] (SS*+) at (8,-2) ;
	
	\draw[->] (SS) edge[bend right=0] node[sloped,above] {$\sum_{\lambda} \beta^{\lambda} \tbracket{\mathrm{S}\link{\omega}\mathrm{S}\link{\lambda}\mathrm{I}}_{t}$} (SI);
	\draw[->] (SS) edge[bend right=0] node[sloped,above] {$\sum_{\lambda} \beta^{\lambda} \tbracket{\mathrm{I}\link{\lambda}\mathrm{S}\link{\omega}\mathrm{S}}_{t}$} (IS);
	
	\draw[->] (SI) edge[bend right=0,wiggle] node[sloped,above] {$\gamma \tbracket{\mathrm{S}\link{\omega}\mathrm{I}}_{t}$} (SR);
	\draw[->] (SI) edge[bend right=10] node[sloped,below] {$\beta^{\omega} \tbracket{\mathrm{S}\link{\omega}\mathrm{I}}_{t}$} (II);
	\draw[->] (SI) edge[bend right=-10] node[sloped,above] {$\sum_{\lambda} \beta^{\lambda} \tbracket{\mathrm{I}\link{\lambda}\mathrm{S}\link{\omega}\mathrm{I}}_{t}$} (II);
	
	\draw[->] (IS) edge[bend right=0,wiggle] node[sloped,above] {$\gamma \tbracket{\mathrm{I}\link{\omega}\mathrm{S}}_{t}$} (RS);
	\draw[->] (IS) edge[bend right=10] node[sloped,above] {$\sum_{\lambda} \beta^{\lambda} \tbracket{\mathrm{I}\link{\omega}\mathrm{S}\link{\lambda}\mathrm{I}}_{t}$} (II);
	\draw[->] (IS) edge[bend right=-10] node[sloped,below] {$\beta^{\omega} \tbracket{\mathrm{I}\link{\omega}\mathrm{S}}_{t}$} (II);
	
	\draw[->] (SR) edge[bend right=0] node[sloped,above] {$\sum_{\lambda} \beta^{\lambda} \tbracket{\mathrm{I}\link{\lambda}\mathrm{S}\link{\omega}\mathrm{R}}_{t}$} (IR);

	\draw[->] (II) edge[bend right=0,wiggle] node[sloped,above] {$\gamma \tbracket{\mathrm{I}\link{\omega}\mathrm{I}}_{t}$} (IR);
	\draw[->] (II) edge[bend right=0,wiggle] node[sloped,above] {$\gamma \tbracket{\mathrm{I}\link{\omega}\mathrm{I}}_{t}$} (RI);
	
	\draw[->] (RS) edge[bend right=0] node[sloped,above] {$\sum_{\lambda} \beta^{\lambda} \tbracket{\mathrm{R}\link{\omega}\mathrm{S}\link{\lambda}\mathrm{I}}_{t}$} (RI);
	
	\draw[->] (IR) edge[bend right=0,wiggle] node[sloped,above] {$\gamma \tbracket{\mathrm{I}\link{\omega}\mathrm{R}}_{t}$} (RR);
	
	\draw[->] (RI) edge[bend right=0,wiggle] node[sloped,above] {$\gamma \tbracket{\mathrm{R}\link{\omega}\mathrm{I}}_{t}$} (RR);
	
	\draw[->] (RS*+) edge[bend right=0,fast wiggle] node[sloped,above] {$\sigma \tbracket{\mathrm{R}\link{\omega}\mathrm{S}}_{t}$} (SS);
	\draw[->] (SR*+) edge[bend right=0,fast wiggle] node[sloped,above] {$\sigma \tbracket{\mathrm{S}\link{\omega}\mathrm{R}}_{t}$} (SS);
	
	\draw[->] (RR*+) edge[bend right=0,fast wiggle] node[sloped,above] {$\sigma \tbracket{\mathrm{R}\link{\omega}\mathrm{R}}_{t}$} (RS);
	\draw[->] (RS) edge[bend right=0,fast wiggle] node[sloped,above] {$\sigma \tbracket{\mathrm{R}\link{\omega}\mathrm{S}}_{t}$} (SS*+);
	
	\draw[->] (RR) edge[bend right=0,fast wiggle] node[sloped,above] {$\sigma \tbracket{\mathrm{R}\link{\omega}\mathrm{R}}_{t}$} (SR*-);
	\draw[->] (RR) edge[bend right=0,fast wiggle] node[sloped,above] {$\sigma \tbracket{\mathrm{R}\link{\omega}\mathrm{R}}_{t}$} (RS*-);
	
	\draw[->] (SR) edge[bend right=0,fast wiggle] node[sloped,above] {$\sigma \tbracket{\mathrm{S}\link{\omega}\mathrm{R}}_{t}$} (SS*-);
	\draw[->] (RR*-) edge[bend right=0,fast wiggle] node[sloped,above] {$\sigma \tbracket{\mathrm{R}\link{\omega}\mathrm{R}}_{t}$} (SR);
	
	\draw[->] (RI*) edge[bend right=0,fast wiggle] node[sloped,above] {$\sigma \tbracket{\mathrm{R}\link{\omega}\mathrm{I}}_{t}$} (SI);
	\draw[->] (IR*) edge[bend right=0,fast wiggle] node[sloped,above] {$\sigma \tbracket{\mathrm{I}\link{\omega}\mathrm{R}}_{t}$} (IS);
	\draw[->] (RI) edge[bend right=0,fast wiggle] node[sloped,above] {$\sigma \tbracket{\mathrm{R}\link{\omega}\mathrm{I}}_{t}$} (SI*);
	\draw[->] (IR) edge[bend right=0,fast wiggle] node[sloped,above] {$\sigma \tbracket{\mathrm{I}\link{\omega}\mathrm{R}}_{t}$} (IS*);

	\draw[->,dashed] (SS)++(-3.5,0) -- node [sloped,above] {$\tparenth{\derivative{t} \norm{p_{t}}^{2}} \tbracket{\mathrm{S}}_{t}^{2} \kroneckerdelta{\omega}{\text{t.}}$} (SS);
	\draw[->,dashed] (SI)++(-3.5,0) -- node [sloped,above] {$\tparenth{\derivative{t} \norm{p_{t}}^{2}} \tbracket{\mathrm{S}}_{t} \tbracket{\mathrm{I}}_{t} \kroneckerdelta{\omega}{\text{t.}}$} (SI);
	\draw[->,dashed] (SR)++(-3.5,0) -- node [sloped,above] {$\tparenth{\derivative{t} \norm{p_{t}}^{2}} \tbracket{\mathrm{S}}_{t} \tbracket{\mathrm{R}}_{t} \kroneckerdelta{\omega}{\text{t.}}$} (SR);
	\draw[->,dashed] (IS)++(3.5,0) -- node [sloped,above]  {$\tparenth{\derivative{t} \norm{p_{t}}^{2}} \tbracket{\mathrm{I}}_{t} \tbracket{\mathrm{S}}_{t} \kroneckerdelta{\omega}{\text{t.}}$} (IS);
	\draw[->,dashed] (II)++(0,-3.5) -- node [sloped,above] {$\tparenth{\derivative{t} \norm{p_{t}}^{2}} \tbracket{\mathrm{I}}_{t}^{2} \kroneckerdelta{\omega}{\text{t.}}$} (II);
	\draw[->,dashed] (IR)++(-3.5,0) -- node [sloped,above] {$\tparenth{\derivative{t} \norm{p_{t}}^{2}} \tbracket{\mathrm{I}}_{t} \tbracket{\mathrm{R}}_{t} \kroneckerdelta{\omega}{\text{t.}}$} (IR);
	\draw[->,dashed] (RS)++(3.5,0) -- node [sloped,above]  {$\tparenth{\derivative{t} \norm{p_{t}}^{2}} \tbracket{\mathrm{R}}_{t} \tbracket{\mathrm{S}}_{t} \kroneckerdelta{\omega}{\text{t.}}$} (RS);
	\draw[->,dashed] (RI)++(3.5,0) -- node [sloped,above]  {$\tparenth{\derivative{t} \norm{p_{t}}^{2}} \tbracket{\mathrm{R}}_{t} \tbracket{\mathrm{I}}_{t} \kroneckerdelta{\omega}{\text{t.}}$} (RI);
	\draw[->,dashed] (RR)++(3.5,0) -- node [sloped,above]  {$\tparenth{\derivative{t} \norm{p_{t}}^{2}} \tbracket{\mathrm{R}}_{t}^{2} \kroneckerdelta{\omega}{\text{t.}}$} (RR);

\end{tikzpicture}
		}
		\\
	\end{grid}

	\caption{\textbf{Mean-field transition diagrams for SIRS-epidemic dynamics in a multiplex network up to pair-motifs.} These diagrams capture the mean-field transition rates between individual- and pair-motifs for SIRS-epidemic dynamics adapted to the case of a multiplex network where in layer $\lambda$ the infection rate along $\mathrm{S}$-$\mathrm{I}$-links is $\beta^{\lambda}$ (cf.~\parencite{kiss2017mathematics}) and where additionally links in a transport layer ($\omega = \text{t.}$) are subject to changes due to the transport process studied in this work. The expected number of (directed) pairs of individuals in state of health $h$ and $h'$ connected via a link in layer $\lambda$ is denoted as $\tbracket{h\link{\lambda}h'}$. Importantly, in this case links whose incident nodes have the same state of health are counted twice. Similarly, the expected number of (directed) triples of individuals in states $h$, $h'$ and $h''$ with the first two connected via a link in layer $\lambda$ and the second two via a link in layer $\lambda'$ is denoted as $\tbracket{h\link{\lambda}h'\link{\lambda'}h''}$. The transport process manifests itself in the additional approximate transition terms $\tparenth{\derivative{t} \norm{p_{t}}^{2}} \tbracket{h}_{t} \tbracket{h'}_{t}$ to every pair of nodes in states $h$ and $h'$ connected via a link in the transport layer ($\omega = \text{t.}$), where $\derivative{t} p_{t} = -\mu \transpose{\Delta} p_{t}$ and $\Delta = \varid - \Rho$ the graph Laplacian of the transport network.
	}
	\label{fig:SIRS-mf-transition-diagrams}
\end{figure}

In order to simplify notation, we adopt the notation of \textcite{kiss2017mathematics} and write $\tbracket{h}_{t}$ for the expected number of individuals in state of health $h$, $\tbracket{h\link{\lambda}h'}_{t}$ for the expected number of pairs of individuals in state of health $h$ and $h'$ connected via a link in layer $\lambda$, as well as $\tbracket{h\link{\lambda}h'\link{\lambda'}h''}$ for the expected number of triples of individuals in state of health $h$, $h'$, and $h''$ connected via links in layer $\lambda$ and $\lambda'$, respectively. As we will show in the following, the transport process and with it the varying number of pairs in certain state of health gives rise to the additional approximate transition terms $\tparenth{\derivative{t} \norm{p_{t}}^{2}} \tbracket{h}_{t} \tbracket{h'}_{t}$ to every pair of individuals in state of health $h$ and $h'$ in the transport layer, where $\derivative{t} p_{t} = -\mu \transpose{\Delta} p_{t}$ and $\Delta = \varid - \Rho$ the graph Laplacian of the transport network. The full transition diagrams for a mean-field description of SIRS-epidemic dynamics from which we will later deduce the mean-field differential equations are thus given by the diagrams in Fig.~\ref{fig:SIRS-mf-transition-diagrams}.

Indeed, denoting the number of individuals in a state of health $h$ occupying a site $x$ at time $t$ as $h_{t}\of{x}$, we have that for any $\tau > 0$ sufficiently small,
\begin{equation}
	\begin{split}
		h_{t + \tau}\of{x} &= h_{t}\of{x} + \sum_{n} \tparenth{\kroneckerdelta{x}{X^{n}_{t + \tau}} - \kroneckerdelta{x}{X^{n}_{t}}} \kroneckerdelta{h}{H^{n}_{t}}
	\end{split}
	\mathpunctuation{.}
	\label{eq:individuals-site-transition}
\end{equation}
Since by assumption, individuals occupying the same site generate links between them in a transitive way, denoting the number of links in the transport layer between individuals in state of health $h$ and $h'$ at site $x$ at time $t$ as $\tset{h\link{\text{t.}}h'}\of{x}$, we consequently also have that
\begin{equation}
	\begin{split}
		\tset{h\link{\text{t.}}h'}_{t + \tau}\of{x} &= \tset{h\link{\text{t.}}h'}_{t}\of{x} + \parenth{h_{t}\of{x} \sum_{n} \tparenth{\kroneckerdelta{x}{X^{n}_{t + \tau}} - \kroneckerdelta{x}{X^{n}_{t}}} \kroneckerdelta{h'}{H^{n}_{t}} + h'_{t}\of{x} \sum_{n} \tparenth{\kroneckerdelta{x}{X^{n}_{t + \tau}} - \kroneckerdelta{x}{X^{n}_{t}}} \kroneckerdelta{h}{H^{n}_{t}}} \\
		&\mathrelphantom{=} {} + \sum_{n, n' : n \neq n'} \tparenth{\kroneckerdelta{x}{X^{n}_{t + \tau}} - \kroneckerdelta{x}{X^{n}_{t}}} \tparenth{\kroneckerdelta{x}{X^{n'}_{t + \tau}} - \kroneckerdelta{x}{X^{n'}_{t}}} \kroneckerdelta{h}{H^{n}_{t}} \kroneckerdelta{h'}{H^{n'}_{t}}
	\end{split}
	\mathpunctuation{.}
	\label{eq:individual-pairs-site-transition}
\end{equation}
Indeed, suppose individual $n$ makes a transition $X^{n}_{t} \rightarrow X^{n}_{t + \tau}$. Then, provided $H^{n}_{t} = h$, $X^{n}_{t} = x$, and $X^{n}_{t + \tau} = x'$, $h_{t}\of{x} \rightarrow h_{t + \tau}\of{x} = h_{t}\of{x} - 1$ and $h_{t}\of{x'} \rightarrow h_{t + \tau}\of{x'} = h_{t}\of{x'} + 1$ while everything else remains constant. Thus the net change of $h_{t}\of{x}$ is $\tparenth{\kroneckerdelta{x}{X^{n}_{t + \tau}} - \kroneckerdelta{x}{X^{n}_{t}}} \kroneckerdelta{h}{H^{n}_{t}}$. Since at every point in time each individual can make such a transition, we sum this across all the individuals and obtain the overall net change as claimed in equation~\eqref{eq:individuals-site-transition}.
For links, note that $\tset{h\link{\text{t.}}h'}_{t}\of{x} = h_{t}\of{x} \parenth{h'_{t}\of{x} - \kroneckerdelta{h}{h'}}$ so that with $\phi\tbracket{h_{t}\of{x}} = \sum_{n} \tparenth{\kroneckerdelta{x}{X^{n}_{t + \tau}} - \kroneckerdelta{x}{X^{n}_{t}}} \kroneckerdelta{h}{H^{n}_{t}}$, we have
\begin{equation}
	\begin{split}
		\tset{h\link{\text{t.}}h'}_{t + \tau}\of{x} &= h_{t + \tau}\of{x} \parenth{h'_{t + \tau}\of{x} - \kroneckerdelta{h}{h'}} \\
		&= \tset{h\link{\text{t.}}h'}_{t + \tau}\of{x} + \parenth{h_{t}\of{x} \phi\tbracket{h'_{t }\of{x}} + h'_{t}\of{x} \phi\tbracket{h_{t }\of{x}}} + \phi\tbracket{h_{t}\of{x}} \parenth{\phi\tbracket{h'_{t }\of{x}} - \kroneckerdelta{h}{h'}} \\
	\end{split}
\end{equation}
and since
\begin{equation}
	\phi\tbracket{h_{t}\of{x}} \phi\tbracket{h'_{t}\of{x}} = \sum_{n, n' : n \neq n'} \tparenth{\kroneckerdelta{x}{X^{n}_{t + \tau}} - \kroneckerdelta{x}{X^{n}_{t}}} \tparenth{\kroneckerdelta{x}{X^{n'}_{t + \tau}} - \kroneckerdelta{x}{X^{n'}_{t}}} \kroneckerdelta{h}{H^{n}_{t}} \kroneckerdelta{h'}{H^{n'}_{t}} + \phi\tbracket{h_{t}\of{x}} \kroneckerdelta{h}{h'}
\end{equation}
we obtain the expressions as claimed in equation~\eqref{eq:individual-pairs-site-transition}. At this point, it might be worthwhile mentioning that we are counting the pairs as directed links following the convention of \textcite{kiss2017mathematics}. In particular, this means that any $h$-$h$-link is counted twice.

Now, recalling that for the random walk every individual performs independently through the transport network we have
\begin{equation}
	\condprobability{X^{n}_{t + \tau} = x'}{X^{n}_{t} = x \wedge H^{n}_{t} = h} = \kroneckerdelta{x}{x'} - \mu \tau \Delta\tof{x,x'} \parenth{1 + \O\tof{\tau}}
\end{equation}
where $\Delta = \varid - \Rho$ is the graph Laplacian of the transport network, we compute, by the law of total expectation,
\begin{equation}
	\begin{split}
		\expectation{\sum_{n} \tparenth{\kroneckerdelta{x}{X^{n}_{t + \tau}} - \kroneckerdelta{x}{X^{n}_{t}}} \kroneckerdelta{h}{H^{n}_{t}}} &= \sum_{n} \sum_{h'} \sum_{x', x''} \condprobability{X^{n}_{t + \tau} = x''}{X^{n}_{t} = x' \wedge H^{n}_{t} = h'} \probability{X^{n}_{t} = x' \wedge H^{n}_{t} = h'} \\ &\mathrelphantom{=} {} \times \condexpectation{\tparenth{\kroneckerdelta{x}{X^{n}_{t + \tau}} - \kroneckerdelta{x}{X^{n}_{t}}} \kroneckerdelta{h}{H^{n}_{t}}}{X^{n}_{t} = x' \wedge H^{n}_{t} = h' \wedge X^{n}_{t + \tau} = x''} \\
		&= \sum_{x', x''} \parenth{\kroneckerdelta{x'}{x''} - \mu \Delta\tof{x',x''} \tau \parenth{1 + \O\tof{\tau}}} \parenth{\kroneckerdelta{x}{x''} - \kroneckerdelta{x}{x'}} \texpectation{h_{t}\tof{x'}} \\
		&= - \mu \tau \parenth{1 + \O\tof{\tau}} \sum_{x', x''} \Delta\tof{x',x''} \parenth{\kroneckerdelta{x}{x''} - \kroneckerdelta{x}{x'}} \texpectation{h_{t}\tof{x'}} \\
		&= - \mu \tau \parenth{1 + \O\tof{\tau}} \parenth{ \sum_{x'} \Delta\tof{x',x} \texpectation{h_{t}\tof{x'}} - \parenth{\sum_{x''} \Delta\tof{x,x''}} \texpectation{h_{t}\tof{x'}}} \\
		&= - \mu \tau \parenth{1 + \O\tof{\tau}} \sum_{x'} \transpose{\Delta}\tof{x,x'} \texpectation{h_{t}\tof{x'}}
	\end{split}
\end{equation}
where in the last step we used that $\sum_{x''} \Delta\tof{x,x''} = 1 - \sum_{x''} \Rho\tof{x,x''} = 0$ since $\Rho$ is a stochastic matrix.

Essentially along the same lines, we also find that
\begin{equation}
	\expectation{h_{t}\tof{x} \sum_{n} \tparenth{\kroneckerdelta{x}{X^{n}_{t + \tau}} - \kroneckerdelta{x}{X^{n}_{t}}} \kroneckerdelta{h'}{H^{n}_{t}}} = - \mu \tau \parenth{1 + \O\tof{\tau}} \texpectation{h_{t}\tof{x}} \sum_{x'} \transpose{\Delta}\tof{x,x'} \texpectation{h'_{t}\tof{x'}}
\end{equation}
and
\begin{equation}
	\expectation{\sum_{n, n' : n \neq n'} \tparenth{\kroneckerdelta{x}{X^{n}_{t + \tau}} - \kroneckerdelta{x}{X^{n}_{t}}} \tparenth{\kroneckerdelta{x}{X^{n'}_{t + \tau}} - \kroneckerdelta{x}{X^{n'}_{t}}} \kroneckerdelta{h}{H^{n}_{t}} \kroneckerdelta{h'}{H^{n'}_{t}}} = \O\tof{\tau^{2}} \mathpunctuation{.}
\end{equation}

Thus,
\begin{equation}
	\begin{split}
		\texpectation{h_{t + \tau}\tof{x}} &= \texpectation{h_{t}\tof{x}} - \mu \tau \parenth{1 + \O\tof{\tau}} \sum_{x'} \transpose{\Delta}\tof{x,x'} \texpectation{h_{t}\tof{x'}} \\
	\end{split}
\end{equation}
and
\begin{equation}
	\begin{split}
		\texpectation{\tset{h\link{\text{t.}}h'}_{t + \tau}\tof{x}} = \texpectation{\tset{h\link{\text{t.}}h}_{t}\tof{x}} - \mu \tau \parenth{1 + \O\tof{\tau}} &\left(\texpectation{h_{t}\tof{x}} \sum_{x'} \transpose{\Delta}\tof{x,x'} \texpectation{h'_{t}\tof{x'}} \right. \\ &\left. {} + \texpectation{h'_{t}\tof{x}} \sum_{x'} \transpose{\Delta}\tof{x,x'} \texpectation{h_{t}\tof{x'}}\right) + \O\tof{\tau^{2}} \\
	\end{split}
\end{equation}
so that after passing to the limit $\tau \to 0$, we finally arrive at
\begin{equation}
	\derivative{t} \texpectation{h_{t}\tof{x}} = - \mu \sum_{x'} \transpose{\Delta}\tof{x,x'} \texpectation{h_{t}\tof{x'}}
	\label{eq:evolution-[h(x)]}
\end{equation}
and
\begin{equation}
	\derivative{t} \texpectation{\tset{h\link{\text{t.}}h'}_{t}\tof{x}} = - \mu \parenth{\texpectation{h_{t}\tof{x}} \sum_{x'} \transpose{\Delta}\tof{x,x'} \texpectation{h'_{t}\tof{x'}} + \texpectation{h'_{t}\tof{x}} \sum_{x'} \transpose{\Delta}\tof{x,x'} \texpectation{h_{t}\tof{x'}}}
	\mathpunctuation{.}
	\label{eq:evolution-[h-h'(x)]}
\end{equation}

Since we are only interested in the overall number of $h$-$h'$-links in the transport layer, we obtain that from the latter by the simply taking the sum across all sites $x$. Hence, with $\tbracket{h\tof{x}}_{t} = \texpectation{h_{t}\tof{x}}$,
\begin{equation}
	\derivative{t} \tbracket{h\link{\text{t.}}h'}_{t} = - \mu \sum_{x, x'} \tparenth{\transpose{\Delta}\tof{x,x'} + \transpose{\Delta}\tof{x',x}} \tbracket{h\of{x}}_{t} \tbracket{h'\of{x'}}_{t}
	\mathpunctuation{.}
	\label{eq:evolution-[h-h']}
\end{equation}

While this equation is exact, it depends on the precise number of individuals at every site and in every state of health which is not suitable for a complete mean-field description at the population level. Thus, assuming the site a given individual is occupying is independent of its state of health, we make the approximation $\tbracket{h\of{x}}_{t} \approx p_{t}\of{x} \tbracket{h}_{t}$, where $p_{t}\of{x}$ is the probability that an individual performing a random walk through the transport network occupies site $x$ at time $t$, and find using equations~\eqref{eq:evolution-[h(x)]} and \eqref{eq:evolution-[h-h']} 
\begin{equation}
	\left\lbrace
	\begin{split}
		\derivative{t} p_{t} &= -\mu \transpose{\Delta} p_{t} \\
		\derivative{t} \tbracket{h\link{\text{t.}}h'}_{t} &\approx \tparenth{\derivative{t} \norm{p_{t}}^{2}} \tbracket{h}_{t} \tbracket{h'}_{t}
	\end{split}
	\right.
	\label{eq:approximate-evolution-[h-h']} 
\end{equation}
to describe the effect of the transport process on the epidemic network. This completes the argument and explains the additional transition terms in the diagrams in Fig.~\ref{fig:SIRS-mf-transition-diagrams}.

From these diagrams, we can now deduce the mean-field equations, which are
\begin{equation}
	\left\lbrace
	\begin{split}
		\derivative{t} p_{t} &= -\mu \transpose{\Delta} p_{t} \\
		\derivative{t} \tbracket{\mathrm{S}}_{t} &= - \sum_{\lambda} \beta^{\lambda} \tbracket{\mathrm{S}\link{\lambda}\mathrm{I}}_{t} + \sigma \tbracket{\mathrm{R}}_{t} \\
		\derivative{t} \tbracket{\mathrm{I}}_{t} &= \sum_{\lambda} \beta^{\lambda} \tbracket{\mathrm{S}\link{\lambda}\mathrm{I}}_{t} - \gamma \tbracket{\mathrm{I}}_{t} \\
		\derivative{t} \tbracket{\mathrm{R}}_{t} &= \gamma \tbracket{\mathrm{I}}_{t} - \sigma \tbracket{\mathrm{R}}_{t} \\
		\derivative{t} \tbracket{\mathrm{S}\link{\omega}\mathrm{S}}_{t} &= - 2 \sum_{\lambda} \beta^{\lambda} \tbracket{\mathrm{S}\link{\omega}\mathrm{S}\link{\lambda}\mathrm{I}}_{t} + 2 \sigma \tbracket{\mathrm{S}\link{\omega}\mathrm{R}}_{t} + \tparenth{\derivative{t} \norm{p_{t}}^{2}} \tbracket{\mathrm{S}}_{t}^{2} \kroneckerdelta{\omega}{\text{t.}} \\
		\derivative{t} \tbracket{\mathrm{S}\link{\omega}\mathrm{I}}_{t} &= \sum_{\lambda} \beta^{\lambda} \tparenth{\tbracket{\mathrm{S}\link{\omega}\mathrm{S}\link{\lambda}\mathrm{I}}_{t} - \tbracket{\mathrm{I}\link{\omega}\mathrm{S}\link{\lambda}\mathrm{I}}_{t}} - \beta^{\omega} \tbracket{\mathrm{S}\link{\omega}\mathrm{I}}_{t} - \gamma \tbracket{\mathrm{S}\link{\omega}\mathrm{I}}_{t} + \sigma \tbracket{\mathrm{I}\link{\omega}\mathrm{R}}_{t} \\ &\mathrelphantom{=} {} + \tparenth{\derivative{t} \norm{p_{t}}^{2}} \tbracket{\mathrm{S}}_{t} \tbracket{\mathrm{I}}_{t} \kroneckerdelta{\omega}{\text{t.}} \\
		\derivative{t} \tbracket{\mathrm{I}\link{\omega}\mathrm{I}}_{t} &= 2 \parenth{ \sum_{\lambda} \beta^{\lambda} \tbracket{\mathrm{I}\link{\omega}\mathrm{S}\link{\lambda}\mathrm{I}}_{t} + \beta^{\omega} \tbracket{\mathrm{S}\link{\omega}\mathrm{I}}_{t} } - 2 \gamma \tbracket{\mathrm{I}\link{\omega}\mathrm{I}}_{t} + \tparenth{\derivative{t} \norm{p_{t}}^{2}} \tbracket{\mathrm{I}}_{t}^{2} \kroneckerdelta{\omega}{\text{t.}} \\
		\derivative{t} \tbracket{\mathrm{S}\link{\omega}\mathrm{R}}_{t} &= - \sum_{\lambda} \beta^{\lambda} \tbracket{\mathrm{R}\link{\omega}\mathrm{S}\link{\lambda}\mathrm{I}}_{t} + \gamma \tbracket{\mathrm{S}\link{\omega}\mathrm{I}}_{t} + \sigma \tparenth{\tbracket{\mathrm{R}\link{\omega}\mathrm{R}}_{t} - \tbracket{\mathrm{S}\link{\omega}\mathrm{R}}_{t}} + \tparenth{\derivative{t} \norm{p_{t}}^{2}} \tbracket{\mathrm{S}}_{t} \tbracket{\mathrm{R}}_{t} \kroneckerdelta{\omega}{\text{t.}} \\
		\derivative{t} \tbracket{\mathrm{I}\link{\omega}\mathrm{R}}_{t} &= \sum_{\lambda} \beta^{\lambda} \tbracket{\mathrm{R}\link{\omega}\mathrm{S}\link{\lambda}\mathrm{I}}_{t} + \gamma \tparenth{\tbracket{\mathrm{I}\link{\omega}\mathrm{I}}_{t} - \tbracket{\mathrm{I}\link{\omega}\mathrm{R}}_{t}} - \sigma \tbracket{\mathrm{I}\link{\omega}\mathrm{R}}_{t} + \tparenth{\derivative{t} \norm{p_{t}}^{2}} \tbracket{\mathrm{I}}_{t} \tbracket{\mathrm{R}}_{t} \kroneckerdelta{\omega}{\text{t.}} \\
		\derivative{t} \tbracket{\mathrm{R}\link{\omega}\mathrm{R}}_{t} &= 2 \gamma \tbracket{\mathrm{I}\link{\omega}\mathrm{R}}_{t} - 2 \sigma \tbracket{\mathrm{R}\link{\omega}\mathrm{R}}_{t} + \tparenth{\derivative{t} \norm{p_{t}}^{2}} \tbracket{\mathrm{R}}_{t}^{2} \kroneckerdelta{\omega}{\text{t.}} \\
	\end{split}
	\right.
	\label{eq:SIRS-mean-field-equations}
\end{equation}
for $\omega \in \set{\text{c.}, \text{t.}}$, where we do not explicitly highlight that the transition terms due to the transport process are only approximations in view of other approximations that are going to be introduced down the line in the form of moment closures.

These equations are valid for any topology of the epidemic network. However, as indicated earlier, these equations are not closed in the sense that the evolution equations for motifs of first or second order, in turn, depend on motifs of the next higher order. This gives rise to a hierarchy of models that will only terminate at motifs of system size and at this point defeating the point of mean-field equations as a low-dimensional description of the macroscopic behaviour of a system~\parencite{house2011insights}. This is where moment closures enter the scene~\parencite{kuehn2016moment}. By expressing higher-order motifs in terms of lower-order ones, one breaks the hierarchy and obtains a closed system of ordinary differential equations.

In this work, we will only focus on first- and second-order mean-field equations, i.e. equations that have been closed at the level of individuals and at the level of pairs, respectively~\parencite{kiss2017mathematics}. In order to do so, we will from now on assume that the community layer of the epidemic network is a $k$-regular network, i.e. that all its nodes have degree $k$, which greatly simplifies the analysis (for the irregular case, see \hyperref[sec:appendix-irregular-networks]{Appendix}).

\subsubsection{First-order mean-field equations}

In order to derive first-order mean-field equations, we will use the well-known pair-closure~\parencite{kiss2017mathematics}.
For that, we first note that, in addition to the community layer, also the transport layer of the epidemic network is in expectation regular with degree $\norm{p_{t}}^{2} \abs{\mathcal{N}}$.

Indeed, one can verify that the probability for any individual $n$ to have degree $k$ in the transport layer is given as $\binomial{\abs{\mathcal{N}} - 1}{k} \sum_{x} p_{t}\of{x}^{k + 1} \parenth{1 - p_{t}\of{x}}^{\abs{\mathcal{N}} - 1 - k}$. From that we conclude that the expected degree is $\norm{p_{t}}^{2} \parenth{\abs{\mathcal{N}} - 1} \approx \norm{p_{t}}^{2} \abs{\mathcal{N}}$ independent of $n$, for $\abs{\mathcal{N}}$ large.

Thus, the pair-closure is given as
\begin{equation}
	\tbracket{\mathrm{S}\link{\omega}\mathrm{I}}_{t} \approx \frac{\kappa^{\omega}\of{p_{t}}}{\abs{\mathcal{N}}} \tbracket{\mathrm{S}}_{t} \tbracket{\mathrm{I}}_{t}
	\label{eq:regular-pair-closure}
\end{equation}
where $\kappa^{\text{c.}}\of{p} = k$ and $\kappa^{\text{t.}}\of{p} = \norm{p}^{2} \abs{\mathcal{N}}$ are the (expected) degree of the networks in the community and transport layer, respectively.

Applying this moment closure to the system of equations~\eqref{eq:SIRS-mean-field-equations}, we obtain
\begin{equation}
	\left\lbrace
	\begin{split}
		\derivative{t} p_{t} &= -\mu \transpose{\Delta} p_{t} \\
		\derivative{t} \tbracket{\mathrm{S}}_{t} &= - \sum_{\lambda} \beta^{\lambda} \frac{\kappa^{\lambda}\tof{p_{t}}}{\abs{\mathcal{N}}} \tbracket{\mathrm{S}}_{t} \tbracket{\mathrm{I}}_{t} + \sigma \tbracket{\mathrm{R}}_{t} \\
		\derivative{t} \tbracket{\mathrm{I}}_{t} &= \sum_{\lambda} \beta^{\lambda} \frac{\kappa^{\lambda}\tof{p_{t}}}{\abs{\mathcal{N}}} \tbracket{\mathrm{S}}_{t} \tbracket{\mathrm{I}}_{t} - \gamma \tbracket{\mathrm{I}}_{t} \\
		\derivative{t} \tbracket{\mathrm{R}}_{t} &= \gamma \tbracket{\mathrm{I}}_{t} - \sigma \tbracket{\mathrm{R}}_{t} \\
	\end{split}
	\right.
	\label{eq:SIRS-1st-order-mean-field-equations}
\end{equation}
as the first-order mean-field equations that describe the dynamics at the level of individuals with initial conditions $p_{0}$ some probability distribution on the transport network and $\tbracket{\mathrm{S}}_{0}$, $\tbracket{\mathrm{I}}_{0}$, and $\tbracket{\mathrm{R}}_{0}$ such that $\abs{\mathcal{N}} = \tbracket{\mathrm{S}}_{0} + \tbracket{\mathrm{I}}_{0} + \tbracket{\mathrm{R}}_{0}$.

Since $\derivative{t} \sum_{h} \tbracket{h}_{t} = 0$,  the total number of individuals $\sum_{h} \tbracket{h}_{t}$ is a conserved quantity. Therefore, these first-order equations are not independent and the total number of individuals as a constant of motion can be used to derive a reduced system of differential equations that equivalently describe the system by eliminating one of the equations for the susceptible, infected, and recovered individuals.

\subsubsection{Second-order mean-field equations}
Since we have described the dynamics up to the level of pair motifs, we can go one step further and also derive the second-order mean-field equations. For that, we need to close the system of equations~\eqref{eq:SIRS-mean-field-equations} at the level of pairs. In order to do so, we will use the closure
\begin{equation}
	\tbracket{h\link{\omega}S\link{\omega'}\mathrm{I}}_{t} \approx \parenth{1 - \frac{\kroneckerdelta{\omega}{\omega'}}{\kappa^{\omega}\tof{p_{t}}}} \frac{\tbracket{\mathrm{S}\link{\omega}h}_{t} \tbracket{\mathrm{S}\link{\omega'}\mathrm{I}}_{t}}{\tbracket{\mathrm{S}}_{t}}
	\label{eq:regular-triple-closure}
\end{equation}
which is frequently used for regular networks~\parencite{kiss2017mathematics} and where as before we have $\kappa^{\text{c.}}\of{p} = k$ and $\kappa^{\text{t.}}\of{p} = \norm{p}^{2} \abs{\mathcal{N}}$.

With that,
\begin{equation}
	\left\lbrace
	\begin{split}
		\derivative{t} p_{t} &= -\mu \transpose{\Delta} p_{t} \\
		\derivative{t} \tbracket{\mathrm{S}}_{t} &= - \sum_{\lambda} \beta^{\lambda} \tbracket{\mathrm{S}\link{\lambda}\mathrm{I}}_{t} + \sigma \tbracket{\mathrm{R}}_{t} \\
		\derivative{t} \tbracket{\mathrm{I}}_{t} &= \sum_{\lambda} \beta^{\lambda} \tbracket{\mathrm{S}\link{\lambda}\mathrm{I}}_{t} - \gamma \tbracket{\mathrm{I}}_{t} \\
		\derivative{t} \tbracket{\mathrm{R}}_{t} &= \gamma \tbracket{\mathrm{I}}_{t} - \sigma \tbracket{\mathrm{R}}_{t} \\
		\derivative{t} \tbracket{\mathrm{S}\link{\omega}\mathrm{S}}_{t} &= - 2 \sum_{\lambda} \beta^{\lambda} \parenth{1 - \frac{\kroneckerdelta{\omega}{\lambda}}{\kappa^{\omega}\tof{p_{t}}}} \tbracket{\mathrm{S}\link{\lambda}\mathrm{I}}_{t} \frac{\tbracket{\mathrm{S}\link{\omega}\mathrm{S}}_{t}}{\tbracket{\mathrm{S}}_{t}} + 2 \sigma \tbracket{\mathrm{S}\link{\omega}\mathrm{R}}_{t} \\ &\mathrelphantom{=} {} + \tparenth{\derivative{t} \norm{p_{t}}^{2}} \tbracket{\mathrm{S}}_{t}^{2} \kroneckerdelta{\omega}{\text{t.}} \\
		\derivative{t} \tbracket{\mathrm{S}\link{\omega}\mathrm{I}}_{t} &= \sum_{\lambda} \beta^{\lambda} \parenth{1 - \frac{\kroneckerdelta{\omega}{\lambda}}{\kappa^{\omega}\tof{p_{t}}}} \tbracket{\mathrm{S}\link{\lambda}\mathrm{I}}_{t} \frac{\tbracket{\mathrm{S}\link{\omega}\mathrm{S}}_{t} - \tbracket{\mathrm{S}\link{\omega}\mathrm{I}}_{t}}{\tbracket{\mathrm{S}}_{t}} - \beta^{\omega} \tbracket{\mathrm{S}\link{\omega}\mathrm{I}}_{t} - \gamma \tbracket{\mathrm{S}\link{\omega}\mathrm{I}}_{t} + \sigma \tbracket{\mathrm{I}\link{\omega}\mathrm{R}}_{t} \\ &\mathrelphantom{=} {} + \tparenth{\derivative{t} \norm{p_{t}}^{2}} \tbracket{\mathrm{S}}_{t} \tbracket{\mathrm{I}}_{t} \kroneckerdelta{\omega}{\text{t.}} \\
		\derivative{t} \tbracket{\mathrm{I}\link{\omega}\mathrm{I}}_{t} &= 2 \parenth{ \sum_{\lambda} \beta^{\lambda} \parenth{1 - \frac{\kroneckerdelta{\omega}{\lambda}}{\kappa^{\omega}\tof{p_{t}}}} \tbracket{\mathrm{S}\link{\lambda}\mathrm{I}}_{t} \frac{\tbracket{\mathrm{S}\link{\omega}\mathrm{I}}_{t}}{\tbracket{\mathrm{S}}_{t}} + \beta^{\omega} \tbracket{\mathrm{S}\link{\omega}\mathrm{I}}_{t} } - 2 \gamma \tbracket{\mathrm{I}\link{\omega}\mathrm{I}}_{t} \\ &\mathrelphantom{=} {} + \tparenth{\derivative{t} \norm{p_{t}}^{2}} \tbracket{\mathrm{I}}_{t}^{2} \kroneckerdelta{\omega}{\text{t.}} \\
		\derivative{t} \tbracket{\mathrm{S}\link{\omega}\mathrm{R}}_{t} &= - \sum_{\lambda} \beta^{\lambda} \parenth{1 - \frac{\kroneckerdelta{\omega}{\lambda}}{\kappa^{\omega}\tof{p_{t}}}} \tbracket{\mathrm{S}\link{\lambda}\mathrm{I}}_{t} \frac{\tbracket{\mathrm{S}\link{\omega}\mathrm{R}}_{t}}{\tbracket{\mathrm{S}}_{t}} + \gamma \tbracket{\mathrm{S}\link{\omega}\mathrm{I}}_{t} + \sigma \tparenth{\tbracket{\mathrm{R}\link{\omega}\mathrm{R}}_{t} - \tbracket{\mathrm{S}\link{\omega}\mathrm{R}}_{t}} \\ &\mathrelphantom{=} {} + \tparenth{\derivative{t} \norm{p_{t}}^{2}} \tbracket{\mathrm{S}}_{t} \tbracket{\mathrm{R}}_{t} \kroneckerdelta{\omega}{\text{t.}} \\
		\derivative{t} \tbracket{\mathrm{I}\link{\omega}\mathrm{R}}_{t} &= \sum_{\lambda} \beta^{\lambda} \parenth{1 - \frac{\kroneckerdelta{\omega}{\lambda}}{\kappa^{\omega}\tof{p_{t}}}} \tbracket{\mathrm{S}\link{\lambda}\mathrm{I}}_{t} \frac{\tbracket{\mathrm{S}\link{\omega}\mathrm{R}}_{t}}{\tbracket{\mathrm{S}}_{t}} + \gamma \tparenth{\tbracket{\mathrm{I}\link{\omega}\mathrm{I}}_{t} - \tbracket{\mathrm{I}\link{\omega}\mathrm{R}}_{t}} - \sigma \tbracket{\mathrm{I}\link{\omega}\mathrm{R}}_{t} \\ &\mathrelphantom{=} {} + \tparenth{\derivative{t} \norm{p_{t}}^{2}} \tbracket{\mathrm{I}}_{t} \tbracket{\mathrm{R}}_{t} \kroneckerdelta{\omega}{\text{t.}} \\
		\derivative{t} \tbracket{\mathrm{R}\link{\omega}\mathrm{R}}_{t} &= 2 \gamma \tbracket{\mathrm{I}\link{\omega}\mathrm{R}}_{t} - 2 \sigma \tbracket{\mathrm{R}\link{\omega}\mathrm{R}}_{t} \\ &\mathrelphantom{=} {} + \tparenth{\derivative{t} \norm{p_{t}}^{2}} \tbracket{\mathrm{R}}_{t}^{2} \kroneckerdelta{\omega}{\text{t.}} \\
	\end{split}
	\right.
	\label{eq:SIRS-2nd-order-mean-field-equations}
\end{equation}
for $\omega \in \set{\text{c.}, \text{t.}}$ are the second-order mean-field equations that describe the dynamics at the level of pairs with initial conditions $p_{0}$ some probability distribution on the transport network and $\tbracket{\mathrm{S}}_{0}$, $\tbracket{\mathrm{I}}_{0}$, and $\tbracket{\mathrm{R}}_{0}$ such that $\abs{\mathcal{N}} = \tbracket{\mathrm{S}}_{0} + \tbracket{\mathrm{I}}_{0} + \tbracket{\mathrm{R}}_{0}$, as well as
$\tbracket{\mathrm{S}\link{\omega}\mathrm{S}}_{0} = \frac{\kappa^{\omega}\tof{p_{0}}}{\abs{\mathcal{N}}} \tbracket{\mathrm{S}}_{0}^{2}$,
$\tbracket{\mathrm{S}\link{\omega}\mathrm{I}}_{0} = \frac{\kappa^{\omega}\tof{p_{0}}}{\abs{\mathcal{N}}} \tbracket{\mathrm{S}}_{0} \tbracket{\mathrm{I}}_{0}$,
$\tbracket{\mathrm{I}\link{\omega}\mathrm{I}}_{0} = \frac{\kappa^{\omega}\tof{p_{0}}}{\abs{\mathcal{N}}} \tbracket{\mathrm{I}}_{0}^{2}$,
$\tbracket{\mathrm{S}\link{\omega}\mathrm{R}}_{0} = \frac{\kappa^{\omega}\tof{p_{0}}}{\abs{\mathcal{N}}} \tbracket{\mathrm{S}}_{0} \tbracket{\mathrm{R}}_{0}$,
$\tbracket{\mathrm{I}\link{\omega}\mathrm{R}}_{0} = \frac{\kappa^{\omega}\tof{p_{0}}}{\abs{\mathcal{N}}} \tbracket{\mathrm{I}}_{0} \tbracket{\mathrm{R}}_{0}$, and
$\tbracket{\mathrm{R}\link{\omega}\mathrm{R}}_{0} = \frac{\kappa^{\omega}\tof{p_{0}}}{\abs{\mathcal{N}}} \tbracket{\mathrm{R}}_{0}^{2}$.

As it has been for the first-order mean-field equations, these equations are not independent. Indeed, there are several conservation relations with which one can derive a reduced set of differential equations that equivalently describe the full system.
Besides the total number of individuals also the (expected) total number of links in both layers of the epidemic network is conserved since $\derivative{t} \parenth{\sum_{h, h'} \tbracket{h\link{\omega}h'}_{t} - \kappa^{\omega}\tof{p_{t}} \abs{\mathcal{N}}} = 0$. In addition, we also have pair-conservation so that $\sum_{h'} \tbracket{h\link{\omega}h'}_{t} = \kappa^{\omega}\tof{p_{t}} \tbracket{h}_{t}$ for every $h$. To see that, note that
\begin{equation}
	\derivative{t} \begin{pmatrix} \sum_{h'} \tbracket{\mathrm{S}\link{\omega}h'}_{t} - \kappa^{\omega}\tof{p_{t}} \tbracket{\mathrm{S}}_{t} \\ \sum_{h'} \tbracket{\mathrm{I}\link{\omega}h'}_{t} - \kappa^{\omega}\tof{p_{t}} \tbracket{\mathrm{I}}_{t} \\ \sum_{h'} \tbracket{\mathrm{R}\link{\omega}h'}_{t} - \kappa^{\omega}\tof{p_{t}} \tbracket{\mathrm{R}}_{t} \\ \end{pmatrix} = \begin{pmatrix} -U^{\omega}_{t} & 0 & \sigma \\ U^{\omega}_{t} & -\gamma & 0 \\ 0 & \gamma & -\sigma \\ \end{pmatrix}  \begin{pmatrix} \sum_{h'} \tbracket{\mathrm{S}\link{\omega}h'}_{t} - \kappa^{\omega}\tof{p_{t}} \tbracket{\mathrm{S}}_{t} \\ \sum_{h'} \tbracket{\mathrm{I}\link{\omega}h'}_{t} - \kappa^{\omega}\tof{p_{t}} \tbracket{\mathrm{I}}_{t} \\ \sum_{h'} \tbracket{\mathrm{R}\link{\omega}h'}_{t} - \kappa^{\omega}\tof{p_{t}} \tbracket{\mathrm{R}}_{t} \\ \end{pmatrix}
\end{equation}
where $U^{\omega}_{t} = \frac{1}{\tbracket{\mathrm{S}}_{t}} \sum_{\lambda} \beta^{\lambda} \parenth{1 - \frac{\kroneckerdelta{\omega}{\lambda}}{\kappa^{\omega}\tof{p_{t}}}} \tbracket{\mathrm{S}\link{\lambda}\mathrm{I}}_{t}$ from which the claim follows via standard arguments~\parencite[Proposition~4.2]{kiss2017mathematics}.

\subsubsection{SIS- and SIR-epidemic mean-field equations in a singular limit}

As mentioned in the beginning, we can derive the corresponding mean-field descriptions for SIS- and SIR-epidemic dynamics by taking the limits $\sigma \to \infty$ and $\sigma \to 0$, respectively. While in the case of the latter, one can simply set $\sigma = 0$, the procedure is more involved in the case of the former and to actually perform the limit one might draw on results from geometric singular perturbation theory. More specifically, as $\sigma$ becomes large, we observe a time-scale separation in the system with the transition $\mathrm{R} \rightarrow \mathrm{S}$ occurring on a fast and the remaining transitions on a slow time scale, so that in the limit $\sigma \to \infty$ only the slow dynamics remain. The mean-field equations describing them are then given as (see \hyperref[sec:appendix-singular-SIS-limit]{Appendix})
\begin{equation}
	\left\lbrace
	\begin{split}
		\derivative{t} p_{t} &= -\mu \transpose{\Delta} p_{t} \\
		\derivative{t} \tbracket{\mathrm{S}}_{t} &= - \sum_{\lambda} \beta^{\lambda} \tbracket{\mathrm{S}\link{\lambda}\mathrm{I}}_{t} + \gamma \tbracket{\mathrm{I}}_{t} \\
		\derivative{t} \tbracket{\mathrm{I}}_{t} &= \sum_{\lambda} \beta^{\lambda} \tbracket{\mathrm{S}\link{\lambda}\mathrm{I}}_{t} - \gamma \tbracket{\mathrm{I}}_{t} \\
		\derivative{t} \tbracket{\mathrm{S}\link{\omega}\mathrm{S}}_{t} &= - 2 \sum_{\lambda} \beta^{\lambda} \parenth{1 - \frac{\kroneckerdelta{\omega}{\lambda}}{\kappa^{\omega}\tof{p_{t}}}} \tbracket{\mathrm{S}\link{\lambda}\mathrm{I}}_{t} \frac{\tbracket{\mathrm{S}\link{\omega}\mathrm{S}}_{t}}{\tbracket{\mathrm{S}}_{t}} + 2 \gamma \tbracket{\mathrm{S}\link{\omega}\mathrm{I}}_{t} \\ &\mathrelphantom{=} {} + \tparenth{\derivative{t} \norm{p_{t}}^{2}} \tbracket{\mathrm{S}}_{t}^{2} \kroneckerdelta{\omega}{\text{t.}} \\
		\derivative{t} \tbracket{\mathrm{S}\link{\omega}\mathrm{I}}_{t} &= \sum_{\lambda} \beta^{\lambda} \parenth{1 - \frac{\kroneckerdelta{\omega}{\lambda}}{\kappa^{\omega}\tof{p_{t}}}} \tbracket{\mathrm{S}\link{\lambda}\mathrm{I}}_{t} \frac{\tbracket{\mathrm{S}\link{\omega}\mathrm{S}}_{t} - \tbracket{\mathrm{S}\link{\omega}\mathrm{I}}_{t}}{\tbracket{\mathrm{S}}_{t}} - \beta^{\omega} \tbracket{\mathrm{S}\link{\omega}\mathrm{I}}_{t} - \gamma \tparenth{\tbracket{\mathrm{S}\link{\omega}\mathrm{I}}_{t} - \tbracket{\mathrm{I}\link{\omega}\mathrm{I}}_{t}} \\ &\mathrelphantom{=} {} + \tparenth{\derivative{t} \norm{p_{t}}^{2}} \tbracket{\mathrm{S}}_{t} \tbracket{\mathrm{I}}_{t} \kroneckerdelta{\omega}{\text{t.}} \\
		\derivative{t} \tbracket{\mathrm{I}\link{\omega}\mathrm{I}}_{t} &= 2 \parenth{ \sum_{\lambda} \beta^{\lambda} \parenth{1 - \frac{\kroneckerdelta{\omega}{\lambda}}{\kappa^{\omega}\tof{p_{t}}}} \tbracket{\mathrm{S}\link{\lambda}\mathrm{I}}_{t} \frac{\tbracket{\mathrm{S}\link{\omega}\mathrm{I}}_{t}}{\tbracket{\mathrm{S}}_{t}} + \beta^{\omega} \tbracket{\mathrm{S}\link{\omega}\mathrm{I}}_{t} } - 2 \gamma \tbracket{\mathrm{I}\link{\omega}\mathrm{I}}_{t} \\ &\mathrelphantom{=} {} + \tparenth{\derivative{t} \norm{p_{t}}^{2}} \tbracket{\mathrm{I}}_{t}^{2} \kroneckerdelta{\omega}{\text{t.}} \\
	\end{split}
	\right.
	\label{eq:SIS-2nd-order-mean-field-equations}
\end{equation}

The corresponding first-order mean-field equations can be obtained analogously or, alternatively, by simply applying the pair-closure from before in equation~\eqref{eq:regular-pair-closure} to these equations. We remark that the reduced slow limiting systems are relatively straightforward in our context but far more complicated singular limits have recently emerged as well in epidemic dynamics depending on which singular perturbation parameters are considered~\parencite{jardon-kojakhmetov2021geometric,li2016turning,schecter2021geometric}.

\subsubsection{Numerical comparison of first- and second-order mean-field equations with the stochastic process}

\begin{figure}[tb]
	\centering

	\begin{grid}{p{12cm}p{\linewidth-12cm}}
		{
			\begin{minipage}{\linewidth}
			
				\textbf{\Large A}

				\vspace{-0.5cm}
				\begin{grid}{p{0.7cm}*{3}{>{\centering\arraybackslash}p{3.7cm}}}
					{} & {\small SIS-epidemic} & {\small SIRS-epidemic} & {\small SIR-epidemic} \\
				\end{grid}
		
				\begin{grid}{p{4.4cm}*{2}{p{3.7cm}}}
					\includegraphics[scale=1,trim={0cm 1cm 0 0},clip]{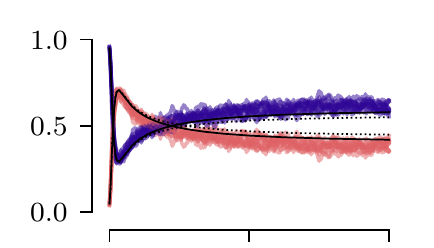}
					&
					\includegraphics[scale=1,trim={0.7cm 1cm 0 0},clip]{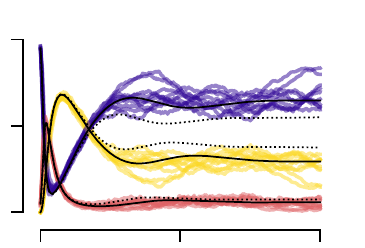}
					&
					\includegraphics[scale=1,trim={0.7cm 1cm 0 0},clip]{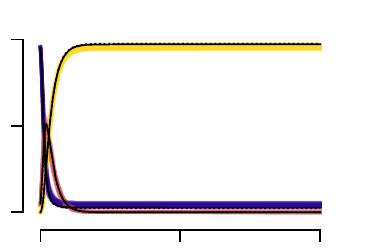}
					\\
				\end{grid}
				
				\vspace{1em}
				
				\textbf{\Large B}
	
				\begin{grid}{p{4.4cm}*{2}{p{3.7cm}}}
					\includegraphics[scale=1,trim={0cm 0cm 0 0},clip]{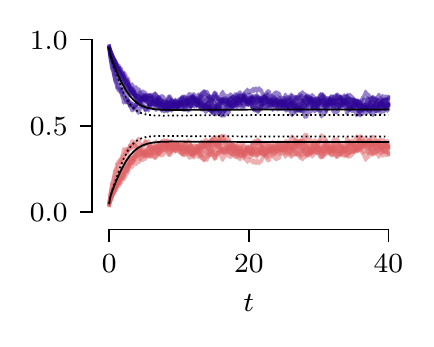}
					&
					\includegraphics[scale=1,trim={0.7cm 0cm 0 0},clip]{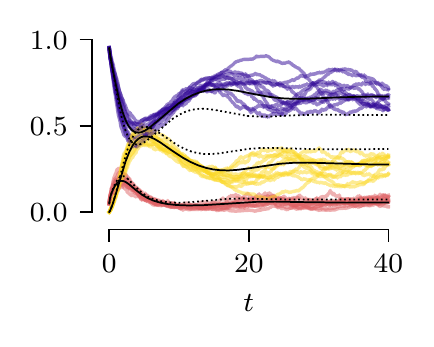}
					&
					\includegraphics[scale=1,trim={0.7cm 0cm 0 0},clip]{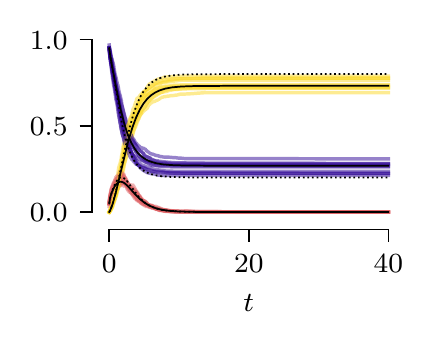}
					\\
				\end{grid}
				
			\end{minipage}
		}
		&
		{
			\vspace*{-3.75cm}
			\begin{minipage}{\linewidth}
				
				\includegraphics[scale=1,trim={0 0.5cm 0 0.5cm},clip]{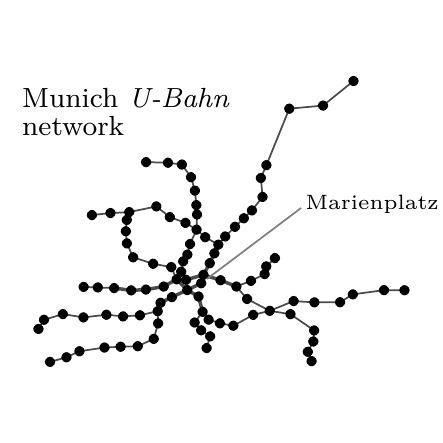}
				
				\includegraphics{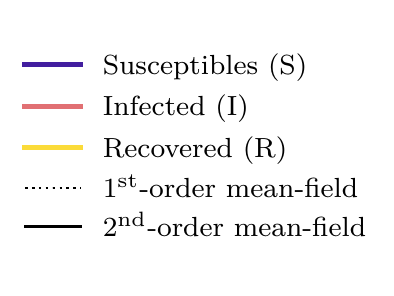}
	
			\end{minipage}
		}
		\\
	\end{grid}

	\caption{\textbf{Comparison between stochastic trajectories of the model and first- and second-order mean-field solutions.} 
	Stochastic trajectories together with their corresponding mean-field solutions for the fractions of susceptibles, infected, and recovered are shown for SIS-, SIRS- and SIR-epidemic dynamics in a population of \num{1000} individuals in a \num{10}-regular network as the community layer of the epidemic network and the Munich \textit{U-Bahn} network that consists of approximately \num{100} sites as the transport network. The epidemic parameters are set to $\beta^{\text{c.}} = \frac{1}{6}$, $\beta^{\text{t.}} = \frac{1}{20} \beta^{\text{c.}}$, $\gamma = 1$, and, in case of SIRS-dynamics, $\sigma = \frac{1}{5}$. The mobility rate is varied, with $\mu = 1$ in \textbf{A} and $\mu = 10$ in \textbf{B}. In each case, the mean initial prevalence is set to \SI{5}{\percent} and, at $t = 0$, the individuals are all located at a single site (\enquote{Marienplatz}) on transport network. The stochastic trajectories are shown without having been time-shifted.}
	\label{fig:numerical-simulation-comparison}
\end{figure}

Having derived mean-field equations to describe the dynamics of the stochastic process at a macroscopic level, obviously raises the question as to how well they describe the full stochastic process. For that, we computed stochastic trajectories by simulating the stochastic process using a discrete-event simulation~\parencite{kiss2017mathematics,law2015simulation}. A numerical comparison between stochastic trajectories and the corresponding first- and second-order mean-field solutions for SIRS-, SIS- and SIR-epidemic dynamics in a population with \num{1000} individuals in a \num{10}-regular network as the community layer of the epidemic network and the Munich \textit{U-Bahn} network which consists of \num{96} sites as the transport network shows overall good agreement between the stochastic trajectories and the mean-field solutions~(Fig.~\ref{fig:numerical-simulation-comparison}). However, as expected, the second-order mean-field solutions are generally in better agreement with the stochastic trajectories.

\subsection{The epidemic threshold in the presence of transport}

\begin{figure}[tb]
	\centering

	\begin{grid}{p{0.31\linewidth}p{0.31\linewidth}p{0.38\linewidth}}
		{
			\textbf{\Large A}
			
			\includegraphics[trim={0.75cm 0.8cm 2.2cm 1.9cm},clip]{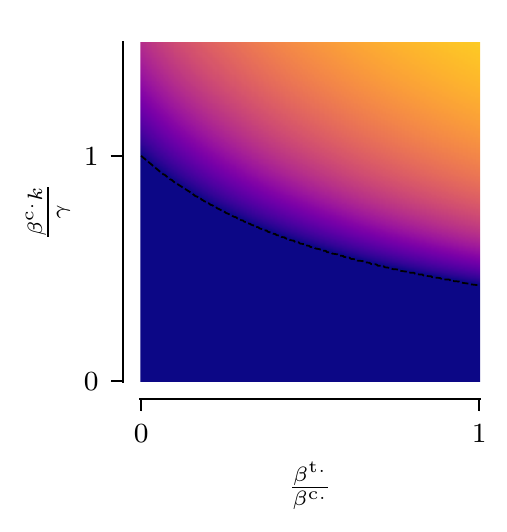}
		}
		&
		{
			\textbf{\Large B}
			
			\includegraphics[trim={0.75cm 0.8cm 2.2cm 1.9cm},clip]{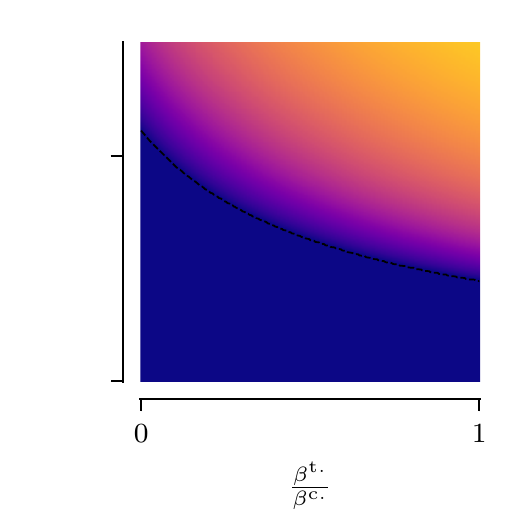}
		}
		&
		{
			\textbf{\Large C}
			
			\includegraphics[trim={0.75cm 0.8cm 1.0cm 1.9cm},clip]{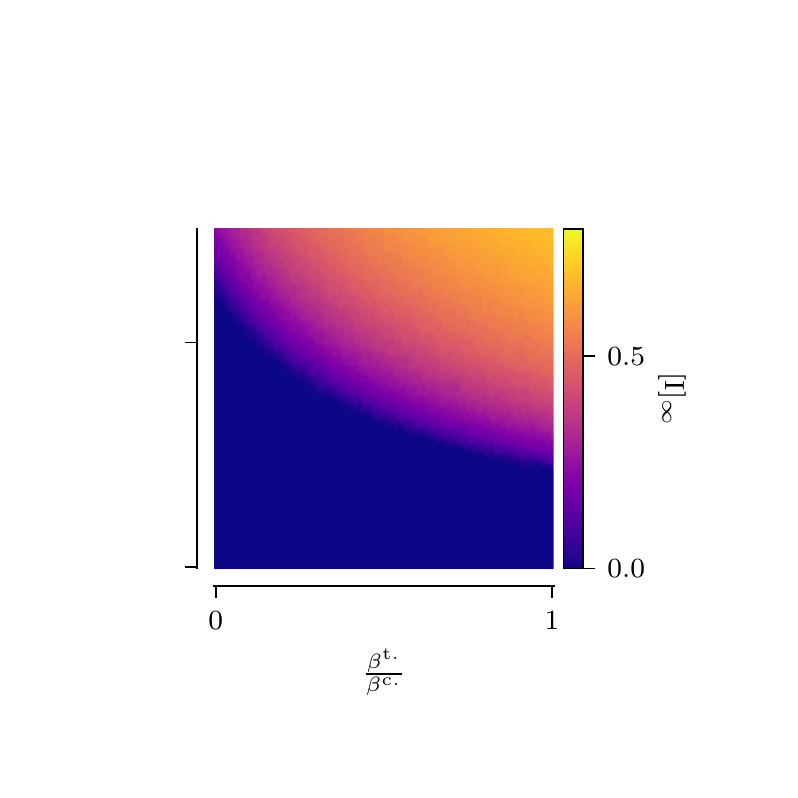}
		}
		\\
	\end{grid}

	\caption{\textbf{The presence of the transport process lowers the epidemic threshold.} The equilibrium prevalence near the epidemic threshold between disease-free and endemic state is shown as a function of the relative strength of the infection rate in the transport layer $\frac{\beta^{\text{t.}}}{\beta^{\text{c.}}}$ and the basic reproduction number in the absence of transport $\frac{\beta^{\text{c.}} k}{\gamma}$ (in first-order approximation) for the first- and second-order mean-field description in \textbf{A} and \textbf{B}, respectively, as well as for the mean of stochastic simulations in \textbf{C}. For simplicity, this is done for SIS-epidemic dynamics (for SIRS-epidemic dynamics one can obtain qualitatively similar plots) in a population of \num{1000} individuals in a \num{10}-regular network as the community layer of the epidemic network and the Munich \textit{U-Bahn} network consisting of approximately \num{100} sites as the transport network. The recovery rate and the mobility rate are set to $\gamma = 1$ and $\mu = 10$, respectively. As the infection rate in the transport layer $\beta^{\text{t.}}$ is increased, an increasing number of infections occurs via the transport layer which in turn has as a consequence that the epidemic threshold is lower.}
	\label{fig:SIS-infectious-pressure_transport-process}
\end{figure}

Due the transport process the individuals are transiently linked to other individuals that they would not have been otherwise and as such exposed to a higher risk of infection. The way the transport process increases the infection pressure can be seen most directly in the first-order mean-field models. In fact, the effective infection rate in the presence of the transport process turns out to be $\sum_{\lambda} \beta^{\lambda} \frac{\kappa^{\lambda}\tof{p_{t}}}{\abs{\mathcal{N}}} = \frac{\beta^{\text{c.}}}{\abs{\mathcal{N}}} \parenth{k + \frac{\beta^{\text{t.}}}{\beta^{\text{c.}}} \norm{p_{t}}^{2} \abs{\mathcal{N}}}$. In other words, the transport process effectively amounts to $\frac{\beta^{\text{t.}}}{\beta^{\text{c.}}} \norm{p_{t}}^{2} \abs{\mathcal{N}} \geq \frac{\beta^{\text{t.}}}{\beta^{\text{c.}}} \frac{\abs{\mathcal{N}}}{\abs{\mathcal{X}}}$ additional contacts for the average individual. These additional, transient contacts can considerably alter the fate of the epidemic. If, e.g. in case of SIS-dynamics, the community layer alone is just sufficiently sparsely connected so that the epidemic cannot be sustained and will die out eventually, the presence of the transport process can lead to an endemic state~(Fig.~\ref{fig:SIS-infectious-pressure_transport-process}).

More precisely, consider the first-order mean-field model in equation~\eqref{eq:SIRS-1st-order-mean-field-equations} and let
\begin{equation}
	\chi\of{p} = \parenth{\beta^{\text{c.}} \frac{k}{\abs{\mathcal{N}}} + \beta^{\text{t.}} \norm{p}^{2}} \frac{\abs{\mathcal{N}}}{\gamma} \mathpunctuation{.}
\end{equation}
Then, the system undergoes a transcritical bifurcation when $\chi\of{p_{\infty}} = 1$ where $p_{\infty}$ is the unique solution to the equation $\transpose{\Delta} p_{\infty} = 0$ with $p_{\infty}\of{x} \geq 0$ and $\sum_{x} p_{\infty}\of{x} = 1$.	
Moreover, for $\chi\of{p_{\infty}} < 1$ and $\chi\of{p_{\infty}} > 1$, the stable equilibrium points are $\parenth{p_{\infty}, \abs{\mathcal{N}}, 0, 0}$  and $\parenth{p_{\infty}, \abs{\mathcal{N}} \frac{1}{\chi\of{p_{\infty}}}, \frac{\sigma}{\gamma + \sigma} \abs{\mathcal{N}} \parenth{1 - \frac{1}{\chi\of{p_{\infty}}}}, \frac{\gamma}{\gamma + \sigma} \abs{\mathcal{N}} \parenth{1 - \frac{1}{\chi\of{p_{\infty}}}}}$ corresponding to the disease-free and the endemic state, respectively.

Indeed, due to the strongly connected transport network there is a unique solution $p_{\infty}$ of $\derivative{t} p_{t} = - \mu \transpose{\Delta} p_{t} = 0$ with $p_{\infty} \geq 0$ and $\sum p_{\infty} = 1$, the equilibrium solution of the random walk on the transport network. From there, a proof of the above statement proceeds entirely analogously to the case of standard SIRS-epidemic dynamics~\parencite{kiss2017mathematics}.

In particular, since $\chi\of{p_{\infty}} \geq \left. \chi\of{p_{\infty}} \right\vert_{\beta^{\text{t.}} = 0}$, in the presence of the transport process and a non-vanishing infection rate in the transport layer, the epidemic threshold towards an endemic state is lower~(Fig.~\ref{fig:SIS-infectious-pressure_transport-process}).

In terms of mitigation strategies for an epidemic, this means that, without comprehensive testing and subsequent quarantining of infected individuals as well as additional hygiene measures on public transport, people that regularly participate in public transport, e.g. by commuting to work, would have to restrict their personal contacts even more drastically than people that do generally avoid public transport and by that keep their number of effective contacts low. Alternatively, the density of people in public transport has to be kept low, which would result in a lower infection rate and similarly keep the overall number of effective contacts low.

Finally, let us also remark that the corresponding results for SIR- and SIS-epidemic dynamics can again be obtained by considering the limits $\sigma \to 0$ and $\sigma \to \infty$, respectively. Unfortunately, a full analytical description of the equilibrium state of the second-order mean-field equations in terms of closed-form expressions is virtually impossible in most cases. Even in the simplest case of the SIS-epidemic dynamics and considering the fully reduced system after making use of all the conservation relations, a description of the endemic state turns out to be very complicated.

\subsection{Non-local, fractional transport dynamics}

So far, we have considered transport under the dynamics of the standard graph Laplacian. In this case, the individuals performing the random walk can only move to sites in the immediate neighbourhood of the graph. However, empirical studies suggest that among other things human mobility patterns are characterised by heavy-tailed distributed jump-lengths~\parencite{brockmann2006scaling} reminiscent of Lévy flights on the network~\parencite{zaburdaev2015levy}. In the following, we will consider a generalisation of our model towards fractional dynamics on the transport network, which are known to give rise to heavy-tailed jump-length distributions~\parencite{riascos2014fractional,michelitsch2017fractional,benzi2020non,michelitsch2019fractional}.

In defining the fractional dynamics for an exponent $0 < \alpha \leq 1$, we follow the approach of \textcite{benzi2020non}. We consider the unnormalised Laplacian $L := K - A$ which is symmetric and non-negative. As such, we can define its fractional power in the usual way: Given the spectral decomposition $L = \sum_{\lambda} \lambda \, \Pi_{\lambda}$, we have $L^{\alpha} = \sum_{\lambda} \lambda^{\alpha} \, \Pi_{\lambda}$. The fractional degree-matrix is then given as $K^{\tof{\alpha}} := \diag\tof{k^{\tof{\alpha}}\of{x}}_{x}$ with $k^{\tof{\alpha}}\of{x} = L^{\alpha}\of{x,x}$ and we obtain the fractional Laplacian $\Delta^{\tof{\alpha}} = K^{\tof{\alpha}}\inverse{{}} L^{\alpha}$ so that the fractional transition matrix is given as $\Rho^{\tof{\alpha}} = \varid - \Delta^{\tof{\alpha}}$. With the fractional transition matrix or Laplacian, we can define our model entirely analogously and also the results we have obtained so far carry over.

\begin{figure}[tb]
	\centering

	\begin{grid}{p{0.31\linewidth}p{0.31\linewidth}p{0.38\linewidth}}
		{
			\textbf{\Large A}
			
			\includegraphics[trim={0.75cm 0.8cm 2.2cm 1.9cm},clip]{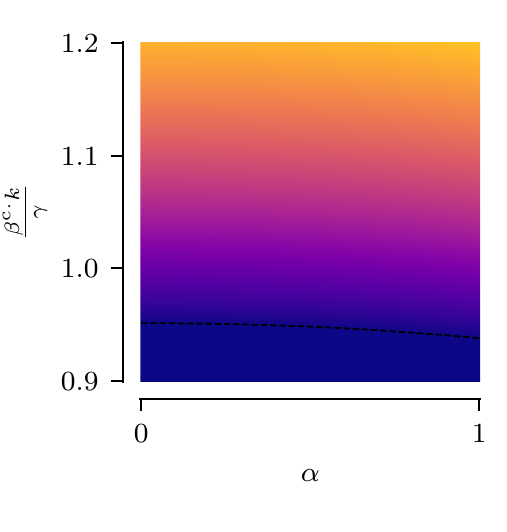}
		}
		&
		{
			\textbf{\Large B}
			
			\includegraphics[trim={0.75cm 0.8cm 2.2cm 1.9cm},clip]{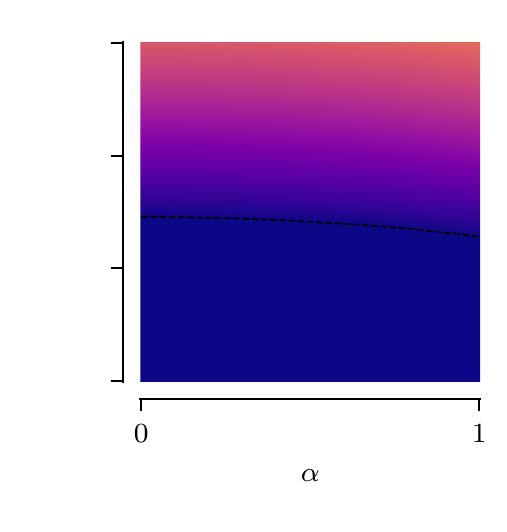}
		}
		&
		{
			\textbf{\Large C}
			
			\includegraphics[trim={0.75cm 0.8cm 1.0cm 1.9cm},clip]{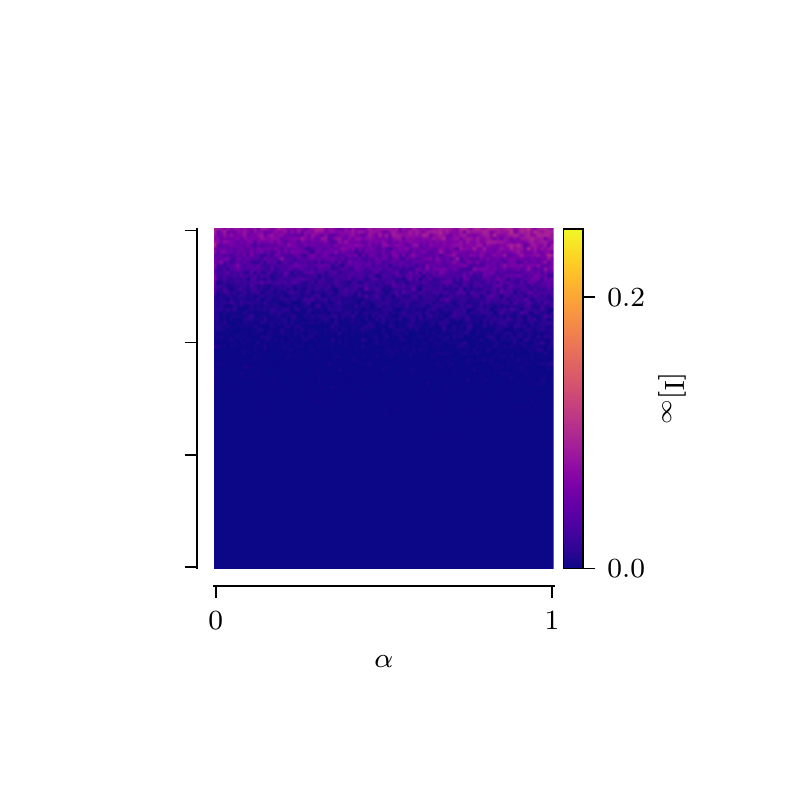}
		}
		\\
	\end{grid}

	\caption{\textbf{Non-local, fractional dynamics in the transport process raise the epidemic threshold.} The equilibrium prevalence near the epidemic threshold between disease-free and endemic state is shown as a function of the fractional exponent $\alpha$ and the basic reproduction number in the absence of transport $\frac{\beta^{\text{c.}} k}{\gamma}$ (in first-order approximation) for the first- and second-order mean-field description in \textbf{A} and \textbf{B}, respectively, as well as for the mean of stochastic simulations in \textbf{C}. For simplicity, this is done for SIS-epidemic dynamics (for SIRS-epidemic dynamics one can obtain qualitatively similar plots) in a population of \num{1000} individuals in a \num{10}-regular network as the community layer of the epidemic network and the Munich \textit{U-Bahn} network consisting of approximately \num{100} sites as the transport network. The infection rate in the transport layer, the recovery rate, and the mobility rate are set to $\beta^{\text{t.}} = \frac{1}{20} \beta^{\text{c.}}$, $\gamma = 1$, and $\mu = 10$, respectively. As the fractional exponent $\alpha$ is lowered, the individuals spread more evenly across the transport network and do not form clusters which in turn has a consequence that the epidemic threshold is higher.}
	\label{fig:SIS-infectious-pressure_fractional-dynamics}
\end{figure}

While for $\alpha = 1$, we trivially recover the dynamics of a classical random walk, for $0 < \alpha < 1$, we obtain superdiffusive behaviour in the transport process as well as an algebraic decay in the jump-length distribution~\parencite{benzi2020non,riascos2014fractional}. One of the consequences that come with the introduction of the fractional dynamics is that it eventually raises the epidemic threshold~(Fig.~\ref{fig:SIS-infectious-pressure_fractional-dynamics}). Under fractional dynamics, individuals spread more evenly on the transport network facilitated by long-distance jumps. As such, the equilibrium distribution of the transport process on the network approaches a uniform distribution as the fractional exponent becomes small.

Indeed, it is well-known that the equilibrium distribution for the transport dynamics is given as $p^{\tof{\alpha}}_{\infty} = \parenth{\frac{k^{\tof{\alpha}}\of{x}}{\sum_{x} k^{\tof{\alpha}}\of{x}}}_{x}$~\parencite{benzi2020non}, where $k^{\tof{\alpha}}\of{x} = \sum_{\lambda} \lambda^{\alpha} \, \Pi_{\lambda}\of{x,x}$ for any $0 < \alpha \leq 1$. On one hand, considering the case $\alpha = 1$, we immediately see that in equilibrium there is cluster formation at well-connected sites. On the other hand, if we let $\alpha$ tend to $0$ these clusters dissolve. More specifically, for $\alpha > 0$, $\sum_{\lambda} \lambda^{\alpha} \, \Pi_{\lambda}\of{x,x} = \sum_{\lambda \suchthat \lambda \neq 0} \lambda^{\alpha} \, \Pi_{\lambda}\of{x,x}$ so that $k^{\tof{0}}\of{x} := \lim_{\alpha \downarrow 0} k^{\tof{\alpha}}\of{x} = 1 - \Pi_{0}\of{x,x} = 1 - \frac{1}{\abs{\mathcal{X}}}$ where we have used that $\sum_{\lambda} \Pi_{\lambda} = \varid$ and that the eigenvector corresponding to the eigenvalue $0$ is constant. Hence, as $\alpha$ approaches $0$, the fractional degrees at every site converge to a constant and, consequently, the equilibrium distribution is continuously approaching a uniform distribution. In turn, since the term $\norm{p}^{2}$ for any probability distribution $p$ is minimised by the uniform distribution, we have that $\chi\tof{p^{\tof{\alpha}}_{\infty}} \geq \chi\tof{p^{\tof{0}}_{\infty}}$ for any $0 < \alpha \leq 1$. 
However, while one might assume that the dependence in $\alpha$ is monotonic, so that the epidemic threshold towards an endemic state is higher the more $\alpha$ is decreased, this turns out to be false in general. In fact, it depends on the topology of the network and there are counterexamples of networks where the epidemic threshold decreases before it finally increases (see \hyperref[sec:appendix-monotonocity]{Appendix}).

Overall, the effect the fractional dynamics have on the epidemic threshold is rather subtle depending on the topology of the transport network. In fact, the more regular it is, the smaller it is. Conversely, the more irregular it is, the more apparent the difference in the epidemic threshold for fractional and standard random walk ($\alpha = 1$) dynamics is.

\subsection{Temporary instead of permanent participation in the transport-epidemic dynamics}

The transport model we have considered so far has the apparent flaw that every individual is permanently moving through the transport network. In contrast, more realistically, individuals would enter the transport network at one site and leave it again at another site and, therefore, only temporarily move through the transport network participating in the transport-induced epidemic dynamics.

In order to incorporate that into the present model, we extend a given transport network with a set of sites, accessible by some or all other sites, that will be regarded as situated beyond the transport network in the sense that whenever individuals occupy these sites they are not aware of each other and as such do not generate a transient link in the transport layer. This change then shows up in the way the distribution on the extended transport network enters the mean-field description. Specifically, if $\tparenth{\closure{\mathcal{X}}, \closure{\mathcal{A}}}$ is the extended transport network with $\overline{\mathcal{X}} = \mathcal{X} \cup \Omega$ and $\Omega$ the aforementioned set of sites beyond the transport network, then any term $\norm{p_{t}}^{2}$ showing up in the equations is replaced by the term $\norm{p_{t} \indicator{\complement{\Omega}}}^{2} = \sum_{x \in \closure{\mathcal{X}} \setminus \Omega} p_{t}\of{x}^2$.

Indeed, if $\Delta$ is now the graph Laplacian of the extended transport network and starting from equation~\eqref{eq:evolution-[h-h'(x)]}, we obtain the overall number of $h$-$h'$-links in the transport layer in this case by only summing across the sites in $\mathcal{X}$ as opposed to also those in $\Omega$, which yields an exact equation analogous to equation~\eqref{eq:evolution-[h-h']}. After making the same approximation as before, $\tbracket{h\of{x}}_{t} \approx p_{t}\of{x} \tbracket{h}_{t}$, we eventually obtain
\begin{equation}
	\left\lbrace
	\begin{split}
		\derivative{t} p_{t} &= -\mu \transpose{\Delta} p_{t} \\
		\derivative{t} \tbracket{h\link{\text{t.}}h'}_{t} &\approx \tparenth{\derivative{t} \sum_{x \in \closure{\mathcal{X}} \setminus \Omega} p_{t}\of{x}^2} \tbracket{h}_{t} \tbracket{h'}_{t} = \tparenth{\derivative{t} \norm{p_{t} \indicator{\complement{\Omega}}}^{2}} \tbracket{h}_{t} \tbracket{h'}_{t}
	\end{split}
	\right.
\end{equation}
which takes the place of equation~\eqref{eq:approximate-evolution-[h-h']} for the extended transport network. Thus, instead of $\norm{p_{t}}^{2}$ we have $\norm{p_{t} \indicator{\complement{\Omega}}}^{2}$ for an extended transport network with void sites $\Omega$. Moreover, note that this reduces to the previous case where there are no void sites and $\Omega = \emptyset$.

By a similar argument where one restricts the sum across the transport network's sites to exclude those in the void, one can show that in expectation the degree of an individual's node in the transport layer is analogously $\norm{p_{t} \indicator{\complement{\Omega}}}^{2} \abs{\mathcal{N}}$ instead of $\norm{p_{t}}^{2} \abs{\mathcal{N}}$.

Overall, the introduction of void sites binds some of the density of individuals on the transport network, reduces the number of links in the transport layer and by that the infectious pressure. Consequently, this raises the epidemic threshold but otherwise does not qualitatively change the behaviour of the dynamics.

\subsection{Numerical investigation of non-equilibrium dynamics}

So far, we have exclusively focused on the equilibrium of the dynamics. In the following, we will discuss two other aspects regarding dynamics that do not necessarily admit an equilibrium on one hand and dynamics before reaching equilibrium on the other hand.

\subsubsection{Time-dependent transport networks}

\begin{figure}[tb]
	\centering

	\begin{grid}{p{12cm}p{\linewidth-12cm}}
		{
			\begin{minipage}{\linewidth}
			
				\textbf{\Large A}

				\vspace{-0.5cm}
				\begin{grid}{p{0.7cm}*{3}{>{\centering\arraybackslash}p{3.7cm}}}
					{} & {\small SIS-epidemic} & {\small SIRS-epidemic} & {\small SIR-epidemic} \\
				\end{grid}
		
				\begin{grid}{p{4.4cm}*{2}{p{3.7cm}}}
					\includegraphics[scale=1,trim={0cm 1cm 0 0},clip]{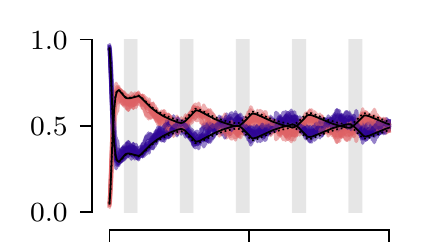}
						&
					\includegraphics[scale=1,trim={0.7cm 1cm 0 0},clip]{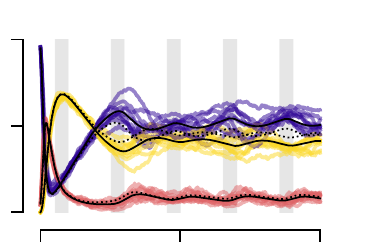}
						&
					\includegraphics[scale=1,trim={0.7cm 1cm 0 0},clip]{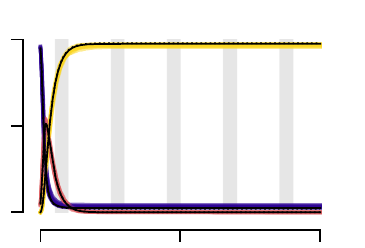}
						\\
				\end{grid}
				
				\vspace{1em}
				
				\textbf{\Large B}
	
				\begin{grid}{p{4.4cm}*{2}{p{3.7cm}}}
					\includegraphics[scale=1,trim={0cm 0cm 0 0},clip]{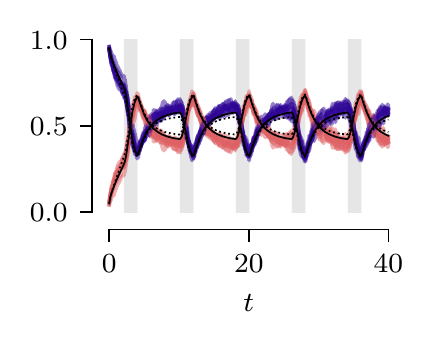}
					&
					\includegraphics[scale=1,trim={0.7cm 0cm 0 0},clip]{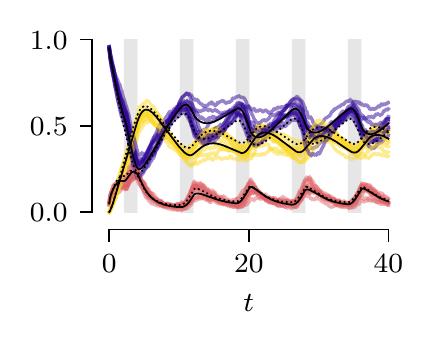}
					&
					\includegraphics[scale=1,trim={0.7cm 0cm 0 0},clip]{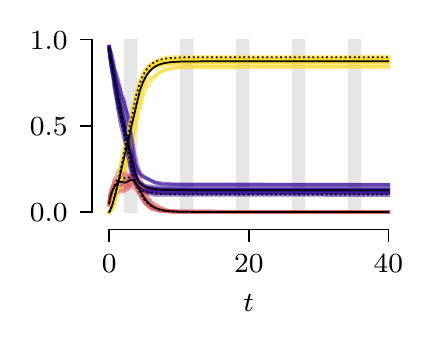}
					\\
				\end{grid}
				
			\end{minipage}
		}
		&
		{
			\vspace*{-3.75cm}
			\begin{minipage}{\linewidth}
				
				\includegraphics[scale=1,trim={0 0.5cm 0 0.5cm},clip]{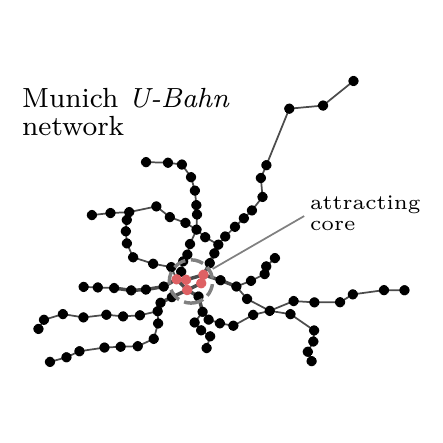}
				
				\includegraphics{figures/3-L}
	
			\end{minipage}
		}
		\\
	\end{grid}

	\caption{\textbf{Comparison between stochastic trajectories of the model and first- and second-order mean-field solutions for a time-varying transport network.}
	Stochastic trajectories together with their corresponding mean-field solutions for the fractions of susceptibles, infected, and recovered are shown for SIS-, SIRS- and SIR-epidemic dynamics in a population of \num{1000} individuals in a \num{10}-regular network as the community layer of the epidemic network and the Munich \textit{U-Bahn} network that consists of approximately \num{100} sites as the transport network. The dynamics switch between the transport network as considered before and one with an attracting core that arises from the former after resolving any link as two directed links in either direction and then deleting any link that strictly increases the number of steps required to reach the core and by that creating a time-varying transport network. The epidemic parameters are set to $\beta^{\text{c.}} = \frac{1}{6}$, $\beta^{\text{t.}} = \frac{1}{20} \beta^{\text{c.}}$, $\gamma = 1$, and, in case of SIRS-dynamics, $\sigma = \frac{1}{5}$. The mobility rate is varied, with $\mu = 1$ in \textbf{A} and $\mu = 10$ in \textbf{B}. In each case, the mean initial prevalence is set to \SI{5}{\percent} and, at $t = 0$, the individuals are all located at a single site (\enquote{Marienplatz}) on the transport network. The stochastic trajectories are shown without having been time-shifted. The periods when the transport network was attracting are shaded in grey.}
	\label{fig:numerical-simulation-driven-comparison}
\end{figure}

Besides static, undirected transport networks that we have considered so far, it is also interesting to have the network dynamically changing. Inspired by the dynamics induced by the daily commute of people from their home to e.g. the central business district, one might consider a network that is temporarily attracting, drawing a random walker into a predefined core.

One way to model this is to consider, in addition to a base network, a second (directed) transport network, derived from the former, where the strength of the links is biased so that a random walk will concentrate a large proportion of its mass at a small set of nodes, an attracting core. Given a function $f : \left[0, \infty\right[ \rightarrow \left[0, 1\right]$, we can then define transport dynamics on a dynamically changing network, governed by a time-dependent Laplacian $\tilde{\Delta}_{t} = \parenth{1 - f\of{t}} \Delta + f\of{t} \Delta'$ with $\Delta$ and $\Delta'$ the Laplacians of the base network and the attracting network, respectively. Depending on the modulation function $f$ one can model different dynamics. In the case of periodic dynamics such as the initially mentioned daily commute, an obvious choice would be an oscillating function. In contrast, a single event temporarily drawing people to a particular site, can be realised by a function that switches only once between the two networks.

The mean-field description that we have derived readily applies to this case and for a numerical comparison between stochastic trajectories and first- and second-order mean-field solutions we consider again as a transport network the Munich \textit{U-Bahn} network with the sites in the city centre as an attracting core. The attracting network arises from the base model after resolving every link between any two sites as two directed links in either direction and then deleting any link that strictly increases the number of steps required to reach the core. As before, we find overall good agreement between the stochastic trajectories and the corresponding first- and second-order mean-field solutions~(Fig.~\ref{fig:numerical-simulation-driven-comparison}).

However, it should come as no surprise that in general, the transport process and with it the whole system does not approach an equilibrium anymore. Moreover, it is not even possible to derive bounds on the epidemic dynamics in terms of the dynamics on one or the other transport network.

\subsubsection{Surges in infections as a consequence of an accumulation of individuals at a single site}

While the rate $\mu$ at which the individuals move through the transport network does not influence the equilibrium of the whole system, it can hugely influence the trajectory towards the equilibrium.
Specifically, this is most profound when a considerable proportion of individuals accumulate at a single site in the transport network. In that case, the accumulation site has to be evacuated sufficiently fast, i.e. $\mu$ has to be large, in order for this to not lead to a surge in infections and a rise in prevalence. 

\begin{figure}[tb]
	\centering

	\begin{grid}{p{0.666\linewidth}p{0.333\linewidth}}
		{
			\textbf{\Large A}
		
			\includegraphics{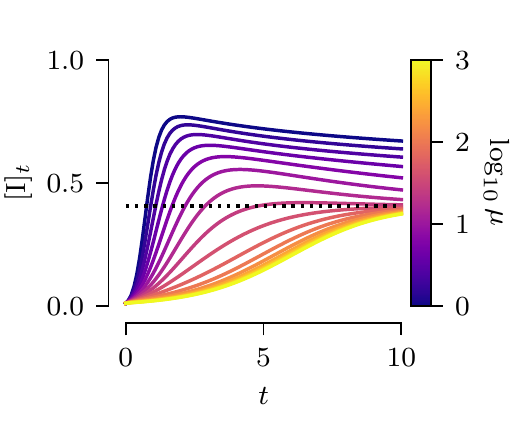}
			\includegraphics{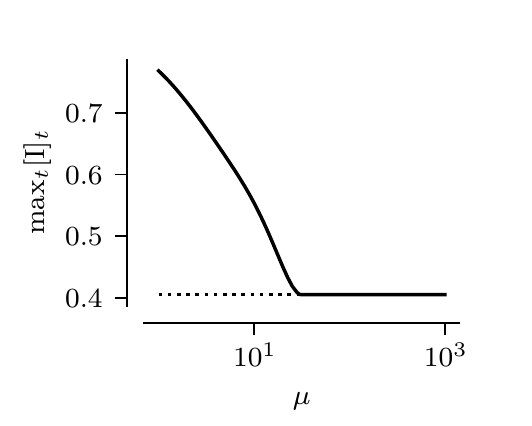}
		}
		&
		{
			\textbf{\Large B}
			
			\includegraphics{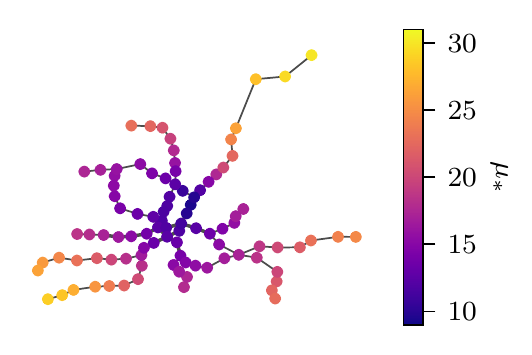}
		}
		\\
	\end{grid}

	\caption{\textbf{Surge of infections due to accumulation of individuals at a single site.} While the mobility rate does not affect the equilibrium prevalence, it can alter the epidemic in the short term. For that consider SIS-epidemic dynamics in a population of \num{1000} individuals in a \num{10}-regular network as the community layer of the epidemic network. As before, the Munich \textit{U-Bahn} network consisting of approximately \num{100} sites is taken as the transport network. The epidemic parameters are set to $\beta^{\text{c.}} = \frac{1}{6}$, $\beta^{\text{t.}} = \frac{1}{20} \beta^{\text{c.}}$, $\gamma = 1$. In \textbf{A}, the trajectories towards equilibrium for different mobility rates are shown when every individual begins at the same initial site. For high mobility rates, the prevalence simply rises from the initial prevalence to equilibrium. However, there is a threshold mobility rate, $\mu_{*}$, under which prevalence rises beyond the equilibrium value. In \textbf{B}, it is shown that this threshold mobility rate depends on the initial site and its neighbourhood in the transport network. Specifically, for sites that are central and well-connected this threshold is lower than for more decentral ones.}
	\label{fig:mobility-dependent-infection-surge}
\end{figure}

Starting with a relatively low prevalence of the epidemic and all the nodes accumulated at a particular site, we may consider the trajectories towards equilibrium under SIS-dynamics for different mobility rates. When the mobility rate is sufficiently high, the accumulation at a single site does not drastically alter the trajectory and the prevalence simply rises to the endemic equilibrium. However, for low mobility rates, one observes the prevalence rising beyond the endemic equilibrium and only slowly recovering to equilibrium afterwards~(Fig.~\ref{fig:mobility-dependent-infection-surge}A). Moreover, the threshold mobility rate for these two extremes critically depends on the initial accumulation site in the transport network. In particular, it turns out that it is not necessarily the degree of the site but rather its centrality in the network that determines this threshold mobility rate. The more central a site is, the lower is this mobility rate~(Fig.~\ref{fig:mobility-dependent-infection-surge}B).

Again, in terms of mitigation strategies, this suggests that for mass events there is a critical time for people to dwell at a particular site. In addition, this time depends on the location. Therefore, for such events to be as safe as possible and not imply a huge number of infections, they should be held at well-connected sites.

\section{Discussion}

In this work, we have introduced a network model that combines epidemic dynamics with a transport process in a multiplex network structure. The latter consists of two layers, a static one and one that is dynamically adapting in response to a transport process on a separate network. We have derived the model's mean-field description and analysed it from a dynamical systems' point of view, characterising its long-term behaviour and specifically focusing on the epidemic threshold between a disease-free and an endemic equilibrium state.
We have shown that the transport process induces additional ways for the contagion to spread and that way lowers the epidemic threshold. We then generalised the transport process to fractional and, in particular, non-local dynamics and have shown that this, conversely, raises the epidemic threshold. The extent of which depends on the topology of the transport network and is more pronounced the more irregular the transport network is. Although we have derived the mean-field description up to second order, our analysis rests upon only the first-order equations. The two-layer multiplex structure of the epidemic network essentially doubles the number of equations required for a second-order mean-field description making it hard to find an analytical description of the endemic equilibrium. Yet, we have shown that first- and second-order numerical solutions are in good agreement so that even the former approximates the dynamics sufficiently well most of the time.

We have presented a generic, conceptual model for epidemic dynamics in the presence of a transport process. In the interest of analytical tractability, we have considered a continuous-time random walk that every individual performs independently on the transport network. However, as has been pointed out elsewhere, these processes do not capture all aspects of human mobility, in particular, since many of those are inherently non-Markovian. This includes behavioural features such as preferential return and recency~\parencite{song2010modelling,barbosa2015effect} as well as more statistical features such as heavy-tailed waiting times~\parencite{brockmann2006scaling}. Intuitively, though, we would expect that these features drive the epidemic by facilitating cluster formations and on average longer waiting times at any site in the transport network.

For the analytical treatment, we have considered static transport networks. However, in reality, one should expect them to dynamically change throughout the day. If we think e.g. about the daily commute, it is obvious that in the morning and evening the flux of individuals moving through the transport network is slightly directed inwards to and outwards from the city centre, respectively. Similarly, events can draw a proportion of individuals towards a certain site. In our model, as a straightforward generalisation, dynamic transport networks are reflected in a dynamic (fractional) Laplacian governing the transport dynamics. The derivation of the mean-field descriptions then carries through analogously. In general though, with a dynamic transport network, the system will not approach an equilibrium anymore and it seems very difficult to prove anything analytically about the epidemic threshold in this case. However, it is conceivable for any such dynamics to temporarily induce cluster formations and therefore again potentially lead to surges in infections. As we have shown, the latter crucially depends on the relation between the characteristic time scales of the transport process and epidemic spreading as well as the accumulation site and the topology of the transport network.

In the context of this work, one potentially interesting idea for future research would be to imagine in addition to the epidemic network having also a multiplex structure on the side of the transport network. Such a structure could then account for different modes of transport so that every individual can switch between the different layers at certain sites or for individual transport networks where every individual moves within its own layer. Another idea would be to consider an adaptive transport process, e.g. by restricting the mobility of individuals in a certain state of health or by avoiding sites with a high local prevalence of the epidemic.

Overall, given the importance transport processes have for the epidemic spreading, we do hope this work is going to inspire further mathematical modelling investigations into the intricate interplay between transport and contact processes that may then in turn inform policies for the mitigation of an epidemic and the design of transport networks in the future.

\section*{Acknowledgements}
CK and JM acknowledge funding from the Deutsche Forschungsgemeinschaft (DFG, German Research Foundation) under project \href{https://gepris.dfg.de/gepris/projekt/458548755}{458548755}. CK acknowledges partial support from the VolkswagenStiftung via a Lichtenberg Professorship.

\clearpage

\printbibliography[heading=bibintoc]

\clearpage

\appendix
\appendixfigures
\appendixequations

\section{Appendix}
\label{sec:appendix}

\subsection{The singular limit \texorpdfstring{$\sigma \to \infty$}{s → ∞} in the second-order mean-field equations of SIRS-dynamics}
\label{sec:appendix-singular-SIS-limit}

In this section, we will demonstrate how to derive the second-order mean-field equations of SIS-epidemic dynamics in the limit $\sigma \to \infty$ from the corresponding second-order mean-field equations of SIRS-epidemic dynamics. As already mentioned, the techniques to perform this limit are provided by geometric singular perturbation theory; see e.g.~\textcite{kuehn2015multiple,wechselberger2020geometric} and references therein. When $\sigma$ becomes large, we eventually observe a time-scale separation in the dynamics with transitions $\mathrm{S} \rightarrow \mathrm{I}$ and $\mathrm{I} \rightarrow \mathrm{R}$ on a comparatively slow and the transition $\mathrm{R} \rightarrow \mathrm{S}$ on a fast time-scale. As such, the idea is to consider the problem on the fast time-scale first, then to determine the critical manifold where the fast dynamics come to a halt, and finally deduce the residual slow dynamics on this critical manifold, which are the only ones remaining in the limit $\sigma \to \infty$.

In order to do that, we start by rescaling time in \eqref{eq:SIRS-2nd-order-mean-field-equations} to the fast time-scale via the transformation $\tau : t \mapsto \sigma t$. Then the SIRS-second-order mean-field equations read
\begin{equation}
	\left\lbrace
	\begin{split}
		\derivative{\tau} p_{\tau} &= - \frac{\mu}{\sigma} \transpose{\Delta} p_{\tau} \\
		\derivative{\tau} \tbracket{\mathrm{S}}_{\tau} &= - \frac{1}{\sigma} \sum_{\lambda} \beta^{\lambda} \tbracket{\mathrm{S}\link{\lambda}\mathrm{I}}_{\tau} + \tbracket{\mathrm{R}}_{\tau} \\
		\derivative{\tau} \tbracket{\mathrm{I}}_{\tau} &= \frac{1}{\sigma} \sum_{\lambda} \beta^{\lambda} \tbracket{\mathrm{S}\link{\lambda}\mathrm{I}}_{\tau} - \frac{\gamma}{\sigma} \tbracket{\mathrm{I}}_{\tau} \\
		\derivative{\tau} \tbracket{\mathrm{R}}_{\tau} &= \frac{\gamma}{\sigma} \tbracket{\mathrm{I}}_{\tau} - \tbracket{\mathrm{R}}_{\tau} \\
		\derivative{\tau} \tbracket{\mathrm{S}\link{\omega}\mathrm{S}}_{\tau} &= - \frac{2}{\sigma} \sum_{\lambda} \beta^{\lambda} \parenth{1 - \frac{\kroneckerdelta{\omega}{\lambda}}{\kappa^{\omega}\tof{p_{\tau}}}} \tbracket{\mathrm{S}\link{\lambda}\mathrm{I}}_{\tau} \frac{\tbracket{\mathrm{S}\link{\omega}\mathrm{S}}_{\tau}}{\tbracket{\mathrm{S}}_{\tau}} + 2 \tbracket{\mathrm{S}\link{\omega}\mathrm{R}}_{\tau} \\ &\mathrelphantom{=} {} + \frac{1}{\sigma} \tparenth{\derivative{\tau} \norm{p_{\tau}}^{2}} \tbracket{\mathrm{S}}_{\tau}^{2} \kroneckerdelta{\omega}{\text{t.}} \\
		\derivative{\tau} \tbracket{\mathrm{S}\link{\omega}\mathrm{I}}_{\tau} &= \frac{1}{\sigma} \sum_{\lambda} \beta^{\lambda} \parenth{1 - \frac{\kroneckerdelta{\omega}{\lambda}}{\kappa^{\omega}\tof{p_{\tau}}}} \tbracket{\mathrm{S}\link{\lambda}\mathrm{I}}_{\tau} \frac{\tbracket{\mathrm{S}\link{\omega}\mathrm{S}}_{\tau} - \tbracket{\mathrm{S}\link{\omega}\mathrm{I}}_{\tau}}{\tbracket{\mathrm{S}}_{\tau}} - \frac{\beta^{\omega}}{\sigma} \tbracket{\mathrm{S}\link{\omega}\mathrm{I}}_{\tau} - \frac{\gamma}{\sigma} \tbracket{\mathrm{S}\link{\omega}\mathrm{I}}_{\tau} + \tbracket{\mathrm{I}\link{\omega}\mathrm{R}}_{\tau} \\ &\mathrelphantom{=} {} + \frac{1}{\sigma} \tparenth{\derivative{\tau} \norm{p_{\tau}}^{2}} \tbracket{\mathrm{S}}_{\tau} \tbracket{\mathrm{I}}_{\tau} \kroneckerdelta{\omega}{\text{t.}} \\
		\derivative{\tau} \tbracket{\mathrm{I}\link{\omega}\mathrm{I}}_{\tau} &= \frac{2}{\sigma} \parenth{ \sum_{\lambda} \beta^{\lambda} \parenth{1 - \frac{\kroneckerdelta{\omega}{\lambda}}{\kappa^{\omega}\tof{p_{\tau}}}} \tbracket{\mathrm{S}\link{\lambda}\mathrm{I}}_{\tau} \frac{\tbracket{\mathrm{S}\link{\omega}\mathrm{I}}_{\tau}}{\tbracket{\mathrm{S}}_{\tau}} + \beta^{\omega} \tbracket{\mathrm{S}\link{\omega}\mathrm{I}}_{\tau} } - \frac{2 \gamma}{\sigma} \tbracket{\mathrm{I}\link{\omega}\mathrm{I}}_{\tau} \\ &\mathrelphantom{=} {} + \frac{1}{\sigma} \tparenth{\derivative{\tau} \norm{p_{\tau}}^{2}} \tbracket{\mathrm{I}}_{\tau}^{2} \kroneckerdelta{\omega}{\text{t.}} \\
		\derivative{\tau} \tbracket{\mathrm{S}\link{\omega}\mathrm{R}}_{\tau} &= - \frac{1}{\sigma} \sum_{\lambda} \beta^{\lambda} \parenth{1 - \frac{\kroneckerdelta{\omega}{\lambda}}{\kappa^{\omega}\tof{p_{\tau}}}} \tbracket{\mathrm{S}\link{\lambda}\mathrm{I}}_{\tau} \frac{\tbracket{\mathrm{S}\link{\omega}\mathrm{R}}_{\tau}}{\tbracket{\mathrm{S}}_{\tau}} + \frac{\gamma}{\sigma} \tbracket{\mathrm{S}\link{\omega}\mathrm{I}}_{\tau} + \tparenth{\tbracket{\mathrm{R}\link{\omega}\mathrm{R}}_{\tau} - \tbracket{\mathrm{S}\link{\omega}\mathrm{R}}_{\tau}} \\ &\mathrelphantom{=} {} + \frac{1}{\sigma} \tparenth{\derivative{\tau} \norm{p_{\tau}}^{2}} \tbracket{\mathrm{S}}_{\tau} \tbracket{\mathrm{R}}_{\tau} \kroneckerdelta{\omega}{\text{t.}} \\
		\derivative{\tau} \tbracket{\mathrm{I}\link{\omega}\mathrm{R}}_{\tau} &= \frac{1}{\sigma} \sum_{\lambda} \beta^{\lambda} \parenth{1 - \frac{\kroneckerdelta{\omega}{\lambda}}{\kappa^{\omega}\tof{p_{\tau}}}} \tbracket{\mathrm{S}\link{\lambda}\mathrm{I}}_{\tau} \frac{\tbracket{\mathrm{S}\link{\omega}\mathrm{R}}_{\tau}}{\tbracket{\mathrm{S}}_{\tau}} + \frac{\gamma}{\sigma} \tparenth{\tbracket{\mathrm{I}\link{\omega}\mathrm{I}}_{\tau} - \tbracket{\mathrm{I}\link{\omega}\mathrm{R}}_{\tau}} - \tbracket{\mathrm{I}\link{\omega}\mathrm{R}}_{\tau} \\ &\mathrelphantom{=} {} + \frac{1}{\sigma} \tparenth{\derivative{\tau} \norm{p_{\tau}}^{2}} \tbracket{\mathrm{I}}_{\tau} \tbracket{\mathrm{R}}_{\tau} \kroneckerdelta{\omega}{\text{t.}} \\
		\derivative{\tau} \tbracket{\mathrm{R}\link{\omega}\mathrm{R}}_{\tau} &= \frac{2 \gamma}{\sigma} \tbracket{\mathrm{I}\link{\omega}\mathrm{R}}_{\tau} - 2 \tbracket{\mathrm{R}\link{\omega}\mathrm{R}}_{\tau} \\ &\mathrelphantom{=} {} + \frac{1}{\sigma} \tparenth{\derivative{\tau} \norm{p_{\tau}}^{2}} \tbracket{\mathrm{R}}_{\tau}^{2} \kroneckerdelta{\omega}{\text{t.}} \\
	\end{split}
	\right.
	\label{eq:SIRS-fast-2nd-order-mean-field-equations}
\end{equation}
which can be written compactly as
\begin{equation}
	\derivative{\tau} \underbrace{\begin{pmatrix} p_{\tau} \\ \tbracket{\mathrm{S}}_{\tau} \\ \tbracket{\mathrm{I}}_{\tau} \\ \tbracket{\mathrm{R}}_{\tau} \\ \tbracket{\mathrm{S}\link{\text{c.}}\mathrm{S}}_{\tau} \\ \tbracket{\mathrm{S}\link{\text{c.}}\mathrm{I}}_{\tau} \\ \tbracket{\mathrm{I}\link{\text{c.}}\mathrm{I}}_{\tau} \\ \tbracket{\mathrm{S}\link{\text{c.}}\mathrm{R}}_{\tau} \\ \tbracket{\mathrm{I}\link{\text{c.}}\mathrm{R}}_{\tau} \\ \tbracket{\mathrm{R}\link{\text{c.}}\mathrm{R}}_{\tau} \\ \tbracket{\mathrm{S}\link{\text{t.}}\mathrm{S}}_{\tau} \\ \tbracket{\mathrm{S}\link{\text{t.}}\mathrm{I}}_{\tau} \\ \tbracket{\mathrm{I}\link{\text{t.}}\mathrm{I}}_{\tau} \\ \tbracket{\mathrm{S}\link{\text{t.}}\mathrm{R}}_{\tau} \\ \tbracket{\mathrm{I}\link{\text{t.}}\mathrm{R}}_{\tau} \\ \tbracket{\mathrm{R}\link{\text{t.}}\mathrm{R}}_{\tau} \\\end{pmatrix}}_{=: \Xi_{\tau}} = \underbrace{\begin{pmatrix} 0 & 0 & 0 & 0 & 0 & 0 & 0 \\ 1 & 0 & 0 & 0 & 0 & 0 & 0 \\ 0 & 0 & 0 & 0 & 0 & 0 & 0 \\ -1 & 0 & 0 & 0 & 0 & 0 & 0 \\ 0 & 2 & 0 & 0 & 0 & 0 & 0 \\ 0 & 0 & 1 & 0 & 0 & 0 & 0 \\ 0 & 0 & 0 & 0 & 0 & 0 & 0 \\ 0 & -1 & 0 & 1 & 0 & 0 & 0 \\ 0 & 0 & -1 & 0 & 0 & 0 & 0 \\ 0 & 0 & 0 & -2 & 0 & 0 & 0 \\ 0 & 0 & 0 & 0 & 2 & 0 & 0 \\ 0 & 0 & 0 & 0 & 0 & 1 & 0 \\ 0 & 0 & 0 & 0 & 0 & 0 & 0 \\ 0 & 0 & 0 & 0 & -1 & 0 & 1 \\ 0 & 0 & 0 & 0 & 0 & -1 & 0 \\ 0 & 0 & 0 & 0 & 0 & 0 & -2 \\ \end{pmatrix}}_{=: N} \underbrace{\begin{pmatrix} \tbracket{\mathrm{R}}_{\tau} \\ \tbracket{\mathrm{S}\link{\text{c.}}\mathrm{R}}_{\tau} \\ \tbracket{\mathrm{I}\link{\text{c.}}\mathrm{R}}_{\tau} \\ \tbracket{\mathrm{R}\link{\text{c.}}\mathrm{R}}_{\tau} \\ \tbracket{\mathrm{S}\link{\text{t.}}\mathrm{R}}_{\tau} \\ \tbracket{\mathrm{I}\link{\text{t.}}\mathrm{R}}_{\tau} \\ \tbracket{\mathrm{R}\link{\text{t.}}\mathrm{R}}_{\tau} \\\end{pmatrix}}_{=: f\of{\Xi_{\tau}}} + \frac{1}{\sigma} G\tof{p_{\tau}, \ldots \tbracket{\mathrm{R}\link{\text{t.}}\mathrm{R}}_{\tau}}
\end{equation}
for some particular choice of $G$.

Since $N$ has full column rank, the requirement $N f\of{\Xi_{\tau}} = 0$ is only fulfilled when $f\of{\Xi_{\tau}} = 0$. Thus, the critical manifold $S$ of this system is given as the zero level-set of $f$, i.e. $S = \set{\xi \suchthat f\of{\xi} = 0}$. In particular, since $\spectrum{D_{\Xi} f\of{\Xi} N} = \set{-1, -2}$, this critical manifold is normally hyperbolic and attracting.

In the singular limit $\sigma \to \infty$, the slow dynamics of the entire system are contained entirely in the critical manifold. On their respective time-scale, they are known to be concisely given as~\parencite[Lemma~3.4]{wechselberger2020geometric}
\begin{equation}
	\derivative{t} \Xi_{t} = \parenth{\varid - N \inverse{\parenth{D_{\Xi}f\of{\Xi} N}} D_{\Xi}f\of{\Xi}} G\of{\Xi}\vert_{S} \mathpunctuation{.}	
\end{equation}

Thus, finally, after performing the algebra, this can be expanded to
\begin{equation}
	\left\lbrace
	\begin{split}
		\derivative{t} p_{t} &= -\mu \transpose{\Delta} p_{t} \\
		\derivative{t} \tbracket{\mathrm{S}}_{t} &= - \sum_{\lambda} \beta^{\lambda} \tbracket{\mathrm{S}\link{\lambda}\mathrm{I}}_{t} + \gamma \tbracket{\mathrm{I}}_{t} \\
		\derivative{t} \tbracket{\mathrm{I}}_{t} &= \sum_{\lambda} \beta^{\lambda} \tbracket{\mathrm{S}\link{\lambda}\mathrm{I}}_{t} - \gamma \tbracket{\mathrm{I}}_{t} \\
		\derivative{t} \tbracket{\mathrm{S}\link{\omega}\mathrm{S}}_{t} &= - 2 \sum_{\lambda} \beta^{\lambda} \parenth{1 - \frac{\kroneckerdelta{\omega}{\lambda}}{\kappa^{\omega}\tof{p_{t}}}} \tbracket{\mathrm{S}\link{\lambda}\mathrm{I}}_{t} \frac{\tbracket{\mathrm{S}\link{\omega}\mathrm{S}}_{t}}{\tbracket{\mathrm{S}}_{t}} + 2 \gamma \tbracket{\mathrm{S}\link{\omega}\mathrm{I}}_{t} \\ &\mathrelphantom{=} {} + \tparenth{\derivative{t} \norm{p_{t}}^{2}} \tbracket{\mathrm{S}}_{t}^{2} \kroneckerdelta{\omega}{\text{t.}} \\
		\derivative{t} \tbracket{\mathrm{S}\link{\omega}\mathrm{I}}_{t} &= \sum_{\lambda} \beta^{\lambda} \parenth{1 - \frac{\kroneckerdelta{\omega}{\lambda}}{\kappa^{\omega}\tof{p_{t}}}} \tbracket{\mathrm{S}\link{\lambda}\mathrm{I}}_{t} \frac{\tbracket{\mathrm{S}\link{\omega}\mathrm{S}}_{t} - \tbracket{\mathrm{S}\link{\omega}\mathrm{I}}_{t}}{\tbracket{\mathrm{S}}_{t}} - \beta^{\omega} \tbracket{\mathrm{S}\link{\omega}\mathrm{I}}_{t} - \gamma \tparenth{\tbracket{\mathrm{S}\link{\omega}\mathrm{I}}_{t} - \tbracket{\mathrm{I}\link{\omega}\mathrm{I}}_{t}} \\ &\mathrelphantom{=} {} + \tparenth{\derivative{t} \norm{p_{t}}^{2}} \tbracket{\mathrm{S}}_{t} \tbracket{\mathrm{I}}_{t} \kroneckerdelta{\omega}{\text{t.}} \\
		\derivative{t} \tbracket{\mathrm{I}\link{\omega}\mathrm{I}}_{t} &= 2 \parenth{ \sum_{\lambda} \beta^{\lambda} \parenth{1 - \frac{\kroneckerdelta{\omega}{\lambda}}{\kappa^{\omega}\tof{p_{t}}}} \tbracket{\mathrm{S}\link{\lambda}\mathrm{I}}_{t} \frac{\tbracket{\mathrm{S}\link{\omega}\mathrm{I}}_{t}}{\tbracket{\mathrm{S}}_{t}} + \beta^{\omega} \tbracket{\mathrm{S}\link{\omega}\mathrm{I}}_{t} } - 2 \gamma \tbracket{\mathrm{I}\link{\omega}\mathrm{I}}_{t} \\ &\mathrelphantom{=} {} + \tparenth{\derivative{t} \norm{p_{t}}^{2}} \tbracket{\mathrm{I}}_{t}^{2} \kroneckerdelta{\omega}{\text{t.}} \\
	\end{split}
	\right.
\end{equation}
and, except for the additional terms we have as a result of the transport process, these equations are essentially the same as the ones that have been derived also elsewhere for SIS-epidemic dynamics~\parencite{kiss2017mathematics}.

\subsection{Monotonicity of the epidemic threshold in the fractional exponent}
\label{sec:appendix-monotonocity}

In this section, we will discuss the question of monotonicity of the epidemic threshold as one varies the fractional exponent of the transport dynamics. This essentially comes down to the question of whether the function $\alpha \mapsto \tnorm{p^{\tof{\alpha}}_{\infty}}^{2}$ with $p^{\tof{\alpha}}_{\infty}$ the equilibrium distribution of the fractional random walk with exponent $0 < \alpha \leq 1$ on some network, which we will assume to be connected, is monotonic. As mentioned in the main text, there is no guarantee for this to be monotonic in general. Specifically, as we will argue, it depends on the topology of the network. In the following, we will first consider networks with a strong block structure and show that these networks violate monotonicity. In contrast to that, we will then consider star networks for which one can explicitly prove monotonicity.

Before moving on, recall that, if $L$ is the Laplacian of some graph with spectral decomposition $L = \sum_{\lambda} \lambda \, \Pi_{\lambda}$ where $\Pi_{\lambda}$ is the orthogonal rank-$1$ spectral projection onto the eigenspace for the eigenvalue $\lambda$, the fractional Laplacian is given as $L^{\alpha} = \sum_{\lambda} \lambda^{\alpha} \, \Pi_{\lambda}$. The equilibrium distribution of the corresponding random walk is then $p^{\tof{\alpha}}_{\infty} = \parenth{\frac{k^{\tof{\alpha}}\of{x}}{\sum_{x} k^{\tof{\alpha}}\of{x}}}_{x}$, where $k^{\tof{\alpha}}\of{x} = L^{\alpha}\of{x,x} = \sum_{\lambda} \lambda^{\alpha} \, \Pi_{\lambda}\of{x,x}$ is the fractional degree at site $x$. Moreover, as $\alpha$ tends to $0$, this distribution becomes uniform and we will denote the limiting distribution as $p^{\tof{0}}_{\infty}$. Hence,
\begin{equation}
    \tnorm{p^{\tof{\alpha}}_{\infty}}^{2} = \frac{\sum_{x} k^{\tof{\alpha}}\of{x}^{2}}{\parenth{\sum_{x'} k^{\tof{\alpha}}\of{x'}}^{2}} = \frac{\sum_{x} L^{\alpha}\of{x,x}^{2}}{\trace\tof{L^{\alpha}}^{2}} \mathpunctuation{,}
\end{equation}
where, in particular, $\tnorm{p^{\tof{\alpha}}_{\infty}}^{2} \geq \tnorm{p^{\tof{0}}_{\infty}}^{2}$ since the uniform distribution among all discrete probability distributions is the unique distribution minimising the $2$-norm. Therefore, if $\alpha \mapsto \tnorm{p^{\tof{\alpha}}_{\infty}}^{2}$ is monotonic, it is necessarily non-decreasing.

Using that $\derivative{\alpha} L^{\alpha} = \frac{1}{\alpha} L^{\alpha} \ln{L^{\alpha}}$, we have that
\begin{equation}
    \begin{split}
        \derivative{\alpha} \tnorm{p^{\tof{\alpha}}_{\infty}}^{2} &= \frac{2}{\alpha \trace\tof{L^{\alpha}}^{3}} \parenth{\trace\tof{L^{\alpha}} \sum_{x} L^{\alpha}\of{x,x} \tparenth{L^{\alpha} \ln{L^{\alpha}}}\of{x,x} - \trace\tof{L^{\alpha} \ln{L^{\alpha}}} \sum_{x} L^{\alpha}\of{x,x}^{2}}
    \end{split}
    \label{eq:equilibrium-distributio-norm-derivative}
\end{equation}
where $\trace\tof{L^{\alpha}} > 0$ so that monotonicity follows if
\begin{equation}
    \trace\tof{L^{\alpha}} \sum_{x} L^{\alpha}\of{x,x} \tparenth{L^{\alpha} \ln{L^{\alpha}}}\of{x,x} - \trace\tof{L^{\alpha} \ln{L^{\alpha}}} \sum_{x} L^{\alpha}\of{x,x}^{2} \geq 0 \mathpunctuation{.}
    \label{eq:monotonicity-trace-inequality}
\end{equation}

\subsubsection*{Non-monotonic behaviour for networks with a block structure}

As it will turn out, networks with a strong block structure produce non-monotonic behaviour and as such violate inequality~\eqref{eq:monotonicity-trace-inequality}. Before presenting an analytical argument, let us first consider a numerical approach that will motivate considering this particular network topology.

For that, we recall that for any Laplacian $L$ and $0 < \alpha \leq 1$, $L^{\alpha}$ is again a Laplacian matrix~\parencite{michelitsch2019fractional}, so that for inequality~\eqref{eq:monotonicity-trace-inequality} to hold it is equivalent to show that for any Laplacian matrix $L$
\begin{equation}
    \trace\tof{L} \sum_{x} L\tof{x,x} \tparenth{L \ln{L}}\tof{x,x} \geq \trace\tof{L \ln{L}} \sum_{x} L\tof{x,x}^{2} \mathpunctuation{.}
    \label{eq:laplace-trace-inequality}
\end{equation}

In fact, without loss of generality, it is enough to consider Laplacian matrices $L$ with $\trace\tof{L} = 1$ only. Indeed, suppose the above has been established for such ones, for arbitrary $L$, let $\hat{L} = \frac{1}{\trace\tof{L}} L$. In that case, $\hat{L} \ln{\hat{L}} = \frac{1}{\trace\tof{L}} L \ln{L} - \frac{\ln{\trace\tof{L}}}{\trace\tof{L}} L$ and $\trace\tof{\hat{L} \ln{\hat{L}}} = \frac{1}{\trace\tof{L}} \trace\tof{L \ln{L}} - \ln{\trace\tof{L}}$. With that, on one hand
\begin{equation}
    \sum_{x} \hat{L}\tof{x,x} \tparenth{\hat{L} \ln{\hat{L}}}\tof{x,x} = \frac{1}{\trace\tof{L}^{2}} \sum_{x} L\tof{x,x} \tparenth{L \ln{L}}\tof{x,x} - \frac{\ln{\trace\tof{L}}}{\trace\tof{L}^{2}} \sum_{x} L\tof{x,x}^{2}
\end{equation}
and on the other hand
\begin{equation}
    \trace\tof{\hat{L} \ln{\hat{L}}} \sum_{x} \hat{L}\tof{x,x}^{2} = \frac{1}{\trace\tof{L}^{3}} \trace\tof{L \ln{L}} \sum_{x} L\tof{x,x}^{2} - \frac{\ln{\trace\tof{L}}}{\trace\tof{L}^{2}} \sum_{x} L\tof{x,x}^{2}
\end{equation}
where by assumption $\sum_{x} \hat{L}\tof{x,x} \tparenth{\hat{L} \ln{\hat{L}}}\tof{x,x} \geq \trace\tof{\hat{L} \ln{\hat{L}}} \sum_{x} \hat{L}\tof{x,x}^{2}$ so that
\begin{equation}
    \frac{1}{\trace\tof{L}^{2}} \sum_{x} L\tof{x,x} \tparenth{L \ln{L}}\tof{x,x} - \frac{\ln{\trace\tof{L}}}{\trace\tof{L}^{2}} \sum_{x} L\tof{x,x}^{2} \geq \frac{1}{\trace\tof{L}^{3}} \trace\tof{L \ln{L}} \sum_{x} L\tof{x,x}^{2} - \frac{\ln{\trace\tof{L}}}{\trace\tof{L}^{2}} \sum_{x} L\tof{x,x}^{2}
\end{equation}
which implies the assertion.

Now, considering only Laplacian matrices $L$ with $\trace\tof{L} = 1$, has the distinctive advantage that by that one can separate the spectra of $L$ and $L \ln{L}$. Indeed, if $\trace\tof{L} = 1$, $\spectrum\tof{L} \subset [0, 1]$ whereas $\spectrum\tof{L \ln{L}} \subset [-\e^{-1}, 0]$ so that $L$ and $L \ln{L}$ are positive and negative semidefinite, respectively. Moreover, it allows for a systematical numerical investigation. For that, observe that the set of $d \times d$ Laplacian matrices with trace $1$ is isomorphic to the standard $\frac{d\parenth{d-1}}{2} - 1$-simplex via the map
\begin{equation}
    \parenth{a_{1}, \ldots a_{\frac{d \parenth{d - 1}}{2}}} \mapsto \frac{1}{2} \begin{pmatrix} * & -a_{1} & \cdots & -a_{\frac{d \parenth{d - 1}}{2}} \\ -a_{1} & * & \ddots & \vdots \\ \vdots & \ddots & \ddots & -a_{d-1} \\ -a_{\frac{d \parenth{d - 1}}{2}} & \cdots & -a_{d-1} & * \\ \end{pmatrix}
\end{equation}
with the diagonal elements set appropriately so that column and row sums are $0$.

\begin{figure}[tbp]
	\centering

	\begin{grid}{p{0.3\linewidth}p{0.7\linewidth}}
		{
			\textbf{\Large A}
		
			\includegraphics[trim={0.0cm 0.6cm 0.8cm 1.2cm},clip]{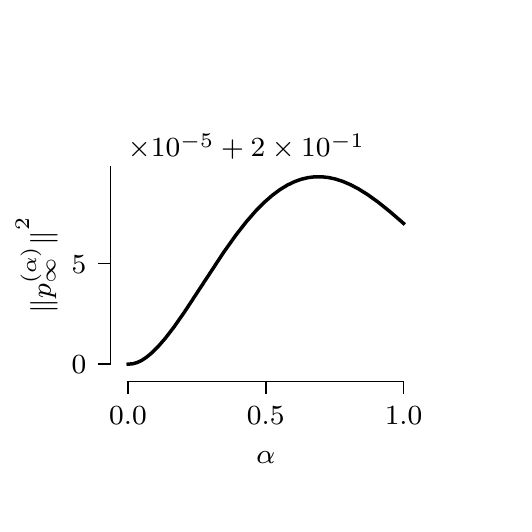}
		}
		&
		{
			\textbf{\Large B}
			
			\includegraphics[trim={0.2cm 0.6cm 1.2cm 1.2cm},clip]{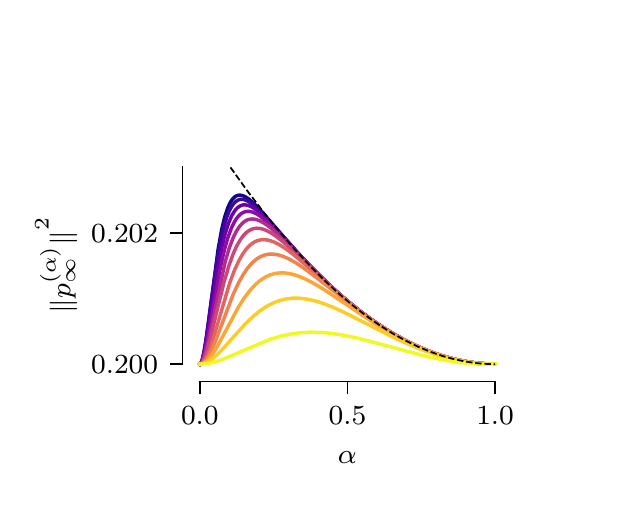}
			\includegraphics[trim={0.2cm 0.6cm 0.1cm 1.2cm},clip]{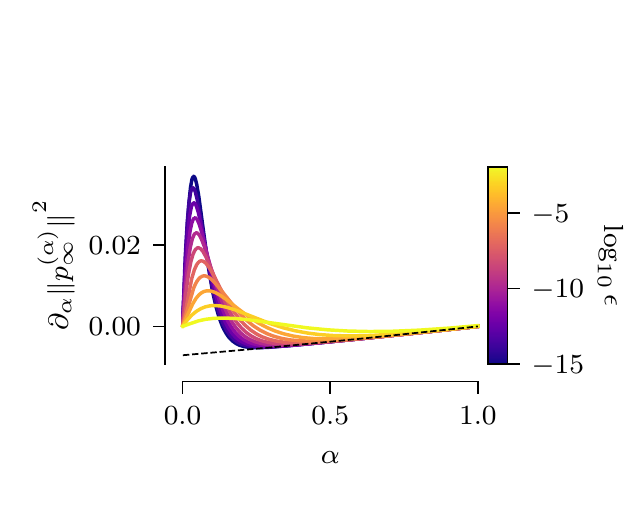}
		}
		\\
	\end{grid}

	\caption{\textbf{Non-monotonic behaviour of the function $\alpha \mapsto \tnorm{p^{\tof{\alpha}}_{\infty}}^{2}$ for networks with strong block structure.} \textbf{A}: The function $\alpha \mapsto \tnorm{p^{\tof{\alpha}}_{\infty}}^{2}$ for the Laplacian matrix in \eqref{eq:monotonicity-numerical-counterexample}. As $\alpha$ is varied from $1$ to $0$, $\tnorm{p^{\tof{\alpha}}_{\infty}}^{2}$ first grows, before it falls to $\frac{1}{5}$ at $\alpha = 0$. \textbf{B}: The function $\alpha \mapsto \tnorm{p^{\tof{\alpha}}_{\infty}}^{2}$ (left) and its derivative (right) for the $(3,2)$-block networks defined in \eqref{eq:block-network-adjacency-matrix} or \eqref{eq:block-network-laplacian-matrix} for different values of $\epsilon$. As in A, $\tnorm{p^{\tof{\alpha}}_{\infty}}^{2}$ first grows, before it eventually falls to $\frac{1}{5}$ at $\alpha = 0$. The dashed line shows the limiting curve for every $\alpha$ as $\epsilon$ approaches $0$.}
	\label{fig:non-monotonicity-block-networks}
\end{figure}

Hence, through sampling the respective standard simplex uniformly, we can numerically test the validity of inequality~\eqref{eq:laplace-trace-inequality} and consequently monotonicity for all possible networks. The first counterexample we found that way was on the standard $9$-simplex, i.e. in dimension $5$, while we were not able to find any counterexamples in lower dimensions. For the matrix
\begin{equation}
    L = \begin{pmatrix} \hphantom{+}0.198 & -0.141 & -0.015 & -0.008 & -0.034 \\ -0.141 & \hphantom{+}0.196 & -0.030 & -0.015 & -0.010 \\ -0.015 & -0.030 & \hphantom{+}0.205 & -0.087 & -0.073 \\ -0.008 & -0.015 & -0.087 & \hphantom{+}0.197 & -0.087 \\ -0.034 & -0.010 & -0.073 & -0.087 & \hphantom{+}0.204 \\ \end{pmatrix}
    \label{eq:monotonicity-numerical-counterexample}
\end{equation}
one easily verifies that it is indeed a Laplacian matrix with $\trace\tof{L} = 1$, whereas
\begin{equation}
    \sum_{x} L\tof{x,x} \tparenth{L \ln{L}}\tof{x,x} \approx -0.260173 \not\geq -0.260112 \approx \trace\tof{L \ln{L}} \sum_{x} L\tof{x,x}^{2} \mathpunctuation{.}
\end{equation}
Thus, inequality~\eqref{eq:laplace-trace-inequality} and therefore also inequality~\eqref{eq:monotonicity-trace-inequality} at $\alpha = 1$ are not satisfied, so that for the corresponding network $\alpha \mapsto \tnorm{p^{\tof{\alpha}}_{\infty}}^{2}$ is not monotonic~(Fig.~\ref{fig:non-monotonicity-block-networks}A).

This counterexample as well as the others we have subsequently found have in common that the underlying networks exhibit a block structure while the degree distribution is approximately uniform. In order to make this precise, we will construct a prototypical example of such a network and demonstrate that these networks do indeed produce non-monotonic behaviour.

Let $d_{1}, d_{2} \geq 2$ and suppose that $d_{1} > d_{2}$. Furthermore, let $\delta = \frac{d_{1} - d_{2}}{d_{1} + d_{2} - 2}$ and define the $\parenth{d_{1},d_{2}}$-block adjacency matrix
\begin{equation}
    A = \begin{pmatrix}
        0 &  & 1-\delta  & \epsilon & \cdots & \epsilon \\
         & \ddots &  & \vdots &  & \vdots \\
        1-\delta &  & 0 & \epsilon & \cdots & \epsilon \\
        \epsilon & \cdots & \epsilon & 0 &  & 1+\delta \\
        \vdots &  & \vdots &  & \ddots &  \\
        \epsilon & \cdots & \epsilon & 1+\delta &  & 0 \\
    \end{pmatrix} \mathpunctuation{.}
    \label{eq:block-network-adjacency-matrix}
\end{equation}

This constitutes an undirected, weighted network with $d_{1} + d_{2}$ nodes. It is connected and the corresponding node degrees in the first an second block are $\parenth{1 - \delta} \parenth{d_{1} - 1} + \epsilon d_{2}$ and $\parenth{1 + \delta} \parenth{d_{2} - 1} + \epsilon d_{1}$, respectively, so that they coincide up to an order of $\epsilon$ since $\parenth{1 - \delta} \parenth{d_{1} - 1} = \parenth{1 + \delta} \parenth{d_{2} - 1}$. Moreover, we can write the corresponding Laplacian matrix as
\begin{equation}
    L = \begin{pmatrix} \parenth{1-\delta} L_{K_{d_{1}}} & 0 \\ 0 & \parenth{1+\delta} L_{K_{d_{2}}} \\ \end{pmatrix} + \epsilon \, L_{K_{d_{1},d_{2}}} =: L_{0} + \epsilon \, L_{K_{d_{1},d_{2}}}
    \label{eq:block-network-laplacian-matrix}
\end{equation}
where $L_{K_{d}}$ and $L_{K_{d,d'}}$ denote the Laplacian matrices corresponding to graphs $K_{d}$ and $K_{d,d'}$, respectively. The former denotes the complete graph on $d$ vertices while the latter denotes the complete bipartite graph on two sets of vertices with size $d$ and $d'$.

Now, for every exponent $0 < \alpha \leq 1$ the map
\begin{equation}
    \Theta_{\alpha} : L \mapsto \trace\tof{L^{\alpha}} \sum_{x} L^{\alpha}\of{x,x} \tparenth{L^{\alpha} \ln{L^{\alpha}}}\of{x,x} - \trace\tof{L^{\alpha} \ln{L^{\alpha}}} \sum_{x} L^{\alpha}\of{x,x}^{2}
\end{equation}
is well-defined and continuous for any self-adjoint matrix $L$ and thus in particular for $L$ a Laplacian matrix. Indeed, the mappings $L \mapsto L^{\alpha}$ and $L \mapsto L^{\alpha} \ln{L^{\alpha}}$ via the functional calculus ($\lambda \mapsto \lambda^{\alpha}$ and $\lambda \mapsto \lambda \ln{ \lambda}$ can be continuously defined on $\reals$ by extending them with $0$ beyond their immediate domain of definition so that one can approximate them via polynomials on a sufficiently large compact set and proceed via a triangle-inequality estimate to show continuity), the trace, the projections $L^{\alpha} \mapsto L^{\alpha}\of{x,x}$ and $L^{\alpha} \ln{L^{\alpha}} \mapsto \parenth{L^{\alpha} \ln{L^{\alpha}}}\of{x,x}$, and, finally, the arithmetic operations are all continuous and therefore also their composition.

With that established, we have that $\Theta_{\alpha}\of{L} \leq \Theta_{\alpha}\of{L_{0}} + \abs{\Theta_{\alpha}\of{L} - \Theta_{\alpha}\of{L_{0}}}$ where, by continuity, the last term can be made arbitrarily small, since $\norm{L - L_{0}} = \epsilon \tnorm{L_{K_{d_{1},d_{2}}}} = \epsilon \max\spectrum{L_{K_{d,d'}}} = \epsilon \parenth{d_{1} + d_{2}}$ \parencite{brouwer2012spectra}. Hence, if $\Theta_{\alpha}\of{L_{0}} < 0$ we also have that $\Theta_{\alpha}\of{L} < 0$ provided that $\epsilon$ is chosen sufficiently small.

Now, observe that, as a consequence of its spectral decomposition, for any complete graph $K_{d}$ and any function $f$ with $f\of{0} = 0$ one has $f\of{\beta L_{K_{d}}} = \frac{f\of{\beta d}}{d} L_{K_{d}}$, so that
\begin{equation}
	f\of{L_{0}} = \begin{pmatrix} f\of{\parenth{1-\delta} L_{K_{d_{1}}}} & 0 \\ 0 & f\of{\parenth{1+\delta} L_{K_{d_{2}}}} \\ \end{pmatrix} = \begin{pmatrix} \frac{f\of{\parenth{1-\delta} d_{1}}}{d_{1}} L_{K_{d_{1}}} & 0 \\ 0 & \frac{f\of{\parenth{1+\delta} d_{2}}}{d_{2}} L_{K_{d_{2}}} \\ \end{pmatrix} \mathpunctuation{.}
\end{equation}
Thus,
\begin{equation}
    \begin{split}
	    \Theta_{\alpha}\of{L_{0}} &= \frac{\alpha}{d_{1} d_{2}}  \parenth{\parenth{1-\delta} d_{1}}^{\alpha} \parenth{d_{1}-1} \parenth{\parenth{1+\delta} d_{2}}^{\alpha} \parenth{d_{2}-1} \\
	    &\mathrelphantom{=} {} \times \parenth{\parenth{\parenth{1-\delta} d_{1}}^{\alpha} \parenth{d_{1}-1} d_{2} - d_{1} \parenth{\parenth{1+\delta} d_{2}}^{\alpha} \parenth{d_{2}-1}} \frac{1}{\alpha} \ln\of{\frac{\parenth{1-\delta} d_{1}}{\parenth{1+\delta} d_{2}}}^{\alpha} \mathpunctuation{.}
	\end{split}
\end{equation}
Here, $\parenth{d_{1}-1} d_{2} - d_{1} \parenth{d_{2}-1} = d_{1} - d_{2} > 0$ so that $\frac{1}{\alpha} \ln\of{\frac{\parenth{1-\delta} d_{1}}{\parenth{1+\delta} d_{2}}}^{\alpha} = - \ln{\frac{\parenth{d_{1}-1} d_{2}}{d_{1} \parenth{d_{2}-1}}} < 0$. Similarly, $\frac{\parenth{d_{1}-1} d_{2} \parenth{\parenth{1-\delta} d_{1}}^{\alpha}}{d_{1} \parenth{d_{2}-1} \parenth{\parenth{1+\delta} d_{2}}^{\alpha}} = \parenth{\frac{\parenth{d_{1}-1} d_{2}}{d_{1} \parenth{d_{2}-1}}}^{1-\alpha} \geq 1$ and, consequently, $\parenth{d_{1}-1} d_{2} \parenth{\parenth{1-\delta} d_{1}}^{\alpha} - d_{1} \parenth{d_{2}-1} \parenth{\parenth{1+\delta} d_{2}}^{\alpha} \geq 0$ with strict inequalities if $\alpha \neq 1$.

Hence, for every $0 < \alpha < 1$ we conclude that $\Theta_{\alpha}\of{L_{0}} < 0$. Now, by continuity, for some $0 < \alpha < 1$ fixed, we get that $\Theta_{\alpha}\of{L} < 0$ for some $\epsilon$ chosen sufficiently small.

By the initial discussion, this shows that there exists an exponent $\alpha$ at which $\alpha \mapsto \tnorm{p^{\tof{\alpha}}_{\infty}}^{2}$ is strictly decreasing contradicting the fact that if it were monotonic it is necessarily non-decreasing and thus demonstrates that indeed monotonicity is violated for networks with a strong block structure~(Fig.~\ref{fig:non-monotonicity-block-networks}B).

\subsubsection*{Monotonic behaviour for star networks}

In contrast to the networks with a block structure, numerical evidence suggests that many other networks indeed produce monotonic behaviour. Here, we will explicitly consider star networks.

Let $d \geq 1$ and consider a star-graph of $d+1$ vertices with adjacency matrix
\begin{equation}
    A = \begin{pmatrix} 0 & 1 & \cdots & 1 \\ 1 & 0 & \cdots & 0 \\ \vdots & \vdots & \ddots & \vdots \\ 1 & 0 & \cdots & 0 \\ \end{pmatrix} \mathpunctuation{.}
\end{equation}

The Laplacian spectrum of such a graph is given as $\set{0, 1, d + 1}$ so that from the corresponding spectral decomposition of the Laplacian matrix we find that for any function $f$
\begin{equation}
    \begin{split}
        f\of{L} &= f\of{0} \Pi_{0} + f\of{1} \parenth{\varid - \Pi_{0} - \Pi_{d+1}} + f\of{d+1} \Pi_{d+1} \\
        &= \parenth{f\of{0} - f\of{1}} \Pi_{0} + f\of{1} \varid + \parenth{f\of{d+1} - f\of{1}} \Pi_{d+1}
    \end{split}
\end{equation}
where $\Pi_{0}$ and $\Pi_{d+1}$ are the orthogonal rank-1 projections onto the linear spaces spanned by $ \parenth{1, \ldots 1}$ and $\parenth{d, -1, \ldots -1}$, the eigenvectors for the eigenvalues $0$ and $d+1$, respectively.

Using that $\Pi_{0}\of{x,x} = \frac{1}{d+1}$ and $\Pi_{d+1}\of{x,x} = \frac{\tparenth{d^{2} - 1} \kroneckerdelta{x}{1} + 1}{d \parenth{d+1}}$, we then have that
\begin{equation}
    \begin{split}
        \Theta_{\alpha}\of{L} &= \trace\tof{L^{\alpha}} \sum_{x} L^{\alpha}\of{x,x} \tparenth{L^{\alpha} \ln{L^{\alpha}}}\of{x,x} - \trace\tof{L^{\alpha} \ln{L^{\alpha}}} \sum_{x} L^{\alpha}\of{x,x}^{2} \\
        &= \frac{\parenth{d-1}^{2}}{\parenth{d+1}^{1-\alpha}} \parenth{\parenth{d+1}^{\alpha} - 1} \ln\of{d+1}^{\alpha} > 0 \\
    \end{split}
\end{equation}
which establishes monotonicity, since, again by the initial discussion, this implies that $\alpha \mapsto \tnorm{p^{\tof{\alpha}}_{\infty}}^{2}$ is non-decreasing.

\subsection{The mean-field equations for irregular networks in the community layer}
\label{sec:appendix-irregular-networks}

When deriving the mean-field equations up to second order, we have assumed that the community layer of the epidemic network is regular, i.e. its nodes all have the same degree. In this section, we will outline how to derive mean-field models in the case when this assumption fails and the community layer is irregular. This can be seen as a generalisation of what we presented so far, however, the analysis also proves slightly more difficult.

As before, the mean-field description of the epidemic dynamics with the transport dynamics frozen in time can be deduced via standard techniques~\parencite{kiss2017mathematics}. Rather than the overall expected number of individuals in a given state of health, one considers the expected number of individuals in a given state of health as well as with a certain degree (in the community layer of the epidemic network) together with the corresponding higher-order motifs. Then, since in the transport process we do not distinguish between the degree of the individuals and their movement is independent of each other, the coupling of the epidemic and the transport dynamics happens entirely analogously to the regular case.

Using the notation of \textcite{kiss2017mathematics}, we write $\tbracket{h_{k}}_{t}$, $\tbracket{h_{k}\link{\lambda}h'_{k'}}_{t}$, and $\tbracket{h_{k}\link{\lambda}h'_{k'}\link{\lambda'}h''_{k''}}_{t}$ for the expected number of individuals in state of health $h$ and degree $k$, the expected number of pairs of individuals in state of health $h$ and $h'$ and degree $k$ and $k'$ connected via a link in layer $\lambda$, and the expected number of triples of individuals in state of health $h$, $h'$, and $h''$ and degree $k$, $k'$, and $k''$ connected via links in layers $\lambda$ and $\lambda'$, respectively. In addition, in order to simplify notation, we set $\tbracket{h_{k}\link{\lambda}h'}_{t} = \sum_{k'} \tbracket{h_{k}\link{\lambda}h'_{k'}}_{t}$ and $\tbracket{h_{k}\link{\lambda}h'_{k'}\link{\lambda'}h''}_{t} = \sum_{k''} \tbracket{h_{k}\link{\lambda}h'_{k'}\link{\lambda'}h''_{k''}}_{t}$. As can be shown along the same lines as before, the transport process amounts to transition terms $\tparenth{\derivative{t} \norm{p_{t}}^{2}} \tbracket{h_{k}}_{t} \tbracket{h'_{k'}}_{t}$ to the expected number of pairs of individuals in state of health $h$ and $h'$ and degree $k$ and $k'$ in the transport layer. Thus, the mean-field equations up to the level of pairs are given as
\begin{equation}
	\left\lbrace
	\begin{split}
		\derivative{t} p_{t} &= -\mu \transpose{\Delta} p_{t} \\
		\derivative{t} \tbracket{\mathrm{S}_{k}}_{t} &= - \sum_{\lambda} \beta^{\lambda} \tbracket{\mathrm{S}_{k}\link{\lambda}\mathrm{I}}_{t} + \sigma \tbracket{\mathrm{R}_{k}}_{t} \\
		\derivative{t} \tbracket{\mathrm{I}_{k}}_{t} &= \sum_{\lambda} \beta^{\lambda} \tbracket{\mathrm{S}_{k}\link{\lambda}\mathrm{I}}_{t} - \gamma \tbracket{\mathrm{I}_{k}}_{t} \\
		\derivative{t} \tbracket{\mathrm{R}_{k}}_{t} &= \gamma \tbracket{\mathrm{I}_{k}}_{t} - \sigma \tbracket{\mathrm{R}_{k}}_{t} \\
		\derivative{t} \tbracket{\mathrm{S}_{k}\link{\omega}\mathrm{S}_{k'}}_{t} &= - \sum_{\lambda} \beta^{\lambda} \tparenth{\tbracket{\mathrm{S}_{k}\link{\omega}\mathrm{S}_{k'}\link{\lambda}\mathrm{I}}_{t} + \tbracket{\mathrm{S}_{k'}\link{\omega}\mathrm{S}_{k}\link{\lambda}\mathrm{I}}_{t}} + \sigma \tparenth{\tbracket{\mathrm{S}_{k}\link{\omega}\mathrm{R}_{k'}}_{t} + \tbracket{\mathrm{S}_{k'}\link{\omega}\mathrm{R}_{k}}_{t}} \\ &\mathrelphantom{=} {} + \tparenth{\derivative{t} \norm{p_{t}}^{2}} \tbracket{\mathrm{S}_{k}}_{t} \tbracket{\mathrm{S}_{k'}}_{t} \kroneckerdelta{\omega}{\text{t.}} \\
		\derivative{t} \tbracket{\mathrm{S}_{k}\link{\omega}\mathrm{I}_{k'}}_{t} &= \sum_{\lambda} \beta^{\lambda} \tparenth{\tbracket{\mathrm{S}_{k}\link{\omega}\mathrm{S}_{k'}\link{\lambda}\mathrm{I}}_{t} - \tbracket{\mathrm{I}_{k'}\link{\omega}\mathrm{S}_{k}\link{\lambda}\mathrm{I}}_{t}} - \beta^{\omega} \tbracket{\mathrm{S}_{k}\link{\omega}\mathrm{I}_{k'}}_{t} - \gamma \tbracket{\mathrm{S}_{k}\link{\omega}\mathrm{I}_{k'}}_{t} + \sigma \tbracket{\mathrm{I}_{k'}\link{\omega}\mathrm{R}_{k}}_{t} \\ &\mathrelphantom{=} {} + \tparenth{\derivative{t} \norm{p_{t}}^{2}} \tbracket{\mathrm{S}_{k}}_{t} \tbracket{\mathrm{I}_{k'}}_{t} \kroneckerdelta{\omega}{\text{t.}} \\
		\derivative{t} \tbracket{\mathrm{I}_{k}\link{\omega}\mathrm{I}_{k'}}_{t} &= \sum_{\lambda} \beta^{\lambda} \tparenth{\tbracket{\mathrm{I}_{k}\link{\omega}\mathrm{S}_{k'}\link{\lambda}\mathrm{I}}_{t} + \tbracket{\mathrm{I}_{k'}\link{\omega}\mathrm{S}_{k}\link{\lambda}\mathrm{I}}_{t}} + \beta^{\omega} \tparenth{\tbracket{\mathrm{S}_{k}\link{\omega}\mathrm{I}_{k'}}_{t} + \tbracket{\mathrm{S}_{k'}\link{\omega}\mathrm{I}_{k}}_{t}} - 2 \gamma \tbracket{\mathrm{I}_{k}\link{\omega}\mathrm{I}_{k'}}_{t} \\ &\mathrelphantom{=} {} + \tparenth{\derivative{t} \norm{p_{t}}^{2}} \tbracket{\mathrm{I}_{k}}_{t} \tbracket{\mathrm{I}_{k'}}_{t} \kroneckerdelta{\omega}{\text{t.}} \\
		\derivative{t} \tbracket{\mathrm{S}_{k}\link{\omega}\mathrm{R}_{k'}}_{t} &= - \sum_{\lambda} \beta^{\lambda} \tbracket{\mathrm{R}_{k'}\link{\omega}\mathrm{S}_{k}\link{\lambda}\mathrm{I}}_{t} + \gamma \tbracket{\mathrm{S}_{k}\link{\omega}\mathrm{I}_{k'}}_{t} + \sigma \tparenth{\tbracket{\mathrm{R}_{k}\link{\omega}\mathrm{R}_{k'}}_{t} - \tbracket{\mathrm{S}_{k}\link{\omega}\mathrm{R}_{k'}}_{t}} \\ &\mathrelphantom{=} {} + \tparenth{\derivative{t} \norm{p_{t}}^{2}} \tbracket{\mathrm{S}_{k}}_{t} \tbracket{\mathrm{R}_{k'}}_{t} \kroneckerdelta{\omega}{\text{t.}} \\
		\derivative{t} \tbracket{\mathrm{I}_{k}\link{\omega}\mathrm{R}_{k'}}_{t} &= \sum_{\lambda} \beta^{\lambda} \tbracket{\mathrm{R}_{k'}\link{\omega}\mathrm{S}_{k}\link{\lambda}\mathrm{I}}_{t} + \gamma \tparenth{\tbracket{\mathrm{I}_{k}\link{\omega}\mathrm{I}_{k'}}_{t} - \tbracket{\mathrm{I}_{k}\link{\omega}\mathrm{R}_{k'}}_{t}} - \sigma \tbracket{\mathrm{I}_{k}\link{\omega}\mathrm{R}_{k'}}_{t} \\ &\mathrelphantom{=} {} + \tparenth{\derivative{t} \norm{p_{t}}^{2}} \tbracket{\mathrm{I}_{k}}_{t} \tbracket{\mathrm{R}_{k'}}_{t} \kroneckerdelta{\omega}{\text{t.}} \\
		\derivative{t} \tbracket{\mathrm{R}_{k}\link{\omega}\mathrm{R}_{k'}}_{t} &= \gamma \tparenth{\tbracket{\mathrm{I}_{k}\link{\omega}\mathrm{R}_{k'}}_{t} + \tbracket{\mathrm{I}_{k'}\link{\omega}\mathrm{R}_{k}}_{t}} - 2 \sigma \tbracket{\mathrm{R}_{k}\link{\omega}\mathrm{R}_{k'}}_{t} + \tparenth{\derivative{t} \norm{p_{t}}^{2}} \tbracket{\mathrm{R}_{k}}_{t} \tbracket{\mathrm{R}_{k'}}_{t} \kroneckerdelta{\omega}{\text{t.}} \\
	\end{split}
	\right.
	\label{eq:SIRS-irregular-mean-field-equations}
\end{equation}
for $\omega \in \set{\text{c.}, \text{t.}}$ and $k$ and $k'$ all possible degree values. For regular networks, this system of equations is the same as the one we have seen earlier in equation~\eqref{eq:SIRS-mean-field-equations}.

\subsubsection*{Moment-closures for first- and second-order mean-field equations}

In order to close this system at the level of pairs and triples to arrive at the first- and second-order mean-field equations, one can apply the following closures. For the first-order equations, the closure relation is given as $\tbracket{\mathrm{S}_{k}\link{\text{c.}}\mathrm{I}_{k'}}_{t} \approx \frac{k k'}{\tanglebr{K} \abs{\mathcal{N}}} \tbracket{\mathrm{S}_{k}}_{t} \tbracket{\mathrm{I}_{k'}}_{t}$ and $\tbracket{\mathrm{S}_{k}\link{\text{t.}}\mathrm{I}_{k'}}_{t} \approx \norm{p_{t}}^{2} \tbracket{\mathrm{S}_{k}}_{t} \tbracket{\mathrm{I}_{k'}}_{t}$ so that
\begin{equation}
	\tbracket{\mathrm{S}_{k}\link{\text{c.}}\mathrm{I}}_{t} \approx \sum_{k'} \frac{k k'}{\tanglebr{K} \abs{\mathcal{N}}} \tbracket{\mathrm{S}_{k}}_{t} \tbracket{\mathrm{I}_{k'}}_{t} \quad \text{and} \quad \tbracket{\mathrm{S}_{k}\link{\text{t.}}\mathrm{I}}_{t} \approx \norm{p_{t}}^{2} \tbracket{\mathrm{S}_{k}}_{t} \tbracket{\mathrm{I}}_{t}
	\label{eq:irregular-pair-closure}
\end{equation}
and for the second-order equations, it is given as
\begin{equation}
	\tbracket{h_{k'}\link{\omega}\mathrm{S}_{k}\link{\omega'}\mathrm{I}}_{t} \approx \parenth{1 - \frac{\kroneckerdelta{\omega}{\omega'}}{\kappa_{k}^{\text{c.}}\of{p}}} \frac{\tbracket{\mathrm{S}_{k}\link{\omega}h_{k'}}_{t} \tbracket{\mathrm{S}_{k}\link{\omega'}\mathrm{I}}_{t}}{\tbracket{\mathrm{S}_{k}}_{t}}
	\label{eq:irregular-triple-closure}
\end{equation}
where $\kappa_{k}^{\text{c.}}\of{p} = k$ and $\kappa_{k}^{\text{t.}}\of{p} = \norm{p}^{2} \abs{\mathcal{N}}$.

For irregular epidemic networks, these closures are well-known~\parencite{kiss2017mathematics} and their generalisation to the multilayer network we are considering here follows along the same lines as before, using that the transport layer is in expectation regular. Note that we denote the moments of the degree distribution as $\anglebr{K^{z}} = \sum_{k} \frac{\abs{\mathcal{N}_{k}}}{\abs{\mathcal{N}}} k^{z}$ where $\mathcal{N}_{k}$ denotes the set of individuals with degree $k$ in the community layer.

\begin{figure}[tb]
	\centering

	\begin{grid}{p{12cm}p{\linewidth-12cm}}
		{
			\begin{minipage}{\linewidth}
			
				\textbf{\Large A}

				\vspace{-0.5cm}
				\begin{grid}{p{0.7cm}*{3}{>{\centering\arraybackslash}p{3.7cm}}}
					{} & {\small SIS-epidemic} & {\small SIRS-epidemic} & {\small SIR-epidemic} \\
				\end{grid}
		
				\begin{grid}{p{4.4cm}*{2}{p{3.7cm}}}
					\includegraphics[scale=1,trim={0cm 1cm 0 0},clip]{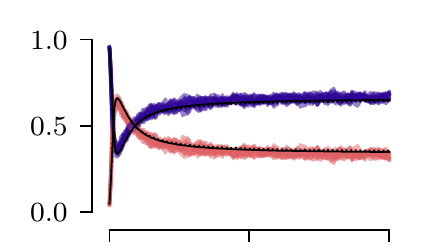}
					&
					\includegraphics[scale=1,trim={0.7cm 1cm 0 0},clip]{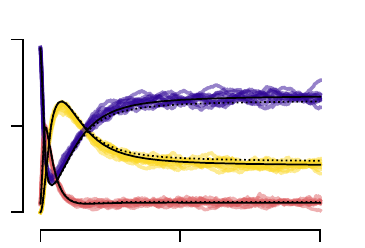}
					&
					\includegraphics[scale=1,trim={0.7cm 1cm 0 0},clip]{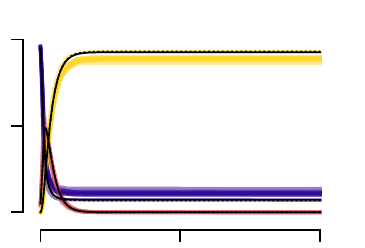}
					\\
				\end{grid}
				
				\vspace{1em}
				
				\textbf{\Large B}
	
				\begin{grid}{p{4.4cm}*{2}{p{3.7cm}}}
					\includegraphics[scale=1,trim={0cm 0cm 0 0},clip]{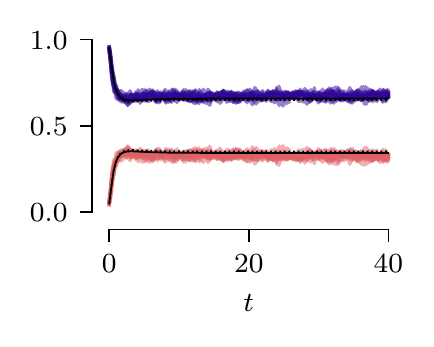}
					&
					\includegraphics[scale=1,trim={0.7cm 0cm 0 0},clip]{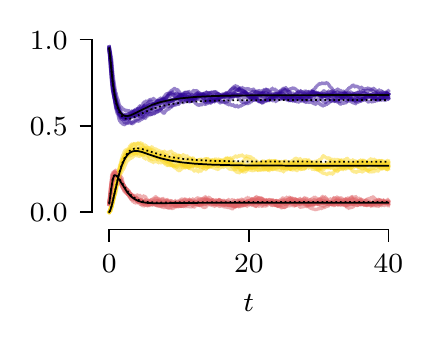}
					&
					\includegraphics[scale=1,trim={0.7cm 0cm 0 0},clip]{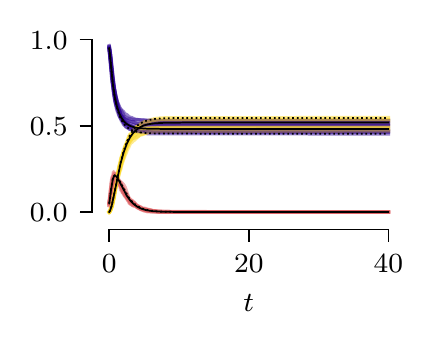}
					\\
				\end{grid}
				
			\end{minipage}
		}
		&
		{
			\vspace*{-3.75cm}
			\begin{minipage}{\linewidth}
				
				\includegraphics[scale=1,trim={0 0.5cm 0 0.5cm},clip]{figures/3-X}
				
				\includegraphics{figures/3-L}
	
			\end{minipage}
		}
		\\
	\end{grid}

	\caption{\textbf{Comparison between stochastic trajectories of the model and first- and second-order mean-field solutions.} 
	Stochastic trajectories together with their corresponding mean-field solutions for the fractions of susceptibles, infected, and recovered are shown for SIS-, SIRS- and SIR-epidemic dynamics in a population of \num{1000} individuals in a network with \num{900} nodes of degree \num{3} and \num{100} nodes of degree \num{73} as the community layer of the epidemic network and the Munich \textit{U-Bahn} network that consists of approximately \num{100} sites as the transport network. The epidemic parameters are set to $\beta^{\text{c.}} = \frac{1}{6}$, $\beta^{\text{t.}} = \frac{1}{20} \beta^{\text{c.}}$, $\gamma = 1$, and, in case of SIRS-dynamics, $\sigma = \frac{1}{5}$. The mobility rate is varied, with $\mu = 1$ in \textbf{A} and $\mu = 10$ in \textbf{B}. In each case, the mean initial prevalence is set to \SI{5}{\percent} and, at $t = 0$, the individuals are all located at a single site (\enquote{Marienplatz}) on transport network. The stochastic trajectories are shown without having been time-shifted.}
	\label{fig:numerical-simulation-irregular-comparison}
\end{figure}

Applying these closure relations, specifically the first-order mean-field equations are given as
\begin{equation}
	\left\lbrace
	\begin{split}
		\derivative{t} p_{t} &= -\mu \transpose{\Delta} p_{t} \\
		\derivative{t} \tbracket{\mathrm{S}_{k}}_{t} &= - \sum_{k'} \parenth{\beta^{\text{c.}} \frac{k k'}{\tanglebr{K} \abs{\mathcal{N}}} + \beta^{\text{t.}} \norm{p_{t}}^2} \tbracket{\mathrm{I}_{k'}}_{t} \tbracket{\mathrm{S}_{k}}_{t} + \sigma \tbracket{\mathrm{R}_{k}}_{t} \\
		\derivative{t} \tbracket{\mathrm{I}_{k}}_{t} &= \sum_{k'} \parenth{\beta^{\text{c.}} \frac{k k'}{\tanglebr{K} \abs{\mathcal{N}}} + \beta^{\text{t.}} \norm{p_{t}}^2} \tbracket{\mathrm{I}_{k'}}_{t} \tbracket{\mathrm{S}_{k}}_{t} - \gamma \tbracket{\mathrm{I}_{k}}_{t} \\
		\derivative{t} \tbracket{\mathrm{R}_{k}}_{t} &= \gamma \tbracket{\mathrm{I}_{k}}_{t} - \sigma \tbracket{\mathrm{R}_{k}}_{t} \\
	\end{split}
	\right.
	\label{eq:SIRS-1st-order-irregular-mean-field-equations}
\end{equation}
for all degree values $k$ with initial conditions $p_{0}$ some probability distribution on the transport network and $\tbracket{\mathrm{S}_{k}}_{0}$, $\tbracket{\mathrm{I}_{k}}_{0}$, and $\tbracket{\mathrm{R}_{k}}_{0}$ such that $\abs{\mathcal{N}_{k}} = \tbracket{\mathrm{S}_{k}}_{0} + \tbracket{\mathrm{I}_{k}}_{0} + \tbracket{\mathrm{R}_{k}}_{0}$ for every degree value $k$. Similar to the regular case, since $\derivative{t} \sum_{h} \tbracket{h_{k}}_{t} = 0$ for every $k$ the total number of individuals is a conserved quantity, so that the total number individuals with degree $k$ as a constant of motion can be used to derive a reduced system of differential equations that equivalently describe the system by eliminating one of the sets of equations for the susceptible, infected, and recovered individuals.

The second-order mean-field equations can be obtained analogously by applying the moment-closure from above. Moreover, as in the regular case, the equations for SIS- and SIR-epidemic dynamics can be derived by taking again the limits $\sigma \to \infty$ and $\sigma \to 0$, respectively.

Finally, a numerical comparison between stochastic trajectories and the corresponding first- and second-order mean-field solutions for SIRS-, SIS- and SIR-epidemic dynamics in a population with \num{1000} individuals in a \num{10}-regular network as the community layer of the epidemic network and the Munich \textit{U-Bahn} network as the transport network shows overall good agreement between the stochastic trajectories and the mean-field solutions~(Fig.~\ref{fig:numerical-simulation-irregular-comparison}).

\subsubsection*{The epidemic threshold for irregular community layer topologies}

As for the epidemic threshold, let
\begin{equation}
	\chi\of{p} = \frac{1}{2} \parenth{ \beta^{\text{c.}} \frac{\tanglebr{K^{2}}}{\tanglebr{K} \abs{\mathcal{N}}} + \beta^{\text{t.}} \norm{p}^{2}  + \sqrt{ \parenth{\beta^{\text{c.}} \frac{\tanglebr{K^{2}}}{\tanglebr{K} \abs{\mathcal{N}}} - \beta^{\text{t.}} \norm{p}^{2}}^{2} + 4 \beta^{\text{c.}} \beta^{\text{t.}} \frac{\tanglebr{K} \norm{p}^{2}}{\abs{\mathcal{N}}} } } \frac{\abs{\mathcal{N}}}{\gamma} \mathpunctuation{.}
\end{equation}	
Then, the system undergoes a transcritical bifurcation when $\chi\of{p_{\infty}} = 1$ where $p_{\infty}$ is the unique solution to the equation $\transpose{\Delta} p_{\infty} = 0$ with $p_{\infty}\of{x} \geq 0$ and $\sum_{x} p_{\infty}\of{x} = 1$. Moreover, for $\chi\of{p_{\infty}} < 1$ the disease-free state $\parenth{p_{\infty}, \ldots, \abs{\mathcal{N}_{k}}, \ldots, 0, 0}$ is a stable equilibrium point.

Indeed, again due to the strongly connected transport network there is a unique solution $p_{\infty}$ of $\derivative{t} p_{t} = - \mu \transpose{\Delta} p_{t} = 0$ with $p_{\infty} \geq 0$ and $\sum p_{\infty} = 1$, the equilibrium solution of the random walk on the transport network. From there, consider the reduced system of mean-field equations that arises from \eqref{eq:SIRS-1st-order-irregular-mean-field-equations} leaving the diffusion equation aside and eliminating the equations for the susceptibles using the relation $\tbracket{\mathrm{S}_{k}}_{t} = \abs{\mathcal{N}_{k}} - \tbracket{\mathrm{I}_{k}}_{t} + \tbracket{\mathrm{R}_{k}}_{t}$ for every $k$. One immediately verifies that the disease-free state with $\tbracket{\mathrm{I}_{k}}_{t} = 0 = \tbracket{\mathrm{R}_{k}}_{t}$ for every $k$ is an equilibrium. At this state, the Jacobian of the (reduced) system takes the form $\parenth{\begin{smallmatrix} \diag\of{\abs{\mathcal{N}_{k}}}_{k} B - \gamma \varid & 0 \\ \gamma \varid & -\sigma \varid \\ \end{smallmatrix}}$ with $B$ such that $B_{k,k'} = \beta^{\text{c.}} \frac{k k'}{\tanglebr{K} \abs{\mathcal{N}}} + \beta^{\text{t.}} \norm{p_{\infty}}^{2}$. Due to its block structure, we have that
\begin{equation}
	\spectrum{\parenth{\begin{matrix} \diag\of{\abs{\mathcal{N}_{k}}}_{k} B - \gamma \varid & 0 \\ \gamma \varid & -\sigma \varid \\ \end{matrix}}} = \set{-\sigma} \cup \parenth{\spectrum{\diag\of{\abs{\mathcal{N}_{k}}}_{k} B} - \gamma} \mathpunctuation{.}
	\label{eq:SIRS-1st-order-irregular-mean-field-jacobian-1}
\end{equation}

In order to compute the remaining eigenvalues, note that for an $m \times m$ square-matrix $W$ with $W_{i,i'} = q_{i} \parenth{a_{i} a_{i'} + c}$ where $q_{i} \geq 0$ for every $i$, $\sum_{i} q_{i} = 1$, and $c > 0$,
\begin{equation}
	\spectrum{W} = \set{0, \frac{1}{2}\parenth{ \parenth{\tanglebr{A^{2}} + c} \pm \sqrt{\parenth{{\tanglebr{A^{2}}} - c}^{2} + 4 c {\tanglebr{A}}^{2}} }} \subset \reals \quad \text{where ${\langle A^{z} \rangle} = \sum_{i} q_{i} a_{i}^{z}$.}
\end{equation}
Indeed, by elementary row manipulations where for every $i = 1, \ldots m - 2$ we add $- \frac{a_{m} - a_{i}}{a_{m} - a_{m-1}} \frac{q_{i}}{q_{m-1}}$ times row $m-1$ and $\frac{a_{m-1} - a_{i}}{a_{m} - a_{m-1}} \frac{q_{i}}{q_{m}}$ times row $m$ to row $i$ and subsequently add $\frac{1}{\mu} q_{m-1} \parenth{a_{m-1} a_{i} + c}$ times row $i$ to row $m-1$ and $\frac{1}{\mu} q_{m} \parenth{a_{m} a_{i} + c}$ times row $i$ to row $m$ we find that
\begin{equation}
	\det\of{W - \mu \varid} = \det{\begin{pmatrix} -\mu \varid & * \\ 0 & \interior{W} - \mu \varid \end{pmatrix}} \quad \text{with $\interior{W} = \tfrac{1}{a_{m} - a_{m-1}} \parenth{\begin{smallmatrix} w_{m-1,m} & -\frac{q_{m-1}}{q_{m}} w_{m-1,m-1} \\ \frac{q_{m}}{q_{m-1}} w_{m,m} & -w_{m,m-1} \end{smallmatrix}} $}
\end{equation}
where $w_{i,i'} = - a_{i} \tanglebr{K^{2}} + \parenth{a_{i} a_{i'} - c} \tanglebr{K} + a_{i'} c$. Thus, one finally has that
\begin{equation}
	\det\of{W - \mu \varid} = \parenth{-\mu}^{m-2} \det\of{\interior{W} - \mu \varid} = \parenth{-\mu}^{m-2} \parenth{\mu^{2} - \parenth{\tanglebr{A^{2}} + c} \mu + \tparenth{\tanglebr{A^{2}} - \tanglebr{A}^{2}} c}
\end{equation}
which yields the assertion upon computing the roots of this polynomial.

Now, observe that the matrix $\diag\of{\abs{\mathcal{N}_{k}}}_{k} B$ appearing in the Jacobian above is of a similar form as the one of the matrix $W$ in the statement above. In fact, $\parenth{\diag\of{\abs{\mathcal{N}_{k}}}_{k} B}_{k,k'} = \frac{\beta^{\text{c.}}}{\tanglebr{K}}  \frac{\abs{\mathcal{N}_{k}}}{\abs{\mathcal{N}}} \parenth{k k' + \frac{\beta^{\text{t.}}}{\beta^{\text{c.}}} \tanglebr{K} \abs{\mathcal{N}} \norm{p_{\infty}}^{2}}$ with $\sum_{k} \frac{\abs{\mathcal{N}_{k}}}{\abs{\mathcal{N}}} = 1$. With that and continuing from equation \eqref{eq:SIRS-1st-order-irregular-mean-field-jacobian-1}, we find that
\begin{equation}
	\begin{split}
		&\spectrum{\parenth{\begin{matrix} \diag\of{\abs{\mathcal{N}_{k}}}_{k} B - \gamma \varid & 0 \\ \gamma \varid & -\sigma \varid \\ \end{matrix}}} = \set{-\sigma} \cup \parenth{\spectrum{\diag\of{\abs{\mathcal{N}_{k}}}_{k} B} - \gamma} \\
		&= \set{-\sigma, -\gamma, \frac{\beta^{\text{c.}} \tanglebr{K^{2}}}{2 \tanglebr{K}} + \frac{\beta^{\text{t.}} \abs{\mathcal{N}} \norm{p_{\infty}}^{2}}{2}  \pm \sqrt{\parenth{\frac{\beta^{\text{c.}} \tanglebr{K^{2}}}{2 \tanglebr{K}} - \frac{\beta^{\text{t.}} \abs{\mathcal{N}} \norm{p_{\infty}}^{2}}{2}}^{2} + \beta^{\text{c.}} \beta^{\text{t.}} \tanglebr{K} \abs{\mathcal{N}} \norm{p_{\infty}}^{2} } - \gamma} \mathpunctuation{.}
	\end{split}
	\label{eq:SIRS-1st-order-irregular-mean-field-jacobian-2}
\end{equation}

These eigenvalues are all real-valued and they are strictly negative as long as
\begin{equation}
	\frac{\beta^{\text{c.}} \tanglebr{K^{2}}}{2 \tanglebr{K}} + \frac{\beta^{\text{t.}} \abs{\mathcal{N}} \norm{p_{\infty}}^{2}}{2} + \sqrt{\parenth{\frac{\beta^{\text{c.}} \tanglebr{K^{2}}}{2 \tanglebr{K}} - \frac{\beta^{\text{t.}} \abs{\mathcal{N}} \norm{p_{\infty}}^{2}}{2}}^{2} + \beta^{\text{c.}} \beta^{\text{t.}} \tanglebr{K} \abs{\mathcal{N}} \norm{p_{\infty}}^{2} } - \gamma < 0
\end{equation}
or equivalently
\begin{equation}
	\frac{1}{2} \parenth{ \beta^{\text{c.}} \frac{\tanglebr{K^{2}}}{\tanglebr{K} \abs{\mathcal{N}}} + \beta^{\text{t.}} \norm{p_{\infty}}^{2}  + \sqrt{ \parenth{\beta^{\text{c.}} \frac{\tanglebr{K^{2}}}{\tanglebr{K} \abs{\mathcal{N}}} - \beta^{\text{t.}} \norm{p_{\infty}}^{2}}^{2} + 4 \beta^{\text{c.}} \beta^{\text{t.}} \frac{\tanglebr{K} \norm{p_{\infty}}^{2}}{\abs{\mathcal{N}}} } } \frac{\abs{\mathcal{N}}}{\gamma} < 1
\end{equation}
which completes the proof of the statement.

Finally, since $\tanglebr{K^{2}} \geq \tanglebr{K}^{2}$ with equality only if the topology is regular we have that $\chi\of{p_{\infty}} \geq \chi\of{p_{\infty}}\restrict{\tanglebr{K^{2}} = \tanglebr{K}^{2}}$.
In terms of a comparison with the case of a regular community layer, this means that an irregular topology lowers the epidemic threshold which is already known in the absence of transport. Moreover, keeping the mean of the degree distribution, $\tanglebr{K}$, constant, we find that $\chi\of{p_{\infty}}$ is in fact monotonic in the variance, $\tanglebr{K^{2}} - \tanglebr{K}^{2}$, since
\begin{equation}
	\derivative{\tanglebr{K^{2}} - \tanglebr{K}^{2}} \chi\of{p_{\infty}} = \frac{1}{2} \frac{\beta^{\text{c.}}}{\gamma \tanglebr{K}} \parenth{1 + \frac{\beta^{\text{c.}} \frac{\tanglebr{K^{2}}}{\tanglebr{K} \abs{\mathcal{N}}} - \beta^{\text{t.}} \norm{p_{\infty}}^{2}}{\sqrt{ \parenth{\beta^{\text{c.}} \frac{\tanglebr{K^{2}}}{\tanglebr{K} \abs{\mathcal{N}}} - \beta^{\text{t.}} \norm{p_{\infty}}^{2}}^{2} + 4 \beta^{\text{c.}} \beta^{\text{t.}} \frac{\tanglebr{K} \norm{p_{\infty}}^{2}}{\abs{\mathcal{N}}} }}} \geq 0 \mathpunctuation{.}
\end{equation}

\clearpage

\end{document}